\documentclass[11pt,a4,twoside]{article}

\usepackage{./styling/style-main}
\graphicspath{{Plots/}}
\usepackage{framed}

\newcommand{\tLO}{\text{LO}}
\newcommand{\tNLO}{\text{NLO}}
\newcommand{\tNNLO}{\text{NNLO}}
\newcommand{\Det}{\text{Det}}

\title{Higher-Order Corrections to the Bubble-Nucleation Rate at Finite Temperature}

\author{Andreas~Ekstedt\thanks{andreas.ekstedt@desy.de}\textsuperscript{~~,\,a}}
\affil{a:~Department of Physics and Astronomy, Uppsala University, P.O. Box 256, SE-751 05 Uppsala, Sweden}

\date{\today}

\begin{document}
	\maketitle
	\thispagestyle{plain}

	\begin{abstract}
In this paper I discuss how to consistently incorporate higher-order corrections to the bubble-nucleation rate at finite temperature. Doing so I examine the merits of different approaches, with the goal of reducing uncertainties for gravitational-wave calculations. To be specific, the region of applicability and accuracy of the derivative expansion is discussed. The derivative expansion is then compared to a numerical implementation of the Gelfand-Yaglom theorem. Both methods are applied to popular first-order phase transition models, like a loop-induced barrier and a SM-EFT tree-level barrier. The results of these calculations are presented in easy-to-use parametrizations that can directly be used in gravitational-wave calculations. In addition, higher-order corrections for models with multiple scalar fields, such as singlet/triplet extensions, are studied. Lastly, the main goal of this paper is to investigate the convergence and uncertainty of all calculation. Doing so I argue that current calculations for the Standard Model with a tree-level barrier are inaccurate.
	\end{abstract}

	\section{Introduction}\label{sec:Introduction}
Gravitational waves offer a new way to study primordial phase transitions. And the area is increasingly active~\cite{Gould:2021oba,Niemi:2021qvp,Schicho:2021gca,Gould:2019qek,Postma:2020toi,Croon:2020cgk,Cutting:2020nla,Schmitz:2020rag,Giese:2020znk,Guo:2021qcq,Guo:2020grp,Cutting:2019zws,Bell:2020gug,Chiang:2017nmu,Giese:2020rtr,Bruggisser:2018mrt,Aggarwal:2020olq,Hindmarsh:2020hop,Hindmarsh:2019phv,Caprini:2019pxz} after the discovery of gravitational waves by the LIGO collaboration~\cite{Abbott:2016blz}. As such, much work is spent looking at the Electroweak phase transition. Being the pivotal moment when the Higgs field broke the Electroweak symmetry. What's more, the Electroweak phase transition can perchance explain the observed Baryon asymmetry. 

Yet little is known about the transition as it took place mere nanoseconds after the Big Bang. Far too early for light to escape. With direct observations tricky, other methods are needed to shed light on the phase transition\te like using gravitational waves.

To wit, if the transition is strongly first-order, then, akin to boiling water, bubbles nucleate and can create turbulence in the primordial plasma: Generating gravitational waves~\cite{Hindmarsh:2020hop,Caprini:2019egz,Hindmarsh:2017gnf,Caprini:2006jb,PhysRevD.92.123009,Jinno:2016vai}. These bubbles are also crucial for generating a Baryon asymmetry. The idea is that, with adequate charge-parity violation, left- and right-handed particles interact with the bubble walls differently. This, together with Baryon violating processes, can create an asymmetry~\cite{Kuzmin:1985mm,Shaposhnikov:1986jp,Shaposhnikov:1987tw,Cohen:1993nk}.

However, this relies on a strong first-order transition in the early universe. But it's unknown if such a transition took place. For one, the Standard Model with the measured Higgs mass can't provide a first-order phase transition by itself~\cite{Kajantie:1996mn,Gurtler:1997hr,Csikor:1998eu}. New physics is needed. For another, given a first-order phase transition, current calculations of the bubble-nucleation rate leave much to be desired\te corrections to the leading-order rate are largely unexplored. These considerations take on a new light since higher-order corrections to the nucleation rate is one of the main uncertainties for gravitational wave production~\cite{Gould:2021oba}. What's more, barring these issues, there are additional uncertainties for predictions of the gravitational wave spectrum~\cite{Guo:2021qcq,Croon:2020cgk,Moore:2000jw,Caprini:2019egz}. Addressing these issues is vital, for if gravitational waves from a phase transition are discovered, accurate predictions are a must to identify the underlying theory.
 
So with major theoretical challenges and upcoming gravitational wave experiments~\cite{Harry:2006fi,Kawamura:2006up,Guo:2018npi,2017arXiv170200786A}, there's a need to push the theory forward.

One of the main hurdles for robust calculations comes from that bubble nucleation is a tunneling process\te intertwined with thermal escape\te and is non-perturbative in nature. Although tunneling is theoretically well-understood in quantum mechanics, much less is known in quantum field theory. In practice the only viable approach is to use saddle-point approximations. In short, the path integral is approximated by a field configuration obeying the classical equations of motion. For tunneling this is the bounce solution~\cite{Coleman:1977py,Callan:1977pt}. To leading order the nucleation rate is controlled by the exponent of the action evaluated on the bounce. Further, Gaussian fluctuations around the bounce results in functional determinants. These determinants give sub-leading corrections to the leading-order bounce action.

These calculations are quite involved even at zero temperature, yet, there are further complications at finite temperature. Because once a bubble has nucleated at zero-temperature, it can grow unimpeded; with a velocity rapidly approaching the speed of light~\cite{Coleman:1977py}. This is not the case at finite temperature. As bubbles nucleate they are slowed down by collisions in the hot plasma\te making the probability for a bubble to nucleate, and grow, depend on dissipation processes~\cite{Moore:2000jw,PhysRevA.8.3230,Langer:1969bc}.

To untangle the calculations it's useful to factorize dissipation effects from pure tunneling ones. This factorization can be achieved if the temperature is sufficiently large. Because temperature effects can then be integrated out~\cite{Kajantie:1995dw,Farakos:1994kx,Braaten:1995cm,PhysRevD.23.2305,GINSP}. In essence the nucleation rate factors into a dynamical and a statistical part. The dynamical factor depends on real-time dynamics and incorporates bubble growth and damping; the statistical one is equivalent to the zero-temperature tunneling calculation in three dimensions. 

Both the statistical and dynamical factors can be studied on the lattice~\cite{Moore:2000jw}. Yet in practice Lattice simulations are, too, slow for large parameter scans. Not to mention that analytical techniques often give further insight into the underlying physics.

Though, analytical methods have their share of problems. For one, calculations are often only performed at leading order. For another, there's an ongoing debate of how to incorporate higher-order corrections; a great many proposals exist in the literature~\cite{Croon:2020cgk,Garny:2012cg,Strumia:1998nf,Baacke:1993ne,PhysRevD.47.5304,Dine:1992wr,Gould:2021ccf}. And it can be hard to gauge the accuracy of these methods.

Accordingly, this paper vies to lay the groundwork for consistently incorporating higher-order corrections to the bubble-nucleation rate. These corrections are described by Feynman diagrams in the bounce background. The bounce solution varies over space-time, and as such it is in general impossible to calculate the diagrams exactly. To combat this, it's possible to either use a derivative expansion or numerical methods. The derivative expansion is intuitive, but in practice often not well-behaved. This, combined with a number of conflicting implementations of the derivative expansion, makes it hard to assess its accuracy. 

So in this paper I discuss when, and how, the derivative expansion can be used. In particular, I compare the derivative expansion with exact numerical computations using the Forman/Gelfand-Yaglom theorem~\cite{Forman1987,Gelfand:1959nq,Kirsten:2004qv,Kirsten:2010eg} for a variety of models. The exact numerical results can be used as inputs for gravitational-wave calculations. To facilitate this, the calculations are summarized in easy-to-use parametrizations. 

Besides numerics, I discuss how perturbative calculations sometimes converge slowly. In short, unless one is careful, large uncertainties might be introduced. 

The paper is organized as follows. In the following section I review effective potential and saddle-point-approximation methods. The next section discusses the derivative expansion and shows when it's applicable. Next, methods for numerical evaluations of functional determinants are discussed and applied to popular Standard Model extensions. The results are then summarized in the discussion. 

    \section{Phase Transitions And Perturbative Methods}\label{sec:PhaseTransitions}
When studying phase transitions the main goal, from the theory side, is to calculate various observables of interest. And there are typically two kinds of observables:  equilibrium and non-equilibrium observables. Or in the language of thermal field theory, imaginary and real-time observables.

Equilibrium calculations, in this context, rely on the effective potential, while non-equilibrium methods are based on semi-classical approximations and kinetic theory~\cite{Arnold:1998cy,Arnold:1997yb,Arnold:1996dy}. Both equilibrium and non-equilibrium processes have several important scales associated with them.

In the case of bubble nucleation, there are two temperature scales of particular importance. The first is the critical temperature, $T_c$, defined as the temperature when two phases coincide in energy. The second is the percolation temperature, $T_p$, which is the temperature where roughly two-thirds of the Universe are in the broken phase.  

Calculations at the critical temperature involves equilibrium physics and can be described by the effective potential. To improve convergence, and avoid large logarithms, it is useful to work with a dimensionally reduced theory in three dimensions. All temperature dependence is here contained in effective couplings~\cite{Niemi:2021qvp,Schicho:2021gca,Gould:2019qek,Kajantie:1995dw,Farakos:1994kx,Karjalainen:1996rk,Laine:1994bf}. Quantities like the latent heat, speed of sound, entropy density, and heat capacity are straightforward to calculate in the effective-field-theory framework.

The situation isn't as clear at the percolation temperature\te owing to the non-perturbative nature of bubble nucleation. The bubble nucleation rate is schematically~\cite{Linde:1981zj,Affleck:1980ac}
\begin{align}
\Gamma=A~e^{-S_3}\left[1+\mathcal{O}\left(\hbar\right)\right],
\end{align}
where $S_3$ is the dimensionally reduced action evaluated on the bounce, and $A$ is the exponential pre-factor. The temperature dependence is here left implicit.

Connected to bubble nucleation is the inverse time-duration of the phase transition evaluated at the percolation temperature:
\begin{align}
\beta_p\approx \left. -H(T) T \frac{d}{dT}\log \Gamma(T)\right\vert_{T=T_p},
\end{align}
 with $H(T)$ being the Hubble parameter. It is further assumed that $\beta_p \gg H(T_p)$. The percolation temperature is approximately given by~\cite{Caprini:2019egz,Guo:2021qcq}
\begin{align}
S_3(T_p)\approx 131+\log\left(\frac{A}{T_p^4}\right)-4\log\left(\frac{T_p}{100~\si{\giga\electronvolt}}\right)-4\log\left(\frac{\beta_p}{100 H_p}\right)+3 \log v_w,
\end{align} 
where $v_w$ is the bubble-wall velocity. Once the percolation temperature is known, it's possible to calculate, amongst other, the latent heat and phase-transition's duration. These observables can then be fed into gravitational-wave simulations.
The type of gravitational wave tends to separated into those coming from bubble collision, sound waves, and magneto-hydrodynamic turbulence \cite{Caprini:2006jb,Hindmarsh:2017gnf,PhysRevD.92.123009,Jinno:2016vai,Hindmarsh:2013xza}

 Since the goal is to have accurate predictions of various observables at the percolation temperature, it is necessary to have robust calculations of the exponential pre-factor and the bounce action $S_3$.

\subsection{The Effective Potential}\label{subsec:EffectivePotential}
The effective potential is a generalization of the classical potential that incorporates thermal and quantum corrections. To make things lucid, start off at zero temperature. The effective potential is of the form
\begin{align}
\mathcal{V}_\text{eff}(\phi)=V_0(\phi)+ V_1(\phi)+ V_2(\phi)+\ldots.
\end{align}
Here $V_0$ is the tree-level potential, $V_1$ is the 1-loop potential, and so forth.
While the perturbative expansion is often organized in powers of loops, this is not always the case. To be sure, there are situations when $V_1(\phi)$ is on-par with $V_0(\phi)$. A famous example is the Coleman-Weinberg model where the scalar quartic coupling, $\lambda$, is related to the vector-coupling via $\lambda \sim g^4$~\cite{PhysRevD.7.1888}. Because $V_0(\phi)\sim \lambda \phi^4$ and $V_1(\phi)\sim g^4 \phi^4$, both terms are of the same order in the power-counting $g$.
As such, I will apply a more apt notation:

\begin{align}
\label{equation:EffectivePotential}
\mathcal{V}_\text{eff}(\phi)=V_\text{LO}(\phi)+x V_\text{NLO}(\phi)+x^2 V_\text{NNLO}(\phi)+\ldots,
\end{align}
where the leading-order result can contain terms from higher loops. Factors of $x$ are only inserted to show the relative size of terms. For example, in the Standard Model the usual power-counting is $x\sim y_t^2\sim g^2\sim g'^2\sim \lambda$.

The situation is similar at finite temperature. Once again the effective potential can be organized as
\begin{align}
\label{equation:EffectivePotential}
\mathcal{V}_\text{eff}(\phi,T)=V_\text{LO}(\phi,T)+x V_\text{NLO}(\phi,T)+x^2 V_\text{NNLO}(\phi,T)+\ldots,
\end{align}
with the crux that each term can depend on the temperature.

This expansion, while consistent, tends to converge slowly at large temperatures. There are several reasons for this. One contributing factor is that, for example, scalar self-energies are of the same order as tree-level masses. This is because these self-energies scale as $g^2 T^2$ for some coupling $g$.  Another problem comes  from large logarithms of the form $\log\left(\frac{T^2}{\mu^2}\right)$ and $\log\left(\frac{m^2}{\mu^2}\right)$; $\mu$ is here the renormalization-group scale and $m$ is an arbitrary mass, for example the field-dependent Higgs mass. If there is an hierarchy between $T$ and $m$ there's no choice of $\mu$ that makes both logarithms small. This is a familiar problem. And the solution is to integrate out the high-energy modes with energy $E\sim T$. This gives an effective field theory describing low-energy modes with a typical energy $E\sim m$. The effective, dimensionally reduced, theory is a regular quantum field theory in three Euclidean dimensions~\cite{Kajantie:1995dw,Farakos:1994kx,Karjalainen:1996rk,Laine:1994bf}. 

Even when using a dimensionally reduced theory, it should be noted that non-perturbative effects are important when the theory contains non-abelian vector bosons\cite{Linde:1980ts}. However, these effects first appear at four loops, and are beyond the scope of this paper.

As an example of dimensional reduction, consider a real-scalar field with potential
\begin{align}
	V(\phi)=\frac{1}{2}m^2_{4d} \phi^2+\frac{1}{4}\lambda_{4d} \phi^4.
\end{align}
Couplings and masses in the three-dimensional theory are different from the original four-dimensional ones. Taking the above potential, the leading-order result is\cite{Gould:2021dzl}
\begin{align}
	m^2_{3d}=m^2_{4d}+\frac{1}{4}\lambda_{4d}T^2, \quad \lambda_{3d}=T \lambda_{4d}.
\end{align}
The relations are similar at higher orders, but also involve logarithms depending on the matching-scale and temperature. Because this effective theory lives in three dimensions, the perturbative expansion is organized differently. For example, in four dimensions the one-loop effective potential scales as $\lambda_{4d}^2$, while in three dimensions it scales as $\lambda_{3d}^{3/2}$.

For the rest of this paper, I assume a dimensionally reduced theory. This means that all temperature dependence is left implicit in couplings. I also assume that all temporal vector-bosons are integrated out. That is to say, I work at the super-soft scale~\cite{Kajantie:1995dw,Farakos:1994kx,Karjalainen:1996rk,Laine:1994bf}. 

Once the effective potential is known, other thermodynamical quantities are easy to find. For example, let's assume two phases (minima of $\mathcal{V}_\text{eff}(\phi,T)$)  with energies~$\mathcal{V}_\text{eff}^{(1)}(T)$ and $\mathcal{V}_\text{eff}^{(2)}(T)$ respectively. The critical temperature is defined by
\begin{align}
\left. \Delta \mathcal{V}(T) \right\vert_{T=T_c}=\left[\mathcal{V}_\text{eff}^{(1)}(T)-\mathcal{V}_\text{eff}^{(2)}(T)\right]_{T=T_c}=0.
\end{align}
Because the temperature dependence is hidden in effective couplings, it's in practice easier to solve for one of the leading-order-potential parameters. Say the critical mass $m_c^2$.

Other important quantities are the difference in entropy (density) $s(T)= -\frac{d}{d T}\Delta \mathcal{V}(T)$ and the trace anomaly $\Delta(T)=\left[T s(T)/4+\Delta \mathcal{V}(T) \right]$.

Note that all quantities derived from the effective potential, including minima of $\mathcal{V}_\text{eff}(\phi,T)$, should be calculated order-by-order for a given power-counting~\cite{Croon:2020cgk,Ekstedt:2020abj}.

From now on I will leave the temperature dependence implicit for all quantities. 
\subsubsection{A Loop-Induced Barrier}
\label{sec:PhaseTransitions}

First-order phase transitions come in all shapes and sizes; the hallmark of a first-order transition is an energy barrier separating the two minima. For some potentials a barrier exists at tree-level, for others not so much. Amongst the possibilities, the loop-induced case is a bit subtle.

To see why, consider a Standard-Model-like potential
\begin{align}
V(\phi)=\frac{1}{2}m^2\phi^2+\frac{1}{4}\lambda \phi^4.
\end{align} 

This potential has no barrier at tree-level and describes a second-order transition. Here the transition occurs when $m^2=0$. Yet a barrier can be induced by loops. To see this, consider a $\text{U}(1)$  gauge theory with vector-boson mass $m_A=e \phi$.
 Integrating out the vector-bosons gives~\cite{Arnold:1992rz}
\begin{align*}
V(\phi)=\frac{1}{2}m^2 \phi^2-\frac{1}{6\pi}e^3 \phi^3+\frac{1}{4}\lambda \phi^4.
\end{align*}

As noted in \cite{Arnold:1992rz}, the phase-transition occurs when all three terms are of comparable size: $\phi\sim \frac{e^3}{\lambda},~\& ~m^2\sim\frac{e^6}{\lambda} $. Also, the perturbative expansion is only consistent when $\lambda\sim e^3$. So focus on that case. The above potential, together with higher-order corrections, can be used as a starting point for looking at phase transition. It should however be stressed that everything hinges on the power-counting used for $\lambda$; in general it's not proper to put loop terms on the same order as tree-level ones.\footnote{The procedure can be streamlined by integrating out the vector bosons and working with an effective theory without vector bosons. See for example \cite{Patel:2017aig,Andreassen:2016cvx,Weinberg:1992ds,Gould:2021ccf}.}
This last point has important consequences for bubble nucleation.

The above discussion, crucially, assumes that we work in a super-soft theory. This means that all temporal-vector bosons are integrated out. For this to work the Debye mass, $m_D^2 \sim e^2$, must be larger than the transverse mass $m_A^2=e^2 \phi^2$. This seems problematic because $\phi\sim 1$. However, paying closer attention to factors of $\pi$ indicates that $\phi \sim \frac{e^3}{\lambda} \pi^{-1}\sim \pi^{-1}$.Thus $m_D^2 m_A^{-2} \sim \pi^2 \gg 1$, and it's proper to integrate out temporal-vector bosons. These considerations can also be checked numerically for given couplings.

\subsection{The Bubble-Nucleation Rate}
While the effective potential captures equilibrium dynamics, a non-equilibrium approach is needed to describe the dynamics of bubble nucleation. The methods used mirror tunneling calculations in four-dimensional quantum field theory. A quantum field theoretical description of tunneling was first given by Coleman and collaborators~\cite{Coleman:1977py,Callan:1977pt,Coleman:1980aw}. These methods were subsequently\te building on Langer's work~\cite{PhysRevA.8.3230,Langer:1969bc}\te extended to finite temperature by Linde and Affleck~\cite{Linde:1981zj,Affleck:1980ac}.  The result is closely related to Kramer's escape problem (see \cite{Berera:2019uyp} for a field-theory version). As before, I assume a dimensionally reduced theory.

The nucleation rate per unit volume is given by
\begin{align}
\label{eq:NucleationProbability}
& \Gamma=A ~e^{-S_3},~\quad A=\frac{\omega_c}{2 \pi}\left(\frac{S_3}{2\pi}\right)^{3/2} \left[\frac{\Det\left(-\nabla^2+V''_{FV}\right)}{\Det'\left(-\nabla^2+V''_B\right)}\right]^{1/2}\left(1+\mathcal{O}\left(x\right)\right).
\end{align}
Here $V''_{FV}$ and $V''_{B}$ is shorthand for $V''(\phi)$ evaluated in the false-vacuum and bounce respectively; $\nabla^2=\partial_\mu \partial^\mu$; and  $\omega_c^2$ denote the magnitude of the only negative eigenvalue of the fluctuation operator $\left(-\nabla^2+V''_B\right)$. In addition to the above formula, there're  determinants for each field that interacts with the Higgs. These determinants are identical to the ones above but with $V''$ replaced by said field's mass. As before $x$ is used to signify suppression according to a particular power-counting. 

Equation \ref{eq:NucleationProbability} is a semi-classical result, based on a saddle-point approximation around the bounce solution. The bounce is a $\text{O}(3)$ symmetric solution to the classical equations of motion. In particular, the bounce satisfies\footnote{To not clutter the pages, I use the notation $\partial \equiv \frac{\partial}{\partial r}$ and $V'(\phi)\equiv \frac{\partial}{\partial \phi}V(\phi)$, where $r$ is the radius $r=|\vec{x}|$.}
\begin{align}
&\partial^2 \phi(r)+\frac{2}{r}\partial \phi(r)=V'(\phi),
\\&\partial \phi(r)\left.\right|_{r=0}=0,
\\&\lim_{r\rightarrow \infty}\phi(r)=\phi_{FV}.
\end{align}
See also~\cite{Espinosa:2018hue,Espinosa:2018voj,Espinosa:2018szu} for an alternative formulation of the bounce.

Physically the bounce starts close to the true-vacuum, crosses the barrier, and settles in the false-vacuum as $r\rightarrow \infty$. For multiple scalar fields the generalization is
\begin{align}
&\partial^2 \phi_i(r)+\frac{2}{r}\partial \phi_i(r)=\frac{\partial}{\partial \phi_i} V,
\\&~\partial \phi_i(r)\left.\right|_{r=0}=0,
\\&~\lim_{r\rightarrow \infty}\phi_i(r)=\phi_{i,FV}.
\end{align}
Analytic bounce solutions are rare, nevertheless, the asymptotic behaviour of the bounce is often known. Take for example a potential
\begin{align}
V(\phi)=\frac{1}{2}m^2\phi^2-\frac{1}{4}\lambda \phi^4+\frac{c_6}{32}\phi^6.
\end{align}
The false-vacuum state is $\phi_{FV}=0$. Then for large $r$ the bounce solution behaves as $\phi(r)\sim \frac{e^{-m r}}{r}$.

The $S_3$ exponent in equation \ref{eq:NucleationProbability} is the action evaluated on the bounce solution; for a real, scalar field this action is
\begin{align}
S_3=\int d^3x\left[\frac{1}{2}\partial_\mu \phi\partial_\mu \phi+V(\phi) \right].
\end{align}
Note that if the false-vacuum has a non-zero potential energy an additional term must be included in $S_3$. 

For practical purposes I use the AnyBubble~\cite{Masoumi:2016wot} package to determine the bounce.

The exponential pre-factor can be factorized into a dynamical and a statistical contribution.\footnote{It is not known if this factorization holds at higher orders.} At leading order these are	
\begin{align}
 &A_\text{dyn}=\frac{\omega_c}{2\pi},
 \\&A_\text{stat}= \left(\frac{S_3}{2\pi}\right)^{3/2} \left[\frac{\Det\left(-\nabla^2+V''_{FV}\right)}{\Det'\left(-\nabla^2+V''_B\right)}\right]^{1/2}.
 \end{align}
The prime on the determinant signifies that no zero-modes should be included.

The statistical pre-factor is really~\cite{PhysRevA.8.3230,Langer:1969bc}
\begin{align}
A_\text{stat}=\frac{2}{\mathcal{V}}\text{Im}\frac{\int \mathcal{D}\psi e^{-S[\phi+\psi]}}{\int \mathcal{D}\psi e^{-S[\phi_{FV}+\psi]}},
\end{align}
where $\mathcal{V}$ is the spatial volume; $\phi$ is the bounce solution, and $\phi_{FV}$ is the false-vacuum solution. In other words, the statistical pre-factor is the usual tunneling pre-factor evaluated in three dimensions. Moreover, the dynamical pre-factor is only an approximation. Indeed, the result $A_\text{dyn}=\frac{\omega_c}{2\pi}$ holds only in the absence of damping.\footnote{At high temperatures the fields follow classical equations of motion with the addition of damping and thermal noise. That is, the fields follow a generalization of the 1-D Langevin equation: $m\ddot{x}=-\eta\dot{x}+\xi(t)$.} Instead for large damping $\eta \gg \omega_c$ it's expected that $A_\text{dyn}\sim \eta^{-1}$~\cite{Moore:2000jw,Arnold:1998cy,Arnold:1997yb,Arnold:1996dy,Moore:1998swa,Moore:1999fs}. While the dynamical pre-factor is important, the statistical pre-factor tends to give the dominant contribution. For simplicity I will assume $A_\text{dyn}=\frac{\omega_c}{2\pi}$ from now on.

\section{The Derivative Expansion}
\label{sec:DerivativeExpansion}

Calculating functional determinants in a bounce background is daunting; leaving the determinants aside, even the bounce has to, more often than not, be found numerically. With analytical results being rare~\cite{Andreassen:2017rzq,Konoplich:1983mn,Konoplich:1986zp,Adams:1993zs,Aravind_2015}, going amain at the problem is not desirable. Being stymied by computational difficulties, one can either resort to numerical computations or analytical approximations. In this paper I compare these two approaches. The numerical framework is presented in Section \ref{sec:NumericalDeterminants}, while this section focuses on the derivative expansion~\cite{Garny:2012cg,Baacke:1993ne,Kripfganz:1994ha,PhysRevD.46.1671,Baacke:2003uw}.

Again, I use the notation $\partial \equiv \frac{\partial}{\partial r}$ and $V'(\phi)\equiv \frac{\partial}{\partial \phi}V(\phi)$, where $r$ is the radius $r=|\vec{x}|$.

 The derivative, or gradient, expansion presupposes that the bounce varies slowly compared to the mass. And the determinants are expanded in inverse powers of said mass.\footnote{The actual expansion parameter, for a particle with a field-dependent mass $m_A(\phi)$, is  $\left(m_A(\phi)\right)^{-6}\left[\partial m_A^2(\phi)\right]^2$ in three dimensions. The exact expansion parameter depends on the quantity and the number of space-time dimensions.} Naturally this expansion is only valid if the mass is "heavy" for the relevant field values.

Let's refine this statement. Recall the bounce equation:
\begin{align}
\partial^2\phi+\frac{2}{r}\partial \phi=V'(\phi).
\end{align}
In addition, because this is a three-dimensional theory the quartic coupling has mass-dimension $1$. Assume that this coupling, $\lambda$, is positive. Then, 
if we take $\lambda$ as the characteristic scale of $V$ we deduce $\left(\partial \phi\right)^2 \sim \lambda$. Likewise, if the field in question is a vector-boson, the mass scales as $m_A^2\sim g^2$. Hence the derivative expansion is only justified if $\frac{\lambda}{g^2}$ is a small number. Note that while both $\lambda$ and $g^2$ have dimensions of mass in three dimensions, their ratio is a dimensionless number.

Before looking at the derivative expansion in depth, I want to separate out the concept of a badly convergent derivative expansion and a badly convergent perturbative expansion. In the current context, the former means that the derivative expansion doesn't adequately reproduce the given functional determinant; while the latter indicates that higher-order corrections to the leading-order bounce action aren't under perturbative control. The derivative expansion can work well while, at the same time, the perturbative expansion isn't. And the other way around. Accordingly, this section only looks at how well the derivative expansion captures the given determinant. For a discussion about the aptness of the perturbative expansion of the nucleation rate as a whole, see Section \ref{sec:EFTvsLoopInduced}.

\subsection{A Loop-Induced Barrier}
\label{sec:DerivativeExpansionLoopInduced}
Let's now consider the derivative expansion for radiative barriers. Take the Standard Model with a small quartic coupling $\lambda \sim g^3$. Where $g$ is the $\text{SU}(2)_L$ gauge-coupling. As seen in Section \ref{sec:PhaseTransitions}, this choice of $\lambda$ should give a loop-induced barrier. And a derivative expansion is applicable since $\frac{\lambda}{g^2}\sim g \ll 1$. The hypercharge coupling is, for simplicity, not included.

There is actually something deeper going on. Because the tree-level Higgs potential doesn't contain a barrier, or a bounce, on its own, a barrier is only generated by vector-bosons loops. 

To see how, recall from Section \ref{sec:PhaseTransitions} that the tree-level potential is of order $V_0(\phi)\sim \lambda\sim g^3$. The leading term from the derivative expansion, which is just the 1-loop effective potential, is of order $m_W^3\sim g^3\sim V_0(\phi)$. Therefore, we must incorporate the vector-contribution already at leading order.
 
 We can take this even further. Assuming  $m_W=\frac{1}{2}g \phi$, the two leading terms in the derivative expansion are\footnote{This result includes summing over polarizations and taking into account all 3 gauge bosons. See \cite{Garny:2012cg,PhysRevD.46.1671,Bodeker:1993kj} for details on the derivative expansion.}
 \begin{align}
 \label{eq:HigherOrderDerivative}
 \delta S_\text{eff}^{\text{vectors}}=-\frac{1}{16\pi}\int d^3x g^3 \phi^3-\frac{11}{32 \pi}\int d^3x \frac{g \left(\partial \phi\right)^2}{ \phi}.
 \end{align}
 
 Note that the term scales as $\sim g^4$; this is also the scaling of the two-loop effective potential $V_2\sim g^2 m_A^2\sim g^4$. Moreover, Goldstone and Higgs functional determinants scale as $\lambda^{3/2}\sim g^{9/2}$. Taking these observations as a whole shows that the effective action evaluated on the bounce solution is
 \begin{align}
 S_\text{eff}=g^3S_\tLO+g^4 S_\tNLO+g^{9/2} S_\tNNLO+\ldots.
 \end{align}

This is quite remarkable. For a loop-induced barrier it's possible to calculate NLO corrections solely using the derivative expansion. And as long as the bounce solution is known, the NLO term is straightforward to calculate. Furthermore, the NLO term is expected to greatly reduces the renormalization-scale dependence of the nucleation rate~\cite{Gould:2021oba,Endo:2015ixx}.

 Further still, the $\text{N}^3\text{LO}$ contribution comes from three-loop diagrams; meaning that the derivative expansion can be used yet again. If $S_\text{NNLO}$ is known, the next order can once again by incorporated through usual effective-potential computations. Barring some sub-leading terms from the derivative expansion. Equivalently, vector-bosons can be integrated out, which was formalized in \cite{Gould:2021ccf}.

 Before moving on, note that only some terms in the derivative expansion should be brought to leading order. For instance, it is tempting to use the double-derivative term in equation \ref{eq:HigherOrderDerivative} to find the leading-order bounce. Yet this is not allowed if we follow a strict perturbative expansion. Instead, this term should be included as a perturbation.

Next, note that the expansion formally breaks down for vector-bosons when $\phi \rightarrow 0$. This complication appears at third order with a term $\sim\int  d^3x \frac{(\partial \phi)^4}{g \phi^5}$. However, for practical purposes the expansion should only be used when $g\gg 1$, so this complication of order $\mathcal{O}\left(\frac{1}{g}\right)$ doesn't' change the numerical results, too, much.

\subsection{The Problem with Scalar Determinants}
While the derivative expansion is applicable for vector-boson determinants, the same is not true for Higgs and Goldstone determinants. Indeed, the Higgs mass scales as $H\sim \lambda$; the expansion parameter is $\frac{(\partial H)^2}{H^3}\sim\frac{\lambda}{\lambda}\sim 1$~\cite{Andreassen:2016cvx,Weinberg:1992ds}. The derivative expansion is not under perturbative control and cannot be trusted.\footnote{This conclusion is modified in the thin-wall limit. For in that case $\partial H\sim 0$. So the derivative expansion is expected to work well.}

This becomes apparent if we include the first two terms in the derivative expansion:
\begin{align}
\delta S_\text{eff}^{\text{Higgs}}=-\frac{1}{12 \pi} \int d^3x H^{3/2}+\frac{1}{384 \pi}\int d^3x\frac{\left[V'''_\tLO(\phi)\right]^2 \left(\partial \phi\right)^2}{H^{3/2}}.
\end{align}

Unfortunately, $H=V''_\text{LO}(\phi)$. So as the bounce moves from the true minimum to the false one it passes through two points where $V''_\text{LO}(\phi)=0$: the derivative expansion does not converge.

A possible solution is to solve the problem exactly close to $V_\tLO''(\phi)=0$. The idea is to use the Green's function representation of the Higgs determinant~\cite{PhysRevD.46.1671}
\begin{align*}
\delta S_\text{eff}^{\text{Higgs}}=\int d^3x \int dm ~m G^{H}(x,x),
\end{align*}
where $m^2$ is the mass appearing in the leading-order potential.

And $G^{H}(x,x')$ is the Higgs Green's function satisfying
\begin{align*}
\left(-\nabla^2+V''_\text{LO}\left[\phi(x)\right]	\right)G^H(x,x')=\delta^{(3)}(x-x').
\end{align*}

Therefore, it is at least feasible to use derivative expansion away from $V''_\text{LO}\left[\phi(r_c)\right]=0$. While close to $V''_\text{LO}(\phi)=0$, say at $r=r_c$, approximate $V''_\text{LO}\left[\phi(r+r_c)\right]\approx V'''_\text{LO}\left[\phi(r_c)\right] r$. The Green's function could then be found exactly and would need to be matched to the derivative expansion akin to the usual WKB matching.

Even if this is done, the procedure is needlessly complicated. Truly, this approach is eclipsed by the simpler, quicker, and more robust method based on the Gelfand-Yaglom theorem that's discussed in Section \ref{sec:NumericalDeterminants}. 

To summarize, using the derivative expansion for the Higgs determinant beyond leading order is doubly-dyed with problems: There's no perturbative control and higher-order terms diverge.
A more realistic approach is to use the leading-order, finite, term from the derivative expansion and accept that this result cannot be systematically improved upon. This is sometimes\te with caveats\te a decent approximation as I will show when comparing to numerical results in Section \ref{sec:Models}.

\subsection{Higher-Order Corrections}

Let's now leave the $\lambda\sim g^3$ scaling behind, and consider a generic model where both scalars and vector-bosons contribute at NLO. The question is whether there is any chance of calculating the two-loop NNLO contribution. It should be mentioned that there might be some corrections due to real-time dynamics at higher orders. I here neglect such effects.

What we are really after is an order-by-order determination of the effective-action. To make this explicit, first define the effective action as 
\begin{align}
&e^{-S_\text{eff}[\phi]}=\int_{1\text{PI}} \mathcal{D}\Phi e^{-S[\phi+\Phi]},
\\&S_\text{eff}[\phi]=S_\text{LO}[\phi]+x S_\text{NLO}[\phi]+\ldots
\end{align}
where powers of $x$ denote suppression according to some power-counting.

 The bounce is a solution of
\begin{align*}
\frac{\delta S_\text{eff}[\phi]}{\delta \phi(x)}=0,
\end{align*}
with appropriate boundary conditions. The leading-order bounce solution satisfies
\begin{align}
\frac{\delta S_\tLO[\phi_\tLO]}{\delta \phi(x)}\equiv\left.\frac{\delta S_\tLO [\phi]}{\delta \phi(x)}\right\vert_{\phi=\phi_\tLO}=0.
\end{align}

At higher-orders the effective action is
\begin{align}
\label{eq:BounceActionHigherOrder}
&S_\text{eff}\left[\phi_\tLO+x \phi_\tNLO+\ldots\right]=S_\tLO\left[\phi_\tLO\right]+x S_\tNLO\left[\phi_\tLO\right]
\\& x^2 \left\lbrace  -\frac{1}{2} \int d^3x \int d^3y \frac{\delta^2 S_\tLO\left[\phi_\tLO\right]}{\delta \phi(x)\delta \phi(y)}\phi_\tNLO(x)\phi_\tNLO(y)+S_\tNNLO\left[\phi_\tLO\right]\right\rbrace+\ldots
\end{align}
The NLO contribution, $ S_\tNLO\left[\phi_\tLO\right]$, is nothing but the functional determinants that we have been discussing so far; $S_\tNNLO\left[\phi_\tLO\right]$ is the sub-leading effective action coming from higher-loop diagrams. As mentioned, both $S_\tNNLO\left[\phi_\tLO\right]$ and $\phi_\tNLO(x)$ are rather difficult to calculate. For example, the NLO correction to the leading-order bounce solution satisfies
\begin{align}
\label{eq:BounceCorrections}
\frac{\delta S_\tNLO[\phi_\tLO]}{\delta \phi(x)}&=-\int d^3 y\frac{\delta^2 S_\tLO\left[\phi_\tLO\right]}{\delta \phi(x)\delta \phi(y)} \phi_\tNLO(x),
\\&=-\left(-\nabla^2+V''_\tLO\left[\phi_\tLO(x)\right]\right)\phi_\tNLO(x),
\end{align}
which can be compared to the effective-potential result $\phi_\tNLO=-\frac{V_\tNLO'(\phi_\tLO)}{V_\tLO''(\phi_\tLO)}$.

While equation \ref{eq:BounceCorrections} can't be solved exactly, some properties of $\phi_\tNLO(x)$ can be deduced. For example, $\phi_\tNLO(x)$ only depends on the radius. This can formally be shown using Green's functions. See appendix \ref{app:BounceCorrection} for the details.

\section{Numerical Evaluations of Functional Determinants}\label{sec:NumericalDeterminants}
In this section I review how to calculate functional determinants numerically. I will show how the generalized Gelfand-Yaglom theorem, or Forman's theorem, can be used to calculate functional determinants for a variety of models, including singlet and triplet extensions. As before $\partial$ stands for the derivative with respect to radius. All the methods in this section are known, but I here review the essentials.

\subsection{The Gelfand-Yaglom Theorem}
Functional determinants can be calculated numerically, and at times analytically, using the Gelfand-Yaglom theorem~\cite{Gelfand:1959nq,Dunne:2005rt,Falco:2017ceh,Kirsten:2010eg,Kirsten:2004qv,Coleman:1978ae}. This theorem has been applied to tunneling problems in both four~\cite{Endo:2017tsz,Baacke:2003uw,Baacke:1999sc,Dunne:2005rt,Coleman:1978ae,Isidori:2001bm,Andreassen:2017rzq,Andreassen:2016cvx}, three~\cite{Baacke:1993ne,Strumia:1998nf}, and two dimension~\cite{Parnachev:2000fz}. While the Gelfand-Yaglom theorem is mainly used for numerical purposes, some analytical results exist in the thin-wall limit~\cite{Konoplich:1983mn,Konoplich:1986zp,Garbrecht:2015oea,Ai:2020sru}.

The standard Gelfand-Yaglom theorem applies to differential equations of the form
\begin{align}
\mathcal{M}\psi_\lambda(x)=(-\partial^2+W\left[x\right])\psi_\lambda(x)=\lambda \psi_\lambda(x),
\end{align}
 with boundary conditions $\psi_\lambda(0)=0,~\partial \psi_\lambda(0)=1$. For our purposes $W[x]$ can be thought of as a field-dependent mass depending on the bounce. Furthermore, $\psi_\lambda(x)$ is a normalized eigenvector and $W[x]$ becomes constant for large radii.
 
  The quantity of interest is
 \begin{align*}
 \Det \mathcal{M}=\Det\left(-\partial^2+W\right)\equiv\prod_i \lambda_i.
\end{align*} 
  For two different "potentials", say $W^{(1)}$ and $W^{(2)}$ , the Gelfand-Yaglom theorem states
\begin{align}
\label{eq:GelfandYaglom}
\frac{\Det\left(-\partial^2+W^{(1)}\right)}{\Det\left(-\partial^2+W^{(2)}\right)}=\frac{\psi_0^{(1)}(\infty)}{\psi_0^{(2)}(\infty)},
\end{align}
with the assumption that $\lambda=0$ is not an eigenvalue of either operator.

In this paper I use a generalization of this theorem that works for matrix operators~\cite{Forman1987,Kirsten:2004qv,Kirsten:2010eg,Falco:2017ceh}. From now on I'll also refer to this generalization as the Gelfand-Yaglom theorem. Since this theorem will be used extensively, it's worth reviewing how it works. So let's prove the theorem by using Coleman's argument~\cite{Coleman:1978ae}.

Consider an eigenvalue equation $\mathcal{L}_{ij}\psi_j=\lambda \psi_i$. Here and henceforth I will leave the $\lambda$ dependence of $\psi_i$ implicit. As an example, for nucleation-rate calculations $\mathcal{L}$ is often of the form $\mathcal{L}_{ij}=\delta_{ij}\left(-\partial^2-2/r \partial+l(l+1)/r^2\right)+V_{ij}(r)$.

Now for the proof. First, write the boundary conditions as~\cite{Kirsten:2004qv,Kirsten:2010eg,Falco:2017ceh}
\begin{align}
\label{eq:BoundaryValue}
&M_{ab} \Psi_b(0)+N_{ab} \Psi_b(\infty)=0,
\\& \Psi_a(r)=\left(\psi(r),\partial \psi(r)\right)_a,
\end{align}
where $M$ and $N$ are $2n\times 2n$ constant matrices and $a,b=1,\ldots 2n$.
Equation \ref{eq:BoundaryValue} can be made more lucid by constructing the fundamental solutions: $\mathcal{L}_{ij}y_{j;a}=\lambda y_{i;a}$. If $\mathcal{L}$ is an $n\times n$ matrix there are $2n$ fundamental solutions, $y_{i;a}$; these can be organized as ($a=1,\ldots 2n$)
\begin{align}
\label{eq:FundamentalSolutions}
y_{1;1}(0)=y_{2;2}(0)=\ldots=y_{n,n}(0)= \partial y_{1,n+1}(0)=\partial y_{2,n+2}(0)=\ldots= \partial y_{n,2n}(0)=1,
\end{align}
and all other components vanish at $r=0$. The idea is that the fundamental solutions form a basis, and any solution to the eigenvalue equation can be written
\begin{align}
\label{eq:FundamentalSolutionMatrix}
&\Psi_a(r)=Y_{a b}(r) \Psi_b(0),
\\& Y_{a b}=\left(y_{*;b},\partial y_{*;b}\right)_a.
\end{align}

So equation \ref{eq:BoundaryValue} is equivalent to
\begin{align}
\left(M+N Y(\infty)\right)_{ab}\Psi_b(0)=0.
\end{align} 
This equation must have a non-trivial solution if $\lambda$ is an eigenvalue, which implies $\Det\left[M+N Y(\infty)\right]=0$.

 Consider now two operators $\mathcal{L}^{(1)}$ and $\mathcal{L}^{(2)}$ with associated fundamental-solution matrices $Y^{(1)}$ and $Y^{(2)}$. The claim is that
\begin{align}
\label{eq:GeneralizedGelfandYaglom}
\frac{\Det\left(\mathcal{L}^{(1)}-\lambda\right)}{\Det\left(\mathcal{L}^{(2)}-\lambda\right)}=\frac{\Det\left[M+N Y^{(1)}_\lambda(\infty)\right]}{\Det\left[M+N Y^{(2)}_\lambda(\infty)\right]}.
\end{align}

Coleman observed that the left-hand side vanishes if $\lambda$ is an eigenvalue of $\mathcal{L}^{(1)}$, and has a pole if $\lambda$ is an eigenvalue of $\mathcal{L}^{(2)}$; which follows from the definition of the functional determinant. The same is true for the right-hand side. Yes, as discussed both $\Det\left[M+N Y^{(1)}_\lambda(\infty)\right]$ and $\Det\left[M+N Y^{(2)}_\lambda(\infty)\right]$ vanish if $\lambda$ is an eigenvalue. Both sides go to $1$ when $\lambda$ goes to infinity along any axis besides the real one (assuming that $V_{ij}(r)$ is well-behaved). So both sides have the same poles, zeroes, and asymptotic behaviour. They are the same. Setting $\lambda=0$ gives the desired ratio of functional determinants.

As an illustration, consider again the one-dimensional Gelfand-Yaglom theorem.
Take the equation $\mathcal{M}\psi^\lambda=(-\partial^2+W)\psi^\lambda=\lambda \psi^\lambda$ with boundary conditions $\psi^\lambda(0)=0,~\partial \psi^\lambda(0)=1$, and $\psi^\lambda(\infty)=0$. In this case
\begin{align}
M=
 \begin{pmatrix}
  1&0\\
  0&0
 \end{pmatrix},~N=
 \begin{pmatrix}
  0&0\\
  1&0
 \end{pmatrix}.
\end{align}

Since $\partial \psi^\lambda(0)=1$ we can directly identify $\psi^\lambda(r)=y^\lambda_{2}(r)$. Putting everything together gives (there is technically a pre-factor that cancels in the ratio of two determinants)
\begin{align}
\Det\left[M+N Y(\infty)\right]=y^0_2(\infty)=\psi^0(\infty).
\end{align}

There is a problem if $\lambda=0$ is an eigenvalue since the determinant vanishes. In most cases the determinant's zero eigenvalue is removed by going to collective coordinates, and the determinant is finite. This procedure is discussed in Section \ref{subsec:ZeroMode}.

\subsection{The Gelfand-Yaglom Theorem in Three Dimensions}
Determinants in the bounce background live in three-dimensions and concerns a spherically symmetric operator. In such a case the solutions can be expanded in spherical harmonics. The determinant of a spherically symmetric operator can then be written~\cite{Andreassen:2016cvx,Dunne:2006ct,Baacke:1999sc,Baacke:1993ne,Baacke:2003uw}
\begin{align}
\Det \mathcal{M}=\prod_{l=0}^\infty \left[ \mathcal{M}^{l}\right]^{2l+1}.
\end{align}

Typical operators are of the form
\begin{align*}
\mathcal{M}^l=-\partial^2-2/r \partial+\frac{l(l+1)}{r^2}+W(r),
\end{align*}

for some function $W(r)$. For instance, the Higgs and Goldstone determinants involve $W(r)=V''\left[\phi(r)\right]$ and $W(r)=\phi^{-1}(r) V'\left[\phi(r)\right]$ respectively. With $\phi(r)$ being the relevant bounce solution. For the tunneling calculation, the false-vacuum operator ($\mathcal{L}^{(2)}$ in equation \ref{eq:GeneralizedGelfandYaglom}), is
\begin{align}
\mathcal{M}^l_{FV}=-\partial^2-2/r \partial+\frac{l(l+1)}{r^2}+W(\infty).
\end{align}
Which for the Standard-Model means $\phi(\infty)=0$.

As boundary conditions for the Gelfand-Yaglom theorem I take $\psi^l(r)\sim r^l$ when $r\rightarrow 0$. These boundary conditions are not of the form given in equation \ref{eq:BoundaryValue}, but can be brought to that form by redefining $\psi^l(r)=\tilde{\psi}^l(r) r^l$. In practice one works with the function $T_l(r)\equiv \frac{\psi^l(r)}{\psi_{FV}^l(r)}$. Hence the above redefinition is unnecessary.

While it is possible to solve 
\begin{align}
\label{eq:SphericallySymmetricFundamentalEq}
\left(-\partial^2-2/r \partial+\frac{l(l+1)}{r^2}+W(r)\right)\psi^l(r)=0
\end{align}
 for each partial-wave $l$, for numerical purposes it is necessary to stop at some large angular cut-off $L$. The idea is that equation \ref{eq:SphericallySymmetricFundamentalEq} can be solved analytically for large $l$\te up to $\mathcal{O}\left(\frac{1}{l^3}\right)$ terms. The sum from $l=L$ to $l=\infty$ can then be done up to $\mathcal{O}\left(\frac{1}{l}\right)$ corrections. See appendix \ref{app:AngCutoff} for the details.

\section{Treatment of Zero-Modes}
\label{sec:ZeroModes}
The Gelfand-Yaglom procedure is problematic when there's a zero eigenvalue since the determinant vanishes. As a rule this zero-mode is removed already in the saddle-point-approximation step by going to collective coordinates. This procedure gives a pre-factor and a determinant with the zero-eigenvalue removed. Take the translational modes. The zero mode is here proportional to $\partial \phi$; going to collective coordinates in this case gives the spatial volume times a factor $\left(\frac{S_3}{2\pi}\right)^{3/2}$~\cite{Callan:1977pt,Andreassen:2017rzq,Isidori:2001bm}. It turns out that the collective-coordinate pre-factor, in this case $\left(\frac{S_3}{2\pi}\right)^{3/2}$, always cancels when calculating the determinant with the zero-eigenvalue removed. I will start off by showing how this works for the Goldstone determinant, before moving on to more complicated cases such as Goldstone-Vector boson mixing and multiple scalar-field tunneling.

In this entire section I will change the range from $r\in \left[0,\infty\right]$ to  $r\in \left[0,R\right]$, with the assumption that $R\rightarrow \infty$ at the end.	
\subsection{Zero-Modes for a Single Field}
\label{sec:GoldStoneZeroMode}
Assume a Goldstone determinant without vector-boson mixing. The zero-eigenvalue of the Goldstone determinant comes from the broken $\mathrm{U}(1)$ symmetry, and the corresponding eigenfunction is proportional to $\phi(x)$. Going to collective coordinates gives
\begin{align}
\Gamma \sim \text{Vol}\left(\mathrm{U}(1)\right)\left(\frac{\int d^3x \phi(x)\phi(x)}{2\pi}\right)^{1/2}\left[\frac{\Det\left(-\nabla^2+G_{FV}\right)}{\Det'\left(-\nabla^2+G\right)}\right]^{1/2}.
\end{align}
Where $\text{Vol}\left(\mathrm{U}(1)\right)=2\pi$ and $G=\frac{1}{\phi(x)}V'\left[\phi(x)\right]$. Let's expand the determinant in partial waves as usual
\begin{align}
&\Det'\left(-\nabla^2+G\right)=\prod_{l=0}^\infty \left[\Det'\left(-\nabla^2_l+G\right)\right]^{2l+1},
\\& \nabla^2_l=\partial^2+2/r \partial-\frac{l(l+1)}{r^2}.
\end{align}

The zero-mode occurs for $l=0$. Indeed, a naive application of the Gelfand-Yaglom theorem would require us to solve
\begin{align}
\left(-\partial^2-\frac{2}{r}\partial+\frac{1}{\phi(r)}V'\left[\phi(r)\right]\right)\psi(r)=0,~ \psi(0)=1,~\partial \psi(0)=0.
\end{align}
Yet from the bounce equation it follows that $\psi(x)=\phi(x)/\phi(0)$ is a normalizable zero-eigenvector. So the determinant is zero.

There are several ways to remove the zero eigenvalue \cite{McKane:1995vp,Endo:2017tsz,Dunne:2005rt,Falco:2017ceh}.\footnote{The modified boundary-value method~\cite{Falco:2017ceh,McKane:1995vp} is quite transparent and powerful, however, there are some subtleties that appears for tunneling problems. So I advice taking care if this is the method of choice.} Here I use the method of \cite{Dunne:2005rt}. 

The idea is to deform the zero-eigenvalue to $\lambda_0= \epsilon$ (not to be confused with the dimensional-regularization $\epsilon$)
\begin{align}
\mathcal{D}^\epsilon\psi^\epsilon(x)\equiv\left(-\partial^2-\frac{2}{r}\partial+\frac{1}{\phi(r)}V'\left[\phi(r)\right]\right)\psi^\epsilon(r)+\epsilon \psi^\epsilon(r)=0,~\partial \psi^\epsilon(0)=0.
\end{align}
Where for general $\epsilon$ the above equation does not have normalized eigenfunction. Specifically, $\psi^\epsilon(r)$ grows exponentially for large $r$, that is, the zero eigenvalue only appears for $\epsilon=0$. Then the determinant without the zero eigenvalue is
\begin{align}
\Det'\left(-\nabla^2_0+G\right)=\lim_{\epsilon\rightarrow 0}\frac{\Det\left(-\nabla^2_0+G+\epsilon\right)}{\epsilon}.
\end{align}

If we use the inner product, $ \left\langle f g\right\rangle\equiv \int d^3x  f(x) g^{\star}(x)$, one finds
\begin{align}
0=\left\langle \phi \mathcal{D}^\epsilon \psi^\epsilon\right\rangle=4\pi\left[r^2(\partial \phi \psi^\epsilon-\phi \partial \psi^\epsilon)\right]_{0}^{R}+\epsilon  \left\langle \phi \psi^\epsilon\right\rangle.
\end{align}

Using the boundary conditions shows that\footnote{The extra factor of 2 comes from that $\phi(R) \sim e^{-a_\phi R}$ and $\psi^\epsilon(R) \sim e^{a_\phi R}+\mathcal{O}(\epsilon)$.}
\begin{align}
\psi^{\epsilon}(R)=\frac{\epsilon  \left\langle \phi \psi^\epsilon\right\rangle}{-8\pi R^2 \partial \phi(R)}.
\end{align}
Finally, plugging everything into the Gelfand-Yaglom theorem gives
\begin{align}
\lim_{\epsilon\rightarrow 0}\frac{\Det\left(-\nabla^2_0+G+\epsilon\right)}{\epsilon}=\lim_{\epsilon\rightarrow 0}\frac{\psi^\epsilon(R)}{\epsilon}=\frac{ \left\langle \phi \phi\right\rangle}{-8\pi R^2 \partial \phi(R) \phi(0)},
\end{align}
where I used that $\lim_{\epsilon\rightarrow 0} \psi^\epsilon(x)=\phi(x)/\phi(0)$. Or, after normalizing with the false-vacuum solution
\begin{align}
\label{eq:GoldStoneZeroMode}
\Det\left(-\nabla^2_0+G_{FV}\right)^{-1}\lim_{\epsilon\rightarrow 0}\frac{\Det\left(-\nabla^2_0+G+\epsilon\right)}{\epsilon}=\frac{ \left\langle \phi \phi\right\rangle}{-8\pi R^2 \partial \phi(R) \phi(0) \psi_{FV}(R)}.
\end{align}
This expression is finite when taking $R\rightarrow \infty$, and $ \left\langle \phi \phi\right\rangle$ cancels the corresponding term coming from going to collective coordinates.

\subsection{Zero-Modes For Multiple-Field Determinants}
\label{subsec:ZeroMode}
Consider now a multi-field zero-mode. A good example is the two-Scalar determinant in equation~\ref{eq:TwoScalarDeterminant} when $l=1$:
\begin{align}
\label{eq:TwoScalarZeroMode}
& -\partial^2 \psi_1-2/r \partial \psi_1+\frac{2}{r^2} \psi_1+V_{11} \psi_1+V_{12}\psi_2=0,
\\& -\partial^2 \psi_2-2/r \partial \psi_2+\frac{2}{r^2} \psi_2+V_{22} \psi_2+V_{12}\psi_1=0.
\end{align}
Or shortened $\mathcal{M} \psi=0$.
Again, a normalizable zero-mode is $\psi_1=\partial \phi,~\psi_2=\partial \sigma$. This eigenfunction satisfies equation \ref{eq:TwoScalarZeroMode} because of the bounce equation \ref{eq:TwoScalarBounceEquation}. 
During the saddle-point approximation, the zero-eigenvalue was removed by going to collective coordinates, and we should really calculate the functional determinant without the zero-eigenvalue.

The idea is the same as for the single-field case. First, the functional determinant is
\begin{align}
\label{eq:MixedScalarsFundamentalSol}
\Det \mathcal{M}= y_{1;3}(R)y_{2;4}(R)-y_{2;3}(R)y_{1;4}(R),
\end{align}
where the fundamental solutions are defined in equation \ref{eq:FundamentalSolutions}. Also, the boundary conditions imply
\begin{align}
\label{eq:MixedScalarsFundamentalSolRelation}
&\partial\phi(x)=\partial^2\phi(0) y_{1;3}(x)+\partial^2\sigma(0) y_{1;4}(x),
\\&\partial\sigma(x)=\partial^2\phi(0) y_{2;3}(x)+\partial^2\sigma(0) y_{2;4.}(x).
\end{align}
Note that the fundamental solutions, by themselves, don't vanish as $R\rightarrow \infty$. In fact, if $V_{11}(\infty)=a_\phi$, $V_{22}(\infty)=a_\sigma$, and $V_{12}(\infty)=0$, the solutions behave as ($R\gg 1$)
\begin{align}
&y_{1;3}(R)=c_{1;3} e^{a_\phi R}/R+ d_{1;3}e^{-a_\phi R}/R,
\\& y_{1;4}(R)=c_{1;4} e^{a_\phi R}/R+ d_{1;4} e^{-a_\phi R}/R.
\end{align}
So equation \ref{eq:MixedScalarsFundamentalSolRelation} means that $c_{1;3} \partial^2 \phi(0)+c_{1;4} \partial^2 \sigma(0)=0$ and $d_{1;3} \partial^2 \phi(0)+d_{1;4} \partial^2 \sigma(0)=-a_\phi\phi_\infty$, where $\partial \phi(R)\sim -a_{\phi} \phi_\infty e^{-a_\phi R}/R$. These, and similar, relations can be used to simplify and rewrite the final result in a number of ways.

 The determinant in equation \ref{eq:MixedScalarsFundamentalSol} vanishes because of the zero-mode. So deform the zero-eigenvalue to $\lambda_0=\epsilon$, and consider instead the operator
 \begin{align}
& -\partial^2 \psi^\epsilon_1-2/r \partial \psi^\epsilon_1+\frac{2}{r^2} \psi^\epsilon_1+V_{11} \psi^\epsilon_1+V_{12}\psi^\epsilon_2+\epsilon \psi_1^\epsilon=0,
\\& -\partial^2 \psi^\epsilon_2-2/r \partial \psi^\epsilon_2+\frac{2}{r^2} \psi^\epsilon_2+V_{22} \psi^\epsilon_2+V_{12}\psi^\epsilon_1+\epsilon \psi^\epsilon_2=0,
\end{align}
or condensed as $\mathcal{D}^\epsilon_{ij}\psi^\epsilon_j=0$.

The determinant with the zero-eigenvalue removed is
\begin{align}
\Det' \mathcal{M}=\lim_{\epsilon\rightarrow 0}\frac{\Det \mathcal{D}^\epsilon}{\epsilon}= \lim_{\epsilon\rightarrow 0}\frac{y^\epsilon_{1;3}(R)y^\epsilon_{2;4}(R)-y^\epsilon_{2;3}(R)y^\epsilon_{1;4}(R)}{\epsilon}.
\end{align}

Call the normalizable eigenfunction $\Psi\equiv \left(\partial \phi,\partial \sigma\right)^t$ and define the inner-product as $ \left\langle f g\right\rangle\equiv \sum_i \int d^3x  f_i(x) g_i^{\star}(x)$. Using $0=\left\langle \Psi \mathcal{D}^\epsilon y^\epsilon_3\right\rangle=\left\langle \Psi \mathcal{D}^\epsilon y^\epsilon_4\right\rangle$ gives
\begin{align}
&8\pi R^2\left[\partial^2 \phi(R) y_{1;3}^\epsilon(R)+\partial^2 \sigma(R) y_{2;3}^\epsilon(R)\right]=-\epsilon \left\langle \Psi y^\epsilon_3\right\rangle,
\\& 8\pi R^2\left[\partial^2 \phi(R) y_{1;4}^\epsilon(R)+\partial^2 \sigma(R) y_{2;4}^\epsilon(R)\right]=-\epsilon \left\langle \Psi y^\epsilon_4\right\rangle.
\end{align}
This, together with $\lim_{\epsilon\rightarrow 0}\left[\partial^2\phi(0) \left\langle \Psi y_3^\epsilon\right\rangle+\partial^2\sigma(0) \left\langle \Psi y^\epsilon_4\right\rangle\right]= \left\langle \Psi \Psi\right\rangle$, gives

\begin{align}
\label{eq:ZeroModeFormulaTwoFields}
\Det' \mathcal{M}=\lim_{\epsilon\rightarrow 0}\frac{y^\epsilon_{1;3}(R)y^\epsilon_{2;4}(R)-y^\epsilon_{2;3}(R)y^\epsilon_{1;4}(R)}{\epsilon}= \frac{\left\langle \Psi \Psi\right\rangle y_{2;4}(R)}{-8\pi R^2 \partial^2\phi(R)\partial^2\phi(0)}.
\end{align}
The above result is finite when normalizing with the false-vacuum solution, and the $\left\langle \Psi \Psi\right\rangle$ factor again cancel the collective-coordinate factor. To be specific, if $\phi(R)=\phi_\infty e^{-a_\phi R}/R$, then the false-vacuum solution behaves as $y_{1;3}^{FV}(R)\sim \frac{3}{2 a_\phi}e^{a_\phi R}/R$. Thus
\begin{align}
\lim_{R\rightarrow \infty} R^2 \partial^2 \phi(R) y_{1;3}^{FV}(R)=\frac{3 a_\phi \phi_\infty}{2}.
\end{align}
The $y_{2;R}(R)\sim e^{a_\sigma R}$ term is likewise finite when normalizing with the false-vacuum solution.

In short, to evaluate the determinant with the zero-eigenvalue removed simply requires knowing the asymptotic behaviour of one fundamental solution.

For completeness, $\left\langle \Psi \Psi\right\rangle=3 S_3$, where $S_3$ is the leading-order bounce action.

The above result can be directly applied to theories where two scalar fields tunnel. Yet there are scenarios with three~\cite{Baum:2020vfl,Athron:2019teq}, or more, tunneling scalar fields. Not to mention models with additional mixing between fields, for example two-Higgs doublet models, singlet extensions, and triplet extensions. See appendix \ref{app:ZeroMode} for the generalization of equation \ref{eq:ZeroModeFormulaTwoFields} to an arbitrary number of scalar fields.

\subsection{Vector-Goldstone mixing}
\label{subsec:VectorGoldstoneMixingZeroMode}
In Fermi gauge the Standard-Model-Goldstone-vector matrix for $l=0$ is~\cite{Andreassen:2017rzq}
\begin{align*}
\mathcal{M}=\begin{pmatrix}
  \frac{1}{\xi}\left(-\partial^2-2/r\partial+\frac{2}{r^2}\right)+ g^2 \phi^2 & g\partial \phi- g\phi \partial\\
  2 g \partial \phi+ g\phi\partial+\frac{2}{r}g \phi &-\partial^2-2/r\partial+\phi^{-1}V'(\phi)
 \end{pmatrix}
\end{align*}
In the absence of mixing we saw in Section \ref{sec:GoldStoneZeroMode} that the Goldstone determinant had a zero mode; something similar happens here.

I will now use the result from the previous subsection but with different boundary conditions. Indeed, $\psi_1= \psi_2\sim r^0$ implies $ \partial \psi_1(0)=\partial \psi_2(0)=0$. And the functional determinant is
\begin{align}
\label{eq:GVMixingDet}
\text{Det}\mathcal{M}=  y_{1;1}(R)y_{2;2}(R)-y_{1;2}(R)y_{2;1}(R),
\end{align}
with, yet again, the fundamental solutions defined according to \ref{eq:FundamentalSolutions}. In the case at hand  $\psi_1=0,~\psi_2=\phi(x)/\phi(0)$ is a normalized zero-eigenvector. Though, equation~\ref{eq:GVMixingDet} is problematic because the $y_1$ fundamental solution does not exist. A way around this is to introduce a radial cut-off at $r=\delta$ and take the $\delta \rightarrow 0$ limit in the end. We don't actually have to find the $y_1$ solution; it is enough to note that
\begin{align}
\lim_{\delta \rightarrow 0}\frac{y_{1;1}(r)}{y^{FV}_{1;1}(r)}=1.
\end{align}

Going through the same steps as in the previous section, defining $\Psi=(0,\phi)^t$ and leaving the $\delta$ dependence implicit gives
\begin{align}
8\pi R^2\left[\partial\phi(R) y_{2;1}^\epsilon(R)-\phi(R) \partial y_{2;1}^\epsilon(R)\right]=-\epsilon \left\langle \Psi y^\epsilon_1\right\rangle,
&\\8\pi R^2\left[\partial\phi(R) y_{2;2}^\epsilon(R)-\phi(R) \partial y_{2;2}^\epsilon(R)\right]=-\epsilon \left\langle \Psi y_2^\epsilon\right\rangle.
\end{align}

Dividing by the false-vacuum, and using $y_{2;2}(x)=\phi(x)/\phi(0)$, we find
\begin{align}
&\frac{\Det' \mathcal{M}}{\Det \mathcal{M}^{FV}}=\lim_{\delta\rightarrow 0}\frac{1}{-8\pi R^2 \partial \phi(R) y_{2;2}^{FV}(R)}\left\langle \Psi y_2\right\rangle \frac{y_{1;1}(R)}{y^{FV}_{1;1}(R)}
\\&=\frac{1}{-8\pi R^2 \partial \phi(R) \phi(0) y_{2;2}^{FV}(R)}\left\langle \phi \phi\right\rangle.
\end{align}
This is the same result as in equation \ref{eq:GoldStoneZeroMode}\te the Goldstone-Vector determinant is just the usual Goldstone determinant without mixing.

Note that these results only holds assuming $\phi(R)\sim \frac{e^{-a_\phi R}}{R}$ for some positive "mass" $a_\phi$.

In summary, for $l=0$ the mixed Vector-Goldstone determinant (not including ghosts or transverse vectors) is just the Goldstone determinant. And for large $l$ the mixed Vector-Goldstone determinant separates into the pure Goldstone determinant times the pure-vector-determinant cubed. These results also hold in four dimensions.

	\section{Specific Models}\label{sec:Models}
I now turn to performing quantitative comparisons of the derivative expansion with the Gelfand-Yaglom approach. Both methods are first applied to the Standard Model with and without higher-order operators. Next, the methods are compared for a model with a two-step phase transition. The results in this section can directly be applied to realistic models, as shown in the next section.

As discussed in Section \ref{sec:DerivativeExpansion}, for scalars one should only use the leading-order term in the derivative expansion.
As such, I use a different number of terms depending on whether the determinant concerns a Higgs/Goldstone field, or a vector field. For scalars I use the leading term~\cite{PhysRevD.46.1671}
\begin{align}
\label{eq:DerivForPlot}
S^{(1)}_\text{deriv}=-\frac{1}{12 \pi}\int d^3x M^3,
\end{align}
and for (transverse) vectors
\begin{align}
\label{eq:DerivForPlot2}
S^{(1)}_\text{deriv}=-\frac{1}{12 \pi}\int d^3x M^3+\frac{1}{384\pi} \int d^3x (\partial_i M^2 )(\partial_i M^2 )M^{-3},
\end{align}
where $M$ is the field-dependent mass. The above expressions also need to be normalized by the false-vacuum contribution.

For vector-bosons I use $g$ to denote the generic (dimensionally reduced) vector-boson coupling; in the Standard Model $g$ is the weak-coupling constant up to a factor of $2$. Note that equation \ref{eq:DerivForPlot2} is the contribution from a \emph{single} transverse	 vector determinant.

In this entire section I take $\phi$ to denote the bounce solution. So when I write out the potentials all other fields are set to zero. To be specific, a model with a Higgs doublet $\Phi=1/\sqrt{2}(\phi_1+i\phi_2,~\phi_3+\phi+i\phi_4)^t$ is taken to have the same potential as a complex singlet $\Phi=1/\sqrt{2}(\phi_1+\phi+i\phi_2)$. However, depending on the field content the functional determinants are different.

\subsection{Loop-Induced Potential}
\label{sec:LoopInduced}
First the loop-induced barrier case, which was discussed in Sections \ref{sec:PhaseTransitions} and \ref{sec:DerivativeExpansionLoopInduced}.

The loop-induced potential is~\cite{Karjalainen:1996rk,Kajantie:1995dw,Arnold:1992rz}
\begin{align}
\label{eq:LoopInducedPotential}
V(\phi)=\frac{1}{2}m^2\phi^2-\frac{1}{2}\eta \phi^3+\frac{\lambda}{8}\phi^4,
\end{align}
where in three dimensions $[\phi]=\frac{1}{2},~[\eta]=\frac{3}{2},~[\lambda]=1$. Note that this potential is identical to the Standard-Model one when vector-bosons (and temporal vectors) are integrated out: $\eta \sim g^3$.

To condense the results it's convenient to rewrite everything in a dimensionless form. Here I use the same notation as \cite{Baacke:1993ne,Baacke:2003uw,Dunne:2005rt,Dine:1992vs,Dine:1992wr}, and redefine
\begin{align}
x\rightarrow m^{-1} x,~\phi\rightarrow m^2 \eta^{-1} \phi.
\end{align}
The action is
\begin{align}
&S_3=\beta \int d^3x \left[ \frac{1}{2}(\partial_\mu \phi)^2+\frac{1}{2}\phi^2-\frac{1}{2}\phi^3+\frac{\alpha}{8}\phi^4\right]\equiv \beta \tilde{S}_3(\alpha),
\\& \alpha=m^2 \lambda \eta^{-2},~\beta=m^3 \eta^{-2}.
\end{align}
Beware that $\alpha$ and $\beta$ have nothing to do with the strength and the characteristic time-scale of the phase transition; they are simply dimensionless parameters describing the potential. I stick to the $\alpha$ and $\beta$ notation here simply because it's prevalent in the literature. Also, for vector-bosons with a coupling $g$ the dimensionless parameter is $\overline{g}^2=\frac{m^2}{\eta^2}g^2$.

The $\alpha \rightarrow 1$ limit corresponds to the critical point\te when the two minima have the same energy\te and $\alpha \rightarrow 0$ corresponds to a second-order phase transition. That is to say, $\alpha=0$ corresponds to $m^2=0$. As we shall see, perturbative corrections are comparable in size to $ \tilde{S}_3(\alpha)$. This means that the expansion is only reliable when $\beta \gg 1$. This observation, together with  $\beta^2=\alpha^3 \left(\frac{\eta^2}{\lambda^3}\right)\sim\alpha^3 \left(\frac{g^6}{\lambda^3}\right)$, implies yet again that perturbation theory only works when $g^2 \gg \lambda$.

Moving on, the dimensionless form of the action greatly simplifies numerical calculations; not only can the leading order-bounce action be parametrized by two parameters, it turns out that functional determinants only depend on $\alpha$.

 In the new variables the nucleation rate is
\begin{align}
\label{eq:NucleationRateLoop}
\Gamma \times m^{-4} =\frac{|\tilde{\omega_c}(\alpha)|}{2\pi}e^{-\beta \tilde{S}_3(\alpha)-S^{(1)}(\alpha)}.
\end{align}
There is also a multiplicative factor $\sqrt{\beta}$ for each distinct zero mode\te three for the Higgs, and one, or more, for the Goldstone field. So for a complex scalar the rate in equation \ref{eq:NucleationRateLoop} should be multiplied by $\beta^2$.
The dimensionless negative-energy eigenvalue ($-\tilde{\omega}_c^2$) is related to the dimensionful one via $\omega_c^2=m^2 \tilde{\omega}_c^2$. And $S^{(1)}(\alpha)$ can be separated according to the field in question.

To be specific, $(-1)\tilde{\omega}_c^2$ is the only negative eigenvalue of the Higgs-fluctuation operator, that is
\begin{align}
\left[-\partial^2-2/r \partial+1-3\phi+\frac{3}{2}\alpha \phi^2\right]\Psi(r)=-\tilde{\omega}_c^2(\alpha) \Psi(r),
\end{align}
where $\Psi(r)$ is a normalizable eigenfunction.

Once $S^{(1)}(\alpha)$ and $\tilde{S}_3(\alpha)$ are known, the nucleation temperature can be evaluated for any temperature (implicit in $m^2,~\eta,~\lambda$).

For parameter scans it is useful to approximate both the tree-level bounce and the functional determinants by functions of the form\footnote{This parametrization is slightly different from that of~\cite{Dine:1992wr}. Here I choose the form of $f(\alpha)$ to give a simultaneous good fit for the bounce action and the determinants.} 
\begin{align}
f(\alpha)=A+B \alpha +C \alpha^2+ D /(1-\alpha)+E/(1-\alpha)^2+F/(1-\alpha)^3.
\end{align}
Using the Gelfand-Yaglom method I find that the tree-level bounce, Higgs determinant, and Goldstone determinants are given according to Table \ref{table:LoopInduced}. The parametrizations given in this table are accurate to $1\%$.  Also, the result in Table \ref{table:LoopInduced} agree with a thin-wall-limit analysis:$ F_3 = 0, E_3 = \frac{32\pi}{81} , D_3 =\frac{ 272\pi}{81} , F_H = 0$,
and $F_G = \frac{8}{243}$ .

\begin{table}
\centering
\begin{tabular}{l*{6}{c}r}
Function              & A & B & C & D & E &F  \\
\hline
$\tilde{S}_3(\alpha)$ & $7.24$ & $5.68$ &$0$& $10.4$ & $1.25$ & $0$\\
$S^{(1)}_H(\alpha)$ & $0.213$ & $4.86$ & $-9.56$ & $3.97$ & $0.505$ &$0$\\
$S^{(1)}_G(\alpha)$ & $0.895$ & $1.08$ & $-1.84$ & $1.51$ & $0.469$ & $0.0329$ \\
\end{tabular}	
\caption{Result for the dimensionless leading-order action and the functional determinants.}
\label{table:LoopInduced}
\end{table}

Mark that the result in Table \ref{table:LoopInduced} is the contribution from a single Goldstone boson. Thus, if $\phi$ is a complex scalar field there should be an additional $\text{Vol}[U(1)]=2\pi$ factor multiplying the nucleation rate. And if $\phi$ is a $\text{SU}(2)$ doublet, $S^{(1)}_G(\alpha)$ should be multiplied by three, and the nucleation probability should be multiplied by $\text{Vol}[\text{SU}(2)]=2 \pi^2$. While for $\text{SU}(2)\times \text{U}(1)\rightarrow \text{U}(1)$ breaking the volume factor is $\pi^2$~\cite{Buchmuller:1993bq}.

The magnitude of the negative eigenvalue can be approximated
\begin{align}
\tilde{\omega}_c^2=(1-\alpha)\left(2.39-0.854 \alpha+2.42 \alpha^2-9.67\alpha^3+5.85 \alpha^4\right).
\end{align}
This approximation holds to $1\%$ and agrees with the result of \cite{Baacke:1993ne}.

All in all, the above results make it easy to study any model whose dimensionally reduced potential is of the form given in equation \ref{eq:LoopInducedPotential}.

\begin{figure}[t!]
     \centering
     \captionsetup[subfigure]{oneside,margin={-0.9cm,0cm}}
     \begin{subfigure}[b]{0.48\textwidth}
         \centering
         \includegraphics[width=\textwidth]{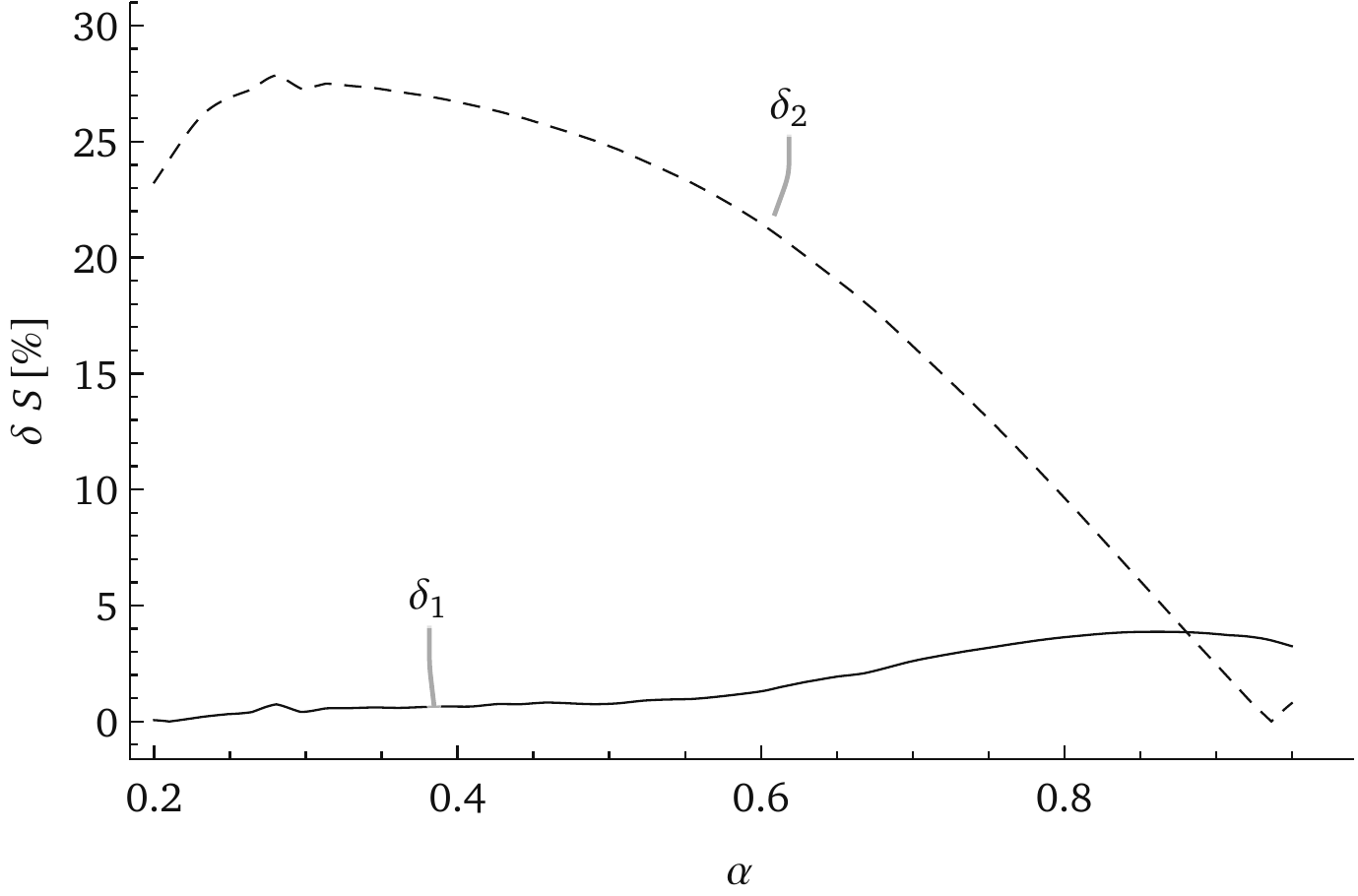}
         \caption{}
         \label{fig:LoopInducedHiggsDeterminant1}
     \end{subfigure}
     \hfill
      \captionsetup[subfigure]{oneside,margin={-0.1cm,0cm}}
     \begin{subfigure}[b]{0.48\textwidth}
         \centering
         \includegraphics[width=\textwidth]{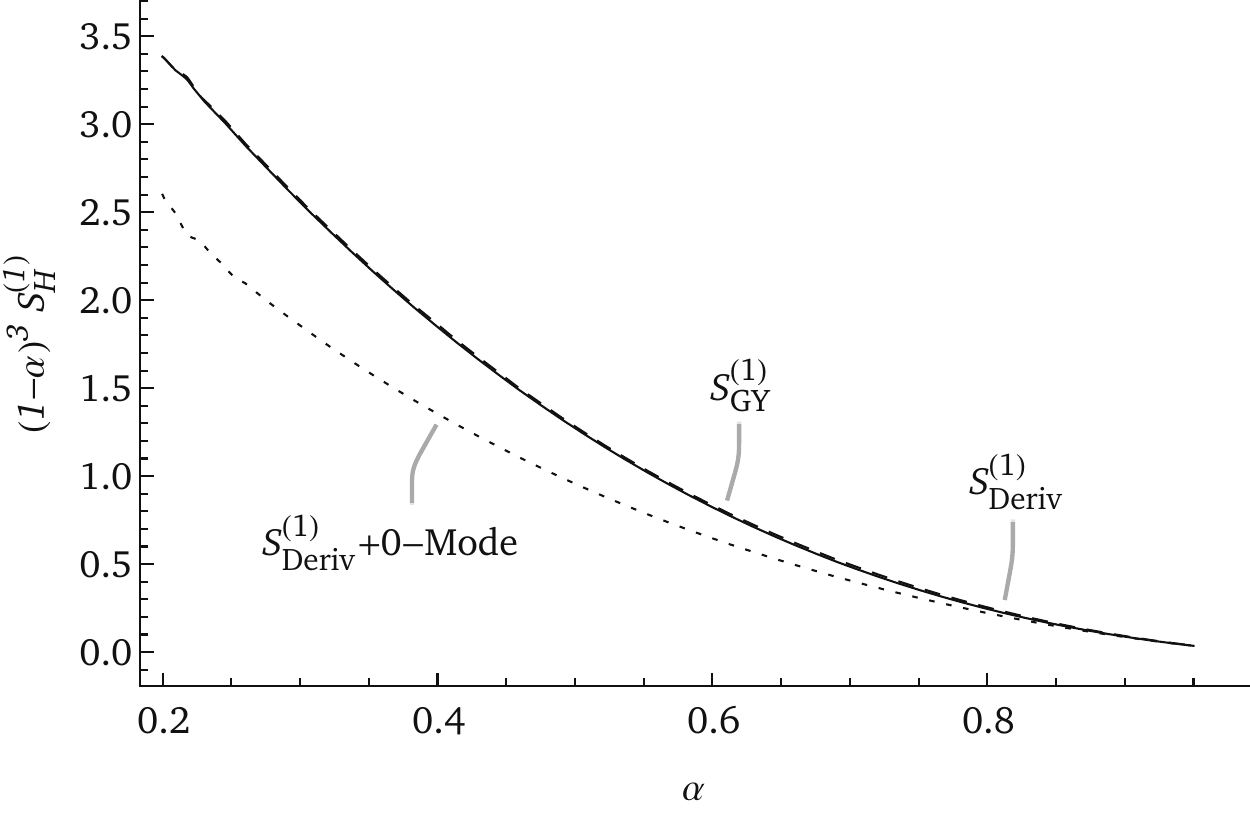}
         \caption{}
         \label{fig:LoopInducedHiggsDeterminant2}
     \end{subfigure}
     \newline
     \captionsetup[subfigure]{oneside,margin={-0.9cm,0cm}}
    \begin{subfigure}[b]{0.48\textwidth}
         \centering
         \includegraphics[width=\textwidth]{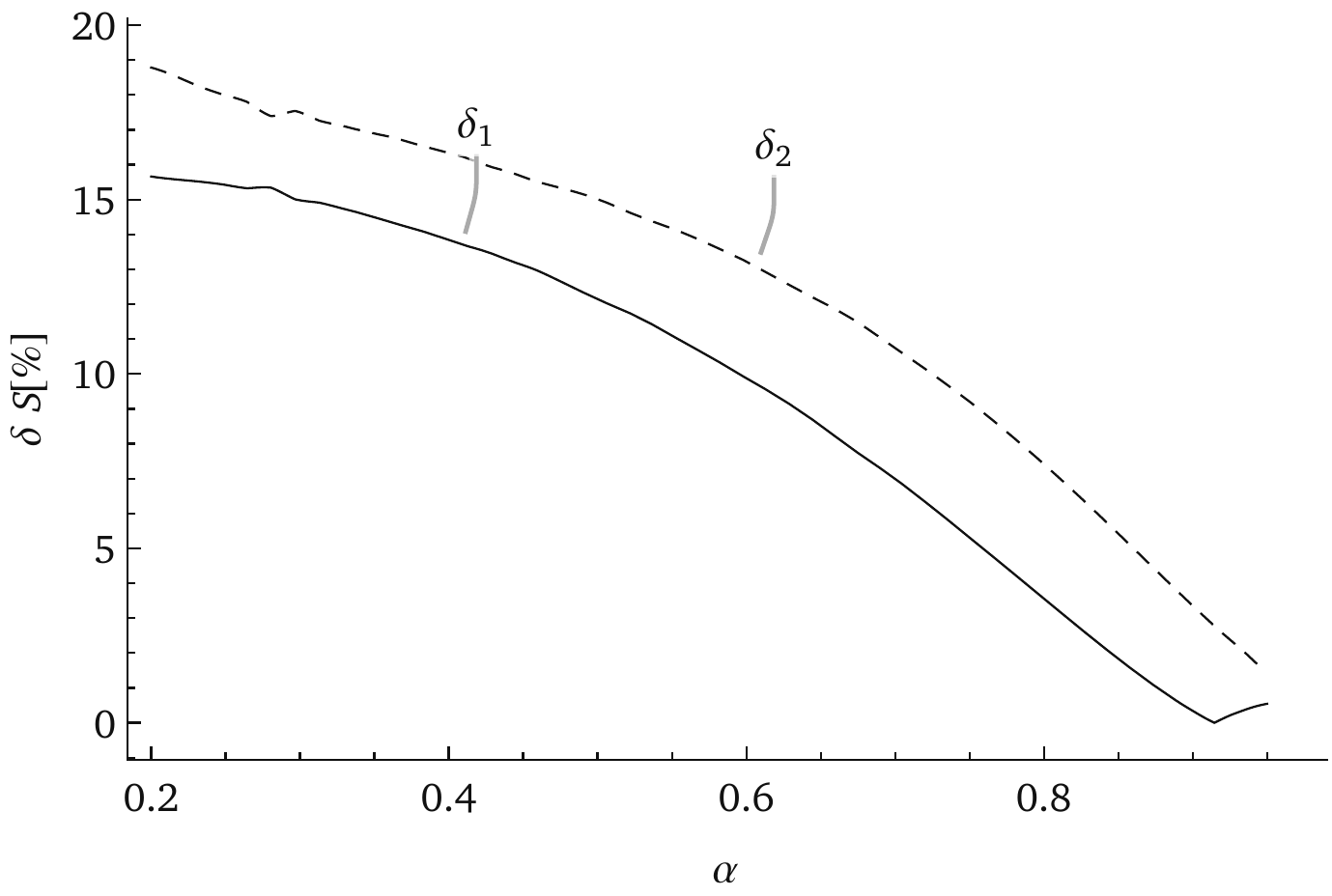}
         \caption{}
         \label{fig:LoopInducedGoldstoneDeterminant1}
     \end{subfigure}
     \hfill
     \captionsetup[subfigure]{oneside,margin={-0.1cm,0cm}}
     \begin{subfigure}[b]{0.48\textwidth}
         \centering
         \includegraphics[width=\textwidth]{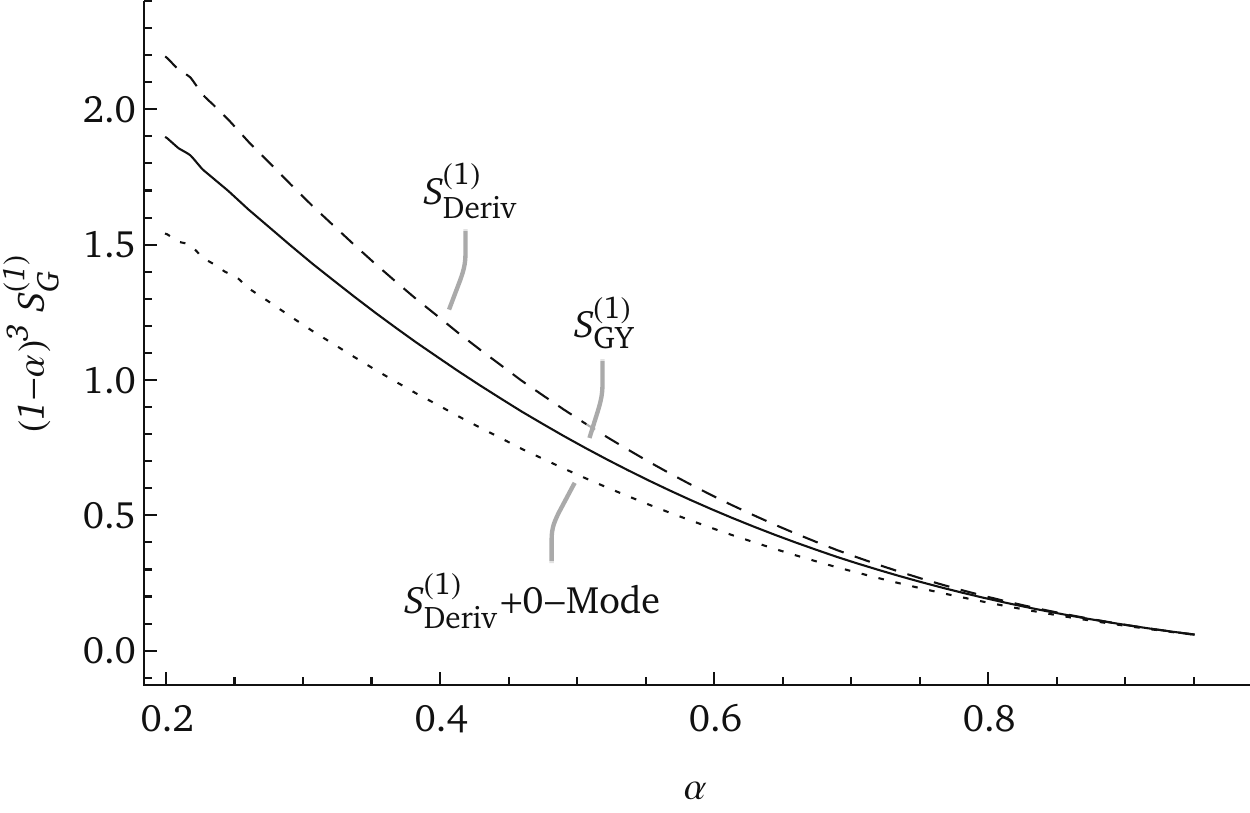}
         \caption{}
         \label{fig:LoopInducedGoldstoneDeterminant2}
     \end{subfigure}
	\caption{\small Comparison of the derivative expansion with the Gelfand-Yaglom method. Figures (a) and (b) concerns the Higgs determinant. Figure (a) compares $\delta_1 \equiv \left|\frac{S^{(1)}_{GY}-S^{(1)}_\text{Deriv}}{S^{(1)}_{GY}}\right|$ with $\delta_2 \equiv \left|\frac{S^{(1)}_{GY}-S^{(1)}_\text{Deriv}-\text{Zero-mode}}{S^{(1)}_{GY}}\right|$, and Figure (b) shows $(1-\alpha)^3 S^{(1)}_H(\alpha)$ for the different methods. Figures (c) and (d) are defined analogously as Figures (a) and (b), but looks instead at the Goldstone determinant. The zero-mode factor for the Higgs determinant is $(-1)\frac{3}{2}\log \frac{\tilde{S}_3}{2\pi}$, and $(-1)\frac{1}{2}\log \frac{\int d^3x \phi^2(x)}{2\pi}$ for the Goldstone determinant.
}
	\label{fig:LoopInducedDeterminant}
\end{figure}

Next let's compare the Gelfand-Yaglom method with the derivative expansion. As discussed in Section \ref{sec:DerivativeExpansion}, this expansion is not convergent, but can nonetheless be used as a proxy. Take the Higgs boson, in this case the leading-order result is
\begin{align}
S^{(1)}_\text{Deriv}=-\frac{1}{12\pi}\int d^3x\left[ (V''[\phi(x)])^{3/2}-V''[0]^{3/2}\right].
\end{align}
The above integral contains spurious imaginary parts, so we should really take the real value. Furthermore, it is not clear how to deal with zero modes. In particular, the Jacobian from going to collective coordinates is $\left(\frac{\tilde{S}_3}{2\pi}\right)^{3/2}$. When using the derivative expansion we are faced with the dilemma of whether this factor should be included or not. Since this factor always cancels when using the Gelfand-Yaglom theorem (see Section \ref{sec:GoldStoneZeroMode}), we might guess that it should be included. 
But this is contrary to what Figure \ref{fig:LoopInducedHiggsDeterminant1} shows. With the $\left(\frac{\tilde{S}_3}{2\pi}\right)^{3/2}$ factor included the error can reach $30\%$ for some $\alpha$ values. While without the error is $0-5\%$. This suggests that the zero-mode factor shouldn't be included. Yet for other models the behaviour can be the opposite as we'll see in the coming sections.

The situation is more sombre for the Goldstone determinant as shown in Figure \ref{fig:LoopInducedGoldstoneDeterminant1}. Indeed, for the Goldstone determinant the derivative expansion can be off by $20~\%$ when the collective-coordinate factor, $\left(\frac{\int d^3x \phi(x)^2}{2\pi}\right)^{1/2}$, is included. While without the aforementioned factor the error is less, and derivative expansion is at worst only off by $\sim 15\%$.

Nevertheless, it should be stressed that $15\%$ error is quite good. Really, we had no reason to trust the derivative expansion in the first place. But we should be wary of that the error can be vastly different depending on which determinant is evaluated. In the end looking at Figures \ref{fig:LoopInducedHiggsDeterminant2} and \ref{fig:LoopInducedGoldstoneDeterminant2} shows that derivative expansion is reasonable regardless if we include the zero-mode factors or not. 

Note that the derivative expansion works well for the loop-induced barrier because the bounce spends the majority of the time in a region where $V''(\phi)$ is negative. Besides, the derivative expansion works well in the thin-wall limit ($\alpha \rightarrow 1$) because the characteristic size of the bounce is large.

In the Standard-Model the potential in equation \ref{eq:LoopInducedPotential} arises when the Higgs quartic is small.\footnote{A small Higgs quartic doesn't necessarily mean that the Higgs is light. See \cite{Camargo-Molina:2021zcp} for an example.} So let's consider such a scenario. As such, assume that the $\eta$ term in equation \ref{eq:LoopInducedPotential} is generated by vector-boson loops. If this is the case, the Higgs and Goldstone determinants constitute but a NNLO correction to the exponent; the derivative expansion from the vectors gives the NLO piece.\footnote{These double-derivative terms were also considered in for example \cite{Moore:2001vf,Bodeker:1993kj,PhysRevD.46.1671,Garny:2012cg}.}

To be specific, consider the Standard Model where we ignore the $g'$ coupling. That is, an $\mathrm{SU}(2)$ theory. In that case the NLO correction to the leading-order action is~(see \cite{Kajantie:1995dw,Farakos:1994kx,PhysRevD.46.1671} for the two-loop term)
\begin{align}\label{eq:NLOCorrection}
 S_\tNLO(\alpha,g,\beta)&=-\frac{11}{32 \pi}\int d^3x \overline{g} \frac{(\partial \phi(x))^2}{\phi(x)}
\\&+\beta^{-1}\frac{1}{1024 \pi^2}\int d^3x \overline{g}^4\phi^2(x) \left\lbrace 51 \log\left(\frac{\mu_3^2}{m^2 \phi^2(x) \overline{g}^2}\right)-126 \log(3/2)+33\right\rbrace,\nonumber
\end{align}
where $\overline{g}^2=\frac{m^2}{\eta^2}g^2$ and the tree-level vector-boson mass is assumed to be $m_W=\frac{1}{2}g \phi$; $\mu_3$ is the renormalization-group scale in the dimensionally reduced theory. The first-term comes from the derivative expansion; the second from the two-loop effective potential. In using this expression we have to assume that $\lambda \ll g^2$. The derivative expansion for the Goldstone-Vector determinant is here apt because the momenta in the Feynman diagrams are of order $m_W \gg (\partial \phi)$.

One might be bothered by that the first term diverges when $\phi\rightarrow 0$. However, for large $r$ the bounce behaves as $\phi(r)\sim e^{-r}/r$, and the integral is convergent.

After finding the bounce numerically the required terms can be approximated to $1\%$
\begin{align*}
&\int d^3x  \frac{(\partial \phi(x))^2}{\phi(x)}\approx 55.1+18.5\alpha-19.4 \alpha^2+53.1/(1-\alpha)+5.59/(1-\alpha)^2,
\\& \int d^3x  \phi^2(x)\approx 29.2+24.3\alpha+24.3 \alpha^2+3.56/(1-\alpha)+20.2/(1-\alpha)^2+4.96/(1-\alpha)^3,
\\&\int d^3x  \phi^2(x)\log \phi^2(x) \approx 14.6+21.4\alpha+41.0 \alpha^2-31.1/(1-\alpha)+18.8/(1-\alpha)^2+6.88/(1-\alpha)^3.
\end{align*}
As before the leading terms, in the $\alpha \rightarrow 1$ limit, can be found from a thin-wall analysis: $\frac{16\pi}{9}$ and $\frac{152\pi}{9}$ for the $\frac{(\partial \phi)^2}{\phi}$ term; $\frac{128\pi}{81}$ and $\frac{512\pi}{81}$ for the $\phi^2$ one; finally $\frac{128\pi}{81}\log4$ for the $\phi^2\log\phi^2$ integral.

The power-counting shows that the $S^{(1)}$ terms in Table \ref{table:LoopInduced} are NNLO corrections.
Using these results, it's easy to evaluate the effective action up to NNLO precision for any parameter. However, note that the derivative expansion, and the result for $S_\tNLO(\alpha,g,\beta)$, are only valid when $\frac{\alpha}{\overline{g}^2}=\frac{\lambda}{g^2}\ll 1$. Note that as long as $\frac{\lambda}{g^2}\ll 1$ holds at $\alpha=1$, the ratio $\frac{\alpha}{\overline{g}^2}\ll 1$ everywhere. This last property is quite handy and makes the loop-induced case particularly well-behaved. Comparing with effective-potential results, the derivative expansion is likely applicable till $\frac{\lambda}{g^2}\sim 0.1$~\cite{Kajantie:1995kf,Rummukainen:1998nu,Moore:2001vf}.

\subsection{Standard Model EFT Potential}
\label{sec:EFTPotential}

Since the Standard Model does not exhibit a first-order phase transition on its own, new physics is required. A model-independent way of parametrizing this is through effective operators. To this end, I consider a potential of the form (hereafter referred to as the SM-EFT potential)
\begin{align}
V(\phi)=\frac{1}{2}m^2\phi^2-\frac{\lambda}{4}\phi^4+\frac{c_6}{32}\phi^6.
\end{align}

Note that, for reasons that'll become apparent, I have normalized the $c_6$ term differently from the literature \cite{Chala:2018ari,Camargo-Molina:2021zcp,Cai:2017tmh,deVries:2017ncy,Croon:2020cgk,Postma:2020toi}. I also assume that all parameters appearing in the potential are positive.

Again, the action can be written in a dimensionless form via the redefinitions
\begin{align}
x\rightarrow m^{-1} x,~\phi \rightarrow \frac{m}{\sqrt{\lambda}}\phi,
\end{align}

giving a leading-order action
\begin{align}
\label{eq:EFTpotentialAction}
&S_3=\beta \int d^3x \left[ \frac{1}{2}(\partial_\mu \phi)^2+\frac{1}{2}\phi^2-\frac{1}{4}\phi^4+\frac{\alpha}{32}\phi^6\right]\equiv \beta \tilde{S}_3(\alpha),
\\& \alpha=c_6 \beta^2,~\beta=m \lambda^{-1}.
\end{align}
The normalization of the potential is chosen so that $\alpha=1$ corresponds to the critical temperature: When the broken and symmetric phase has the same energy. This is also refereed to as the thin-wall limit in this context. Since this is an effective theory, $c_6$ is usually thought of as arising from new physics. This is often parametrized as $c_6\sim T^2 \Lambda^{-2}$, where $\Lambda$ is the new-physics scale and $T$ is the temperature. The factor of $T^2$ appears because we are working in a dimensionally reduced theory~\cite{Croon:2020cgk}.

Just as for the loop-induced potential, the nucleation rate can be written
\begin{align}
\Gamma\times m^{-4}= \frac{\tilde{\omega_c}(\alpha)}{2\pi}e^{-\beta \tilde{S}_3(\alpha)-S^{(1)}(\alpha)},
\end{align}
where $\omega_c^2=m^2 \tilde{\omega}_c^2$, and a multiplicative factor of $\sqrt{\beta}$ should be included for each zero-mode.

The bounce action and the functional derivatives can be parametrized by functions of the form
\begin{align}
f(\alpha)=A+B \alpha+C (1-\alpha)^{-1}+D (1-\alpha)^{-2}+E (1-\alpha)^{-3}.
\end{align} 
 For the SM-EFT potential the determinants can't be approximated by a single function for the entire range. So it's necessary to split up the $\alpha$ range. The results for the Goldstone and Higgs determinants are summarized Table~\ref{table:EFT}; the first number corresponds to $0 \leq \alpha \leq 0.8$, and the one in the parenthesis to $ 1\geq\alpha \geq 0.8$. In addition, the result in Table~\ref{table:EFT} agree with a thin-wall analysis: $D_3=\frac{4\pi}{3}$, $E_3=0$, $E_H=-\frac{7}{9}$. and $E_G=\frac{1}{9}$.

\begin{table}
\begin{tabular}{l*{5}{c}r}
Function              & A & B & C & D & E  \\	
\hline
$\tilde{S}_3(\alpha)$ & $1.76$ & $-0.142$ & $12.6$ & $4.19$ & $0$\\
$S^{(1)}_H(\alpha)$ & $-2.65(+5.30)$ & $5.32(-28.8)$ & $0.533(8.14)$ & $1.25(0.119)$ & $-0.845(-0.765)$ \\
$S^{(1)}_G(\alpha)$ & $-0.289(-0.839)$ & $-0.155(-1.70)$ & $-0.5441(0.304)$ & $0.424(0.296)$ & $0.106(0.113)$ \\
\end{tabular}
\caption{Result for the dimensionless leading-order action and functional determinants. The $\tilde{S}_3(\alpha)$ parametrization holds for $1\geq \alpha \geq 0$. While the $S^{(1)}_H(\alpha)$  and $S^{(1)}_G(\alpha)$ parametrizations are divided into two $\alpha$ ranges. The number not in the parenthesis applies for  $0 \leq \alpha \leq 0.8$; and the number in the parenthesis applies for $ 1\geq\alpha \geq 0.8$. }
\label{table:EFT}
\end{table}

The magnitude of the dimensionless negative eigenvalue can be approximated
\begin{align}
\tilde{\omega}_c^2=
  \begin{cases}
    (1-\alpha)\left( 9.31 \alpha^4-33.4 \alpha^3+46.0 \alpha^2-30.7 \alpha+8.81\right)     & \quad 0.6\leq \alpha \leq 1 \\
    (1-\alpha)\left( 65.6 \alpha^4-167 \alpha^3+163 \alpha^2-75.5 \alpha+15.2\right) & \quad 0.6\geq \alpha \geq 0
  \end{cases}
\end{align}
This approximation is accurate up to $1\%$.

Now for the comparison of the derivative expansion with the Gelfand-Yaglom result. First, I make the comparison for each determinant individually. That is, disregarding any mixing. Afterwards, I take into account the Vector-Goldstone mixing.

\begin{figure}[t!]
     \centering
     \captionsetup[subfigure]{oneside,margin={-0.9cm,0cm}}
     \begin{subfigure}[b]{0.48\textwidth}
         \centering
         \includegraphics[width=\textwidth]{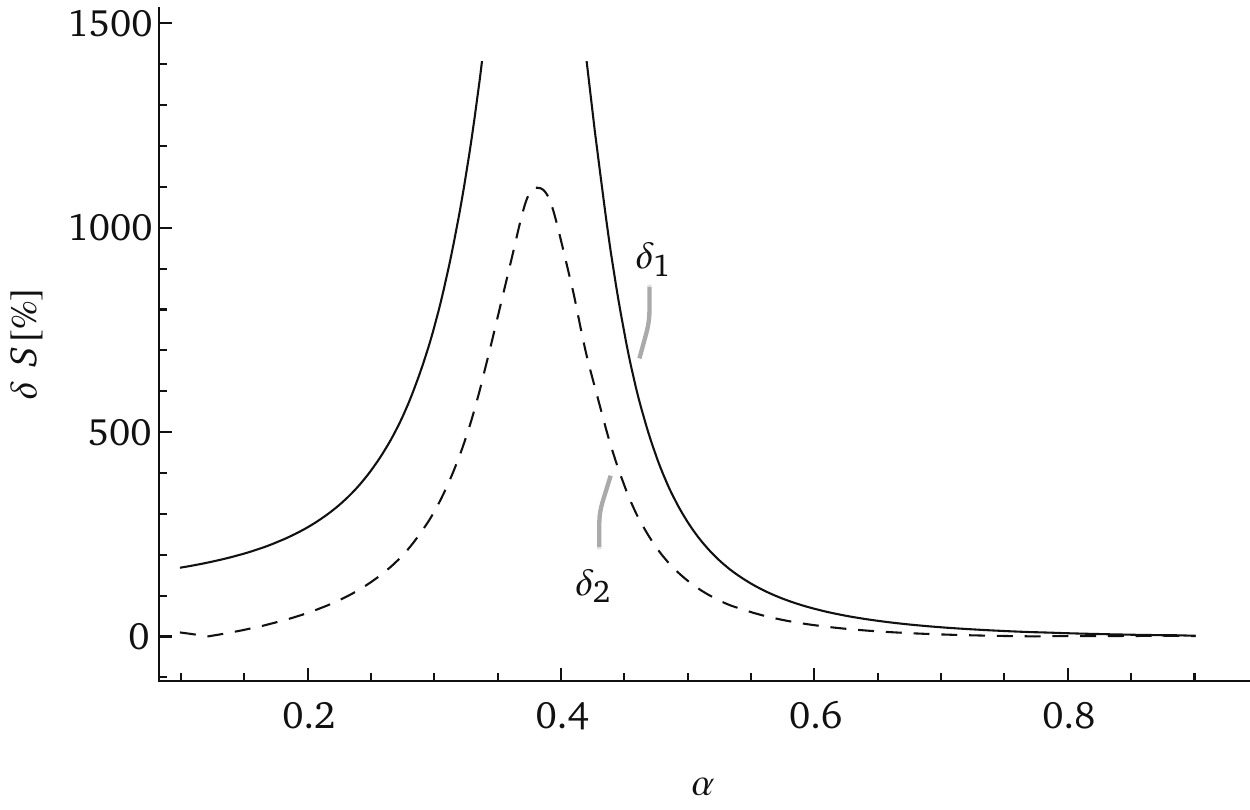}
         \caption{}
         \label{fig:EFTHiggsDeterminant1}
     \end{subfigure}
     \hfill
      \captionsetup[subfigure]{oneside,margin={-0.1cm,0cm}}
     \begin{subfigure}[b]{0.48\textwidth}
         \centering
         \includegraphics[width=\textwidth]{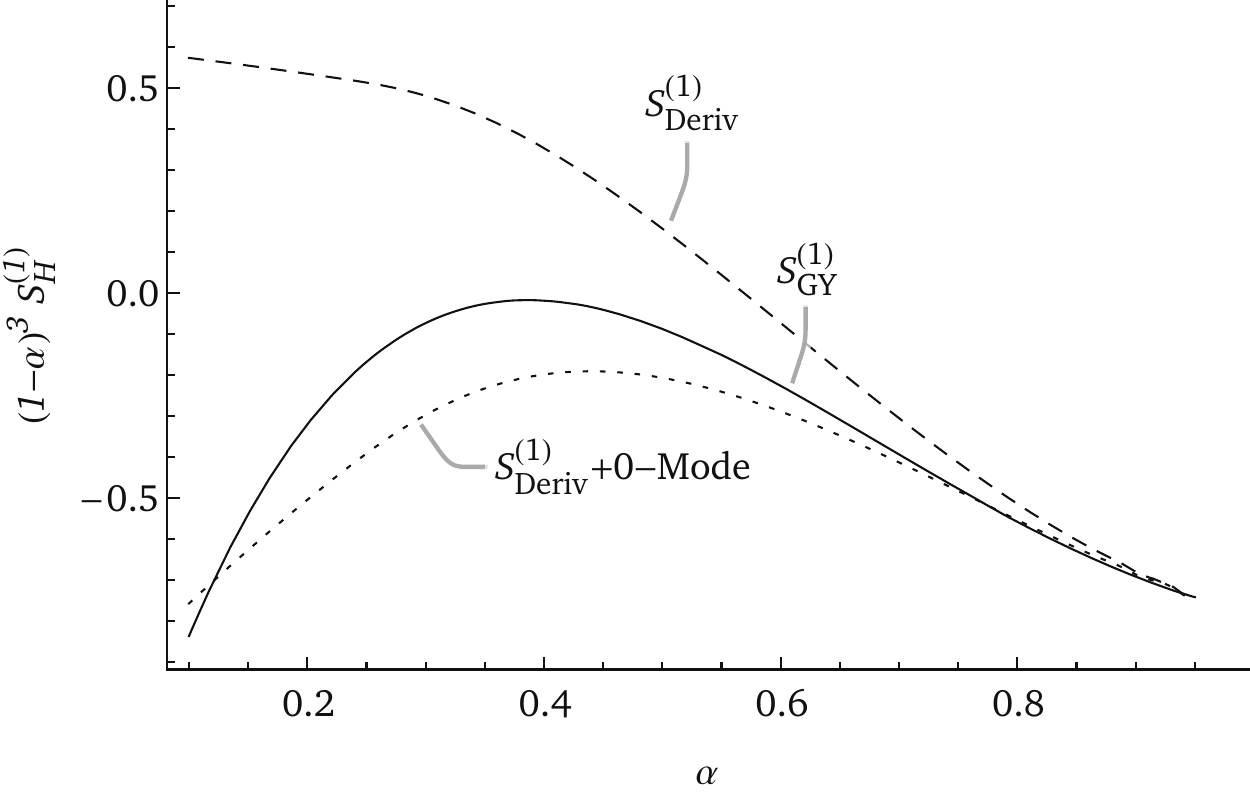}
         \caption{}
         \label{fig:EFTHiggsDeterminant2}
     \end{subfigure}
     \newline
     \captionsetup[subfigure]{oneside,margin={-0.9cm,0cm}}
    \begin{subfigure}[b]{0.48\textwidth}
         \centering
         \includegraphics[width=\textwidth]{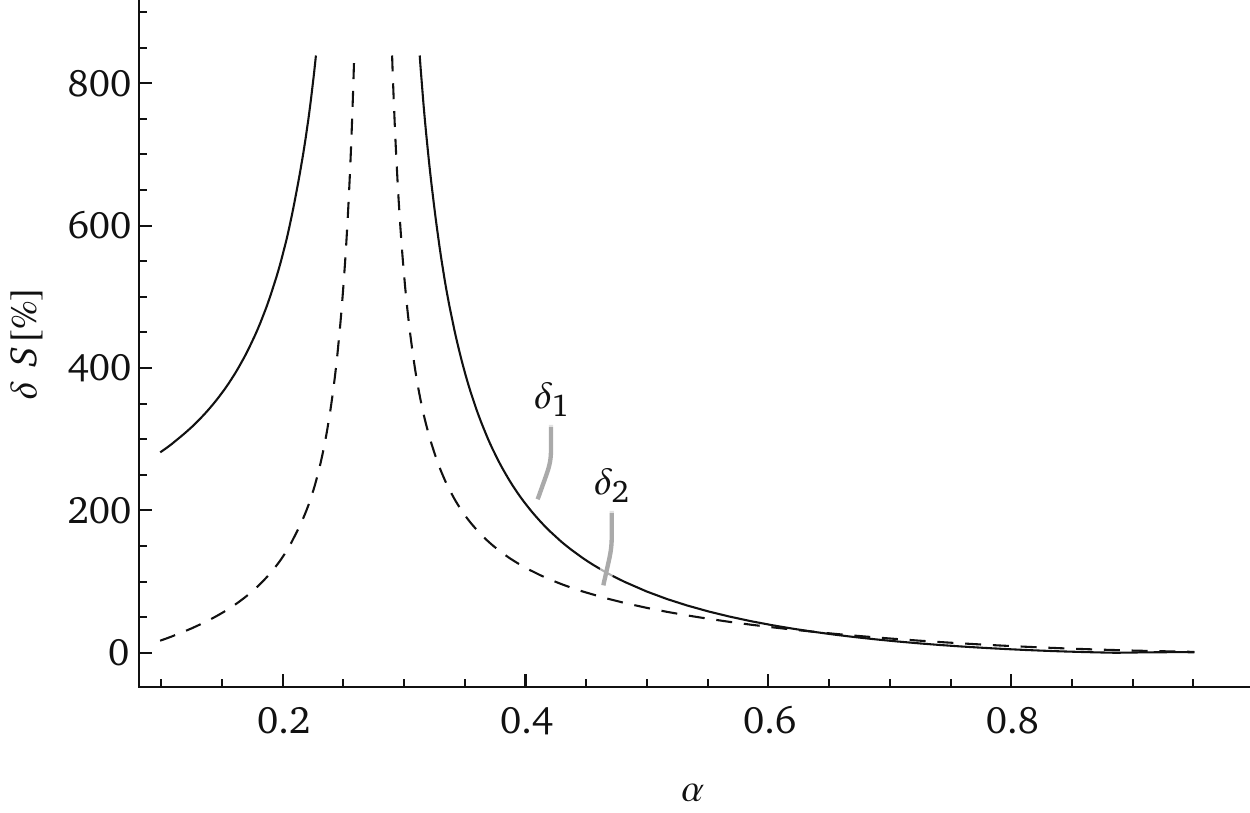}
         \caption{}
         \label{fig:EFTGoldstoneDeterminant1}
     \end{subfigure}
     \hfill
     \captionsetup[subfigure]{oneside,margin={-0.1cm,0cm}}
     \begin{subfigure}[b]{0.48\textwidth}
         \centering
         \includegraphics[width=\textwidth]{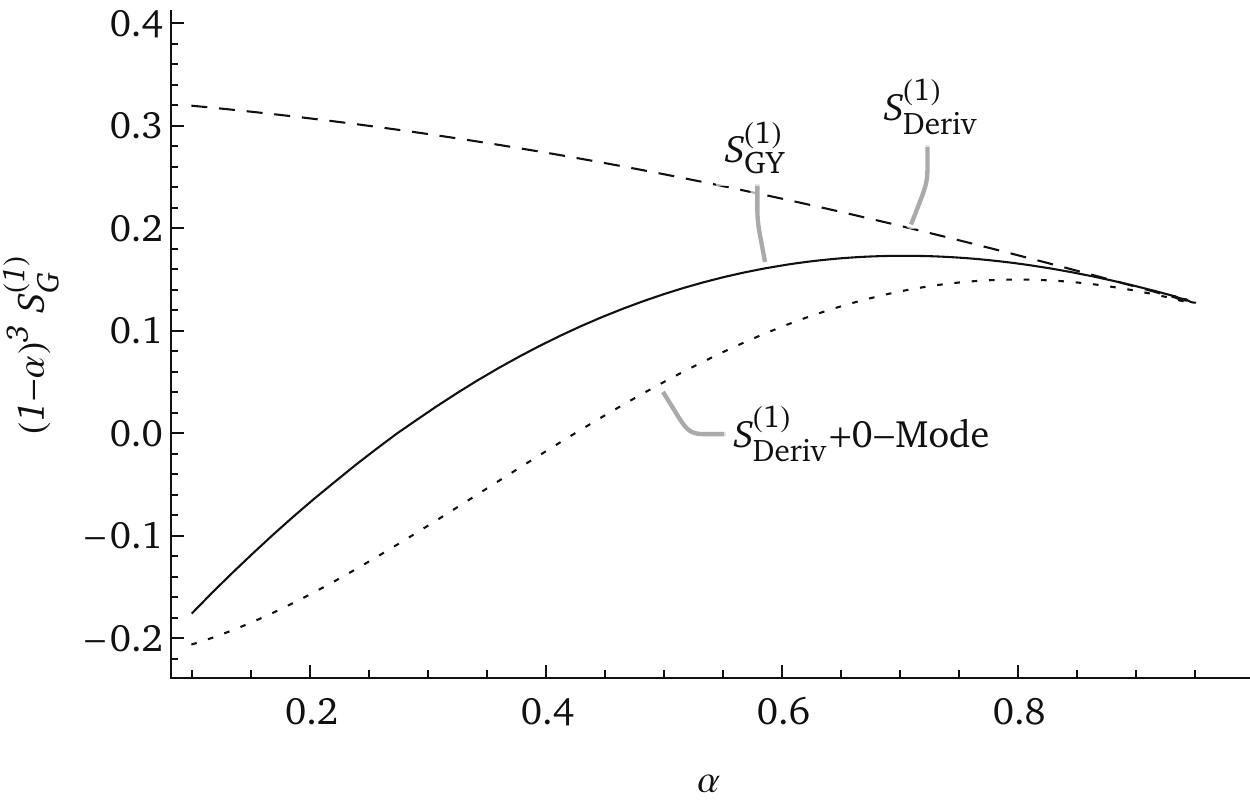}
         \caption{}
         \label{fig:EFTGoldstoneDeterminant2}
     \end{subfigure}
	\caption{\small Comparison of the derivative expansion with the Gelfand-Yaglom method. Figures (a) and (b) concerns the Higgs determinant. Figure (a) compares $\delta_1 \equiv \left|\frac{S^{(1)}_{GY}-S^{(1)}_\text{Deriv}}{S^{(1)}_{GY}}\right|$ with $\delta_2 \equiv \left|\frac{S^{(1)}_{GY}-S^{(1)}_\text{Deriv}-0-\text{Mode}}{S^{(1)}_{GY}}\right|$, and Figure (b) shows $(1-\alpha)^3~S^{(1)}_H(\alpha)$. Figures (c) and (d) are defined analogously as Figures (a) and (b) but looks at the Goldstone determinant. 
}
	\label{fig:EFTDeterminant}
\end{figure}

Let's start with the Higgs and Goldstone determinants. Figure \ref{fig:EFTDeterminant} shows the comparison of the Gelfand-Yaglom method with the derivative expansion. First notice that the derivative expansion behaves worse in this model compared to the loop-induced potential. What's more, it seems that including the zero-mode gives a better agreement than leaving it out; contrary to the previous case. It should be stressed that the large errors around $\alpha\approx 0.3-0.4$ in Figure \ref{fig:EFTDeterminant} are a bit misleading. The big jump comes from that the $S^{(1)}_\text{GY}$ curve gets close to zero. Overall, Figures \ref{fig:EFTHiggsDeterminant2} and \ref{fig:EFTGoldstoneDeterminant2} show that the derivative expansion is quite decent.

\begin{figure}[t!]
     \centering
     \captionsetup[subfigure]{oneside,margin={-0.9cm,0cm}}
     \begin{subfigure}[b]{0.48\textwidth}
         \centering
         \includegraphics[width=\textwidth]{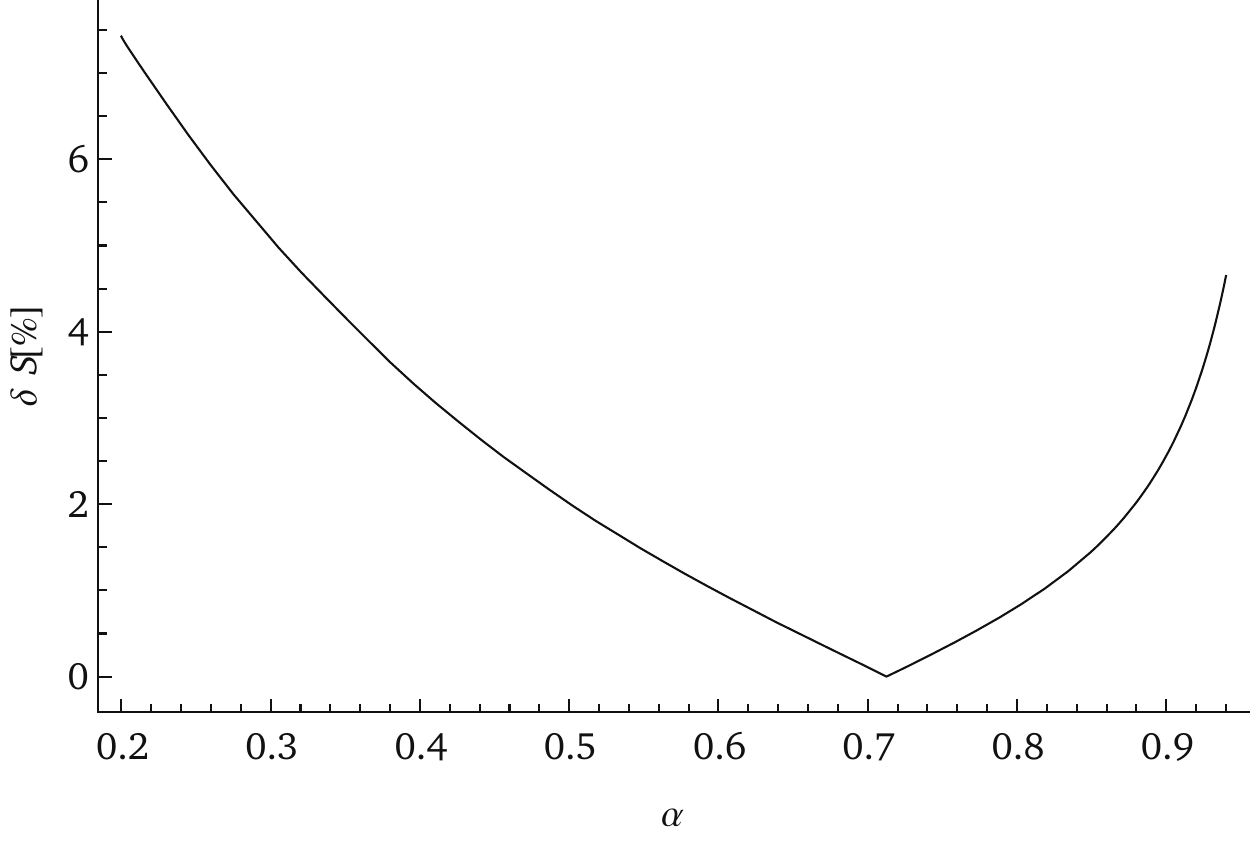}
         \caption{}
         \label{fig:EFTVectorDeterminant}
     \end{subfigure}
     \hfill
      \captionsetup[subfigure]{oneside,margin={-0.1cm,0cm}}
     \begin{subfigure}[b]{0.48\textwidth}
         \centering
         \includegraphics[width=\textwidth]{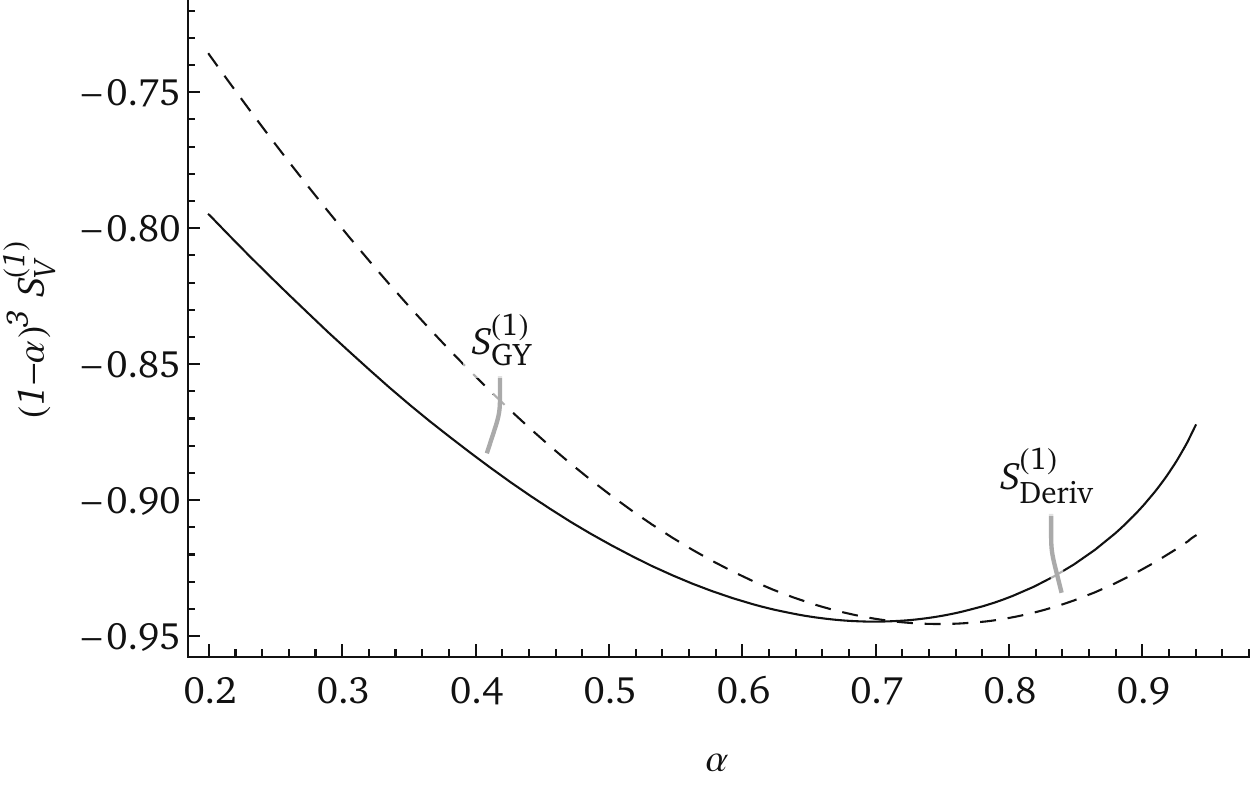}
         \caption{}
         \label{fig:EFTVectorDeterminantPlot}
     \end{subfigure}
     \newline
        \captionsetup[subfigure]{oneside,margin={-0.9cm,0cm}}
    \begin{subfigure}[b]{0.48\textwidth}
         \centering
         \includegraphics[width=\textwidth]{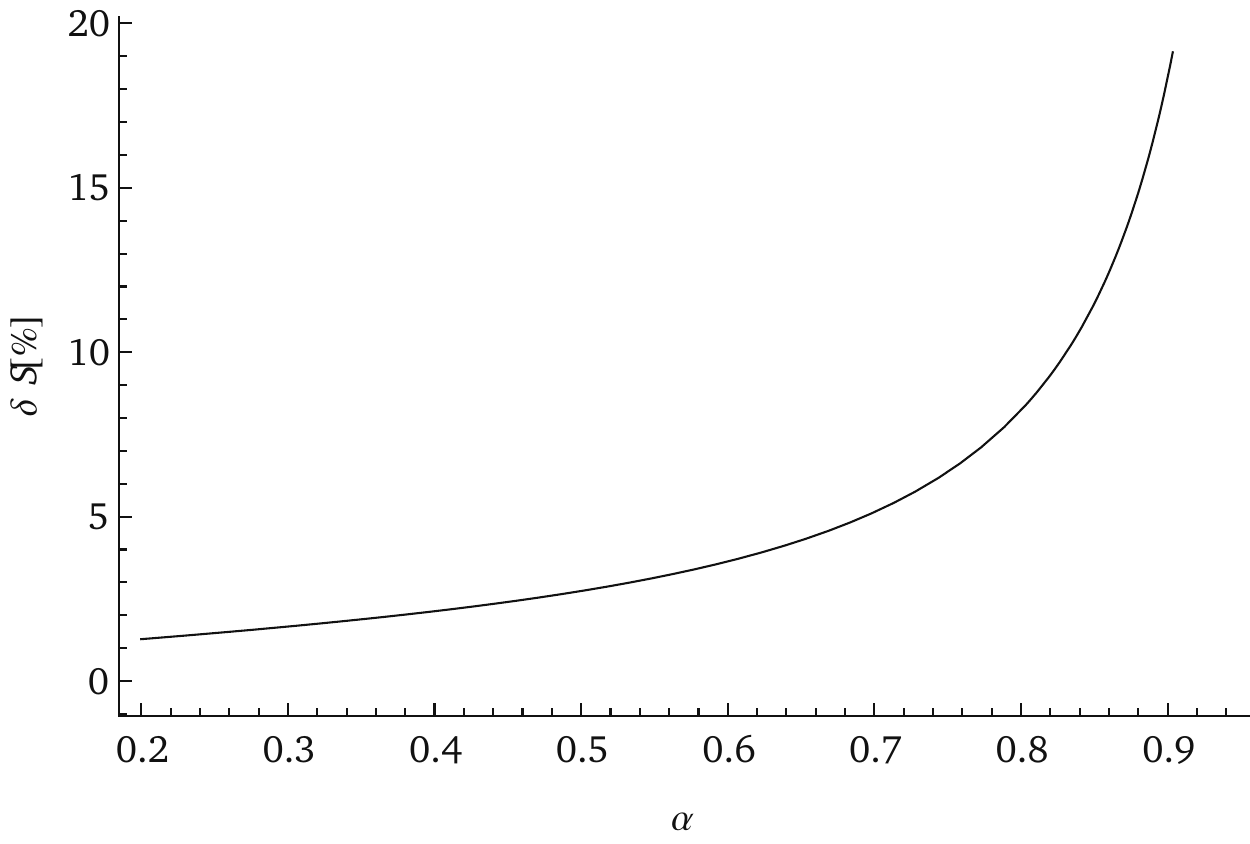}
         \caption{}
         \label{fig:EFTVectorDeterminantLarge}
     \end{subfigure}
     \hfill
     \captionsetup[subfigure]{oneside,margin={-0.1cm,0cm}}
     \begin{subfigure}[b]{0.48\textwidth}
         \centering
         \includegraphics[width=\textwidth]{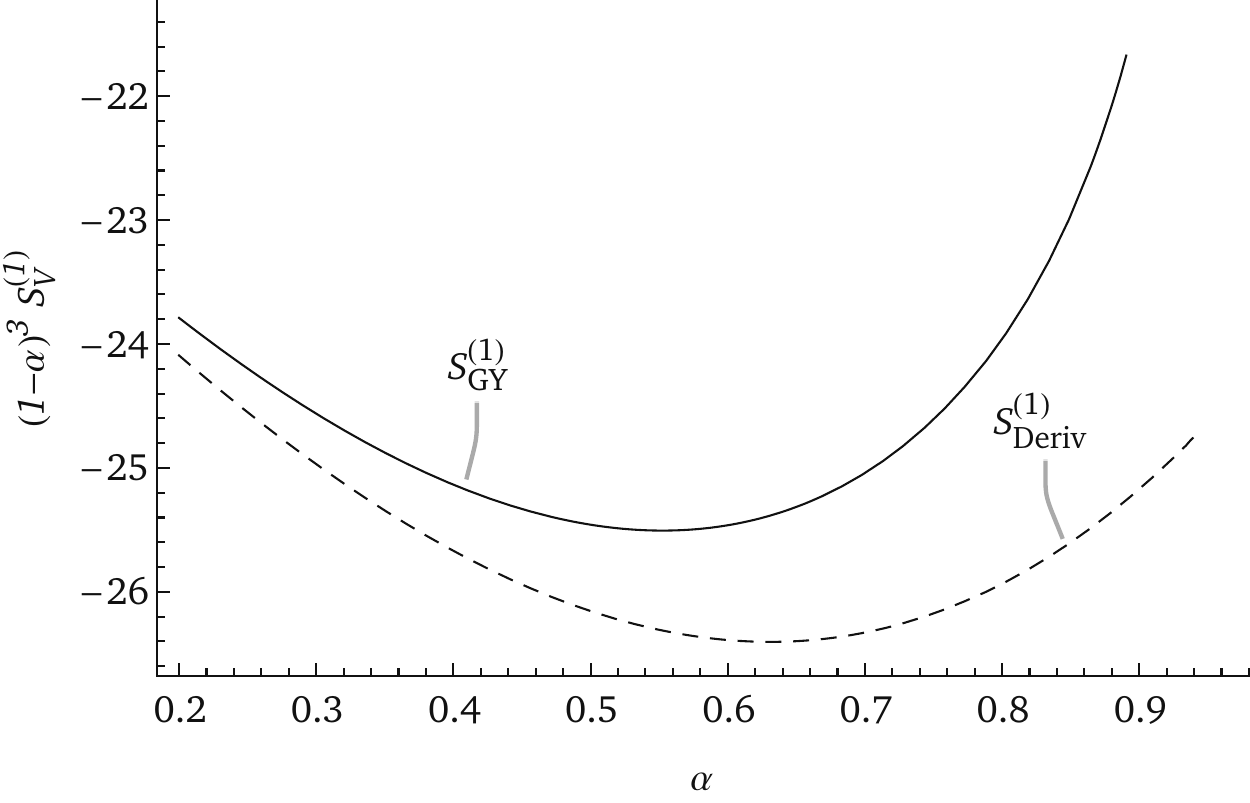}
         \caption{}
         \label{fig:EFTVectorDeterminantLargePlot}
     \end{subfigure}
      \newline
        \captionsetup[subfigure]{oneside,margin={-0.9cm,0cm}}
    \begin{subfigure}[b]{0.48\textwidth}
         \centering
         \includegraphics[width=\textwidth]{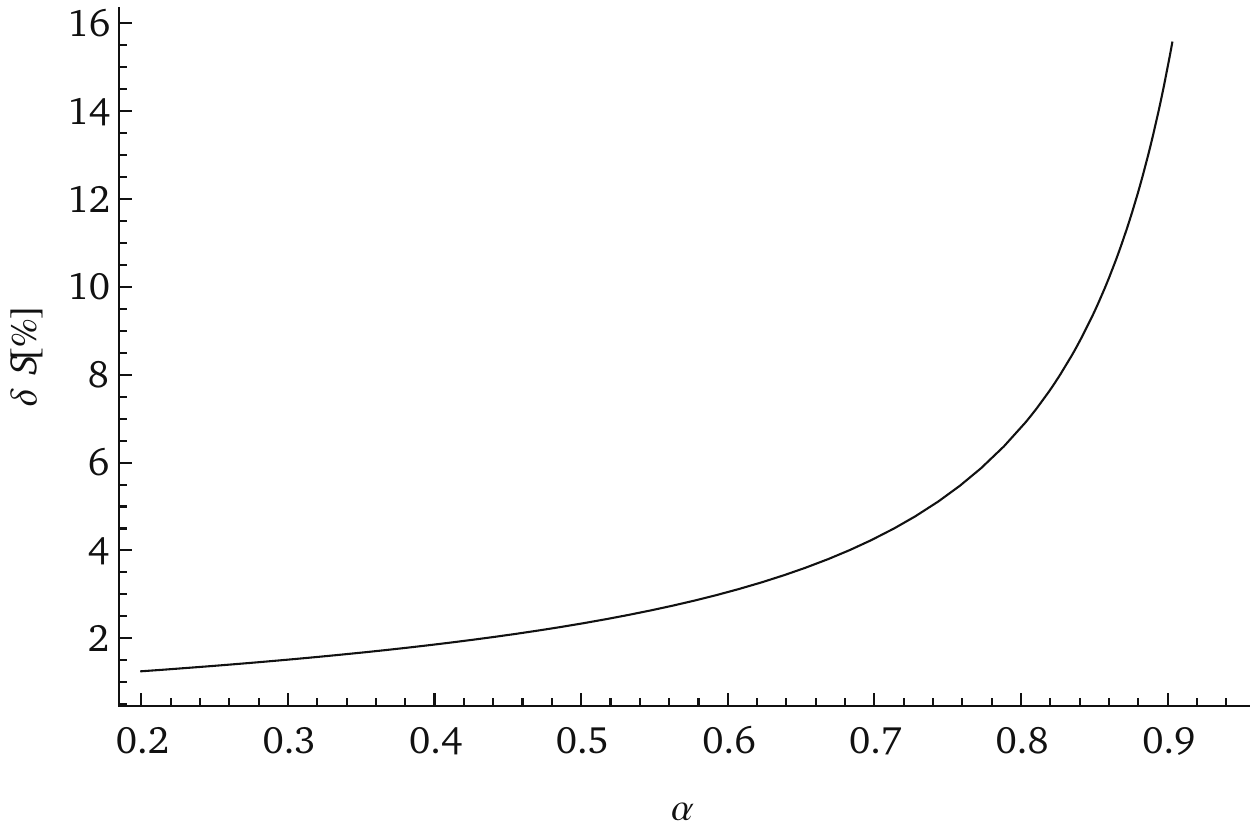}
         \caption{}
         \label{fig:EFTVectorDeterminantLarge2}
     \end{subfigure}
     \hfill
     \captionsetup[subfigure]{oneside,margin={-0.1cm,0cm}}
     \begin{subfigure}[b]{0.48\textwidth}
         \centering
         \includegraphics[width=\textwidth]{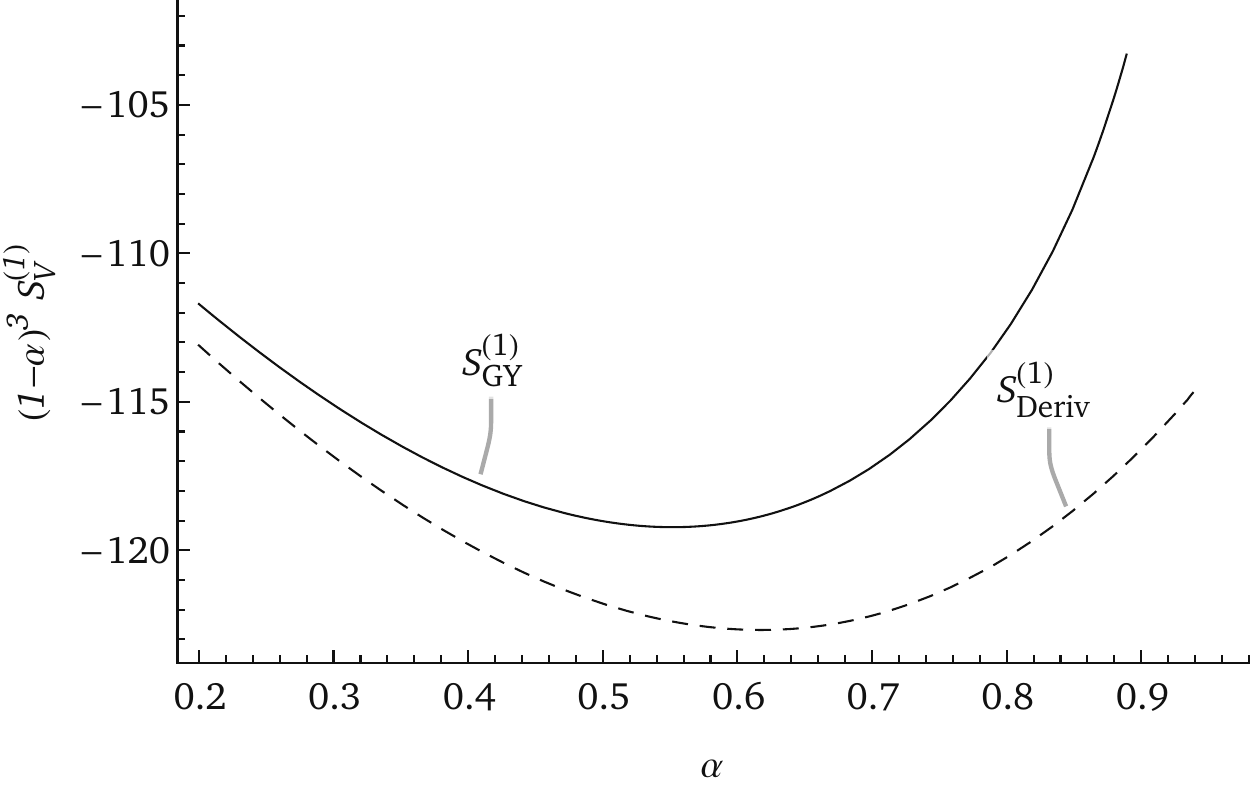}
         \caption{}
         \label{fig:EFTVectorDeterminantLargePlot2}
     \end{subfigure}
	\caption{\small Comparison of the derivative expansion with the Gelfand-Yaglom method for the vector determinant. Figures (a) and (b) assumes $\overline{g}= 1$ . Figure (a) shows $\delta_1 \equiv \left|\frac{S^{(1)}_{GY}-S^{(1)}_\text{Deriv}}{S^{(1)}_{GY}}\right|$, and Figure (b) shows $(1-\alpha)^3~S^{(1)}_V(\alpha)$. Figures (c) and (d) are defined analogously as Figures (a) and (b) but with $\overline{g}= 3$; and Figures (e) and (f) assume $\overline{g}= 5$. 
}
	\label{fig:EFTVector}
\end{figure}

Next the vector bosons. The dimensionless vector-boson coupling is $\overline{g}^2=\frac{g^2}{\lambda}$ where I assume that the vector-boson mass is $m_A= g \phi$. As before, $g$ is equivalent to the weak coupling constant up to a factor of $2$. A priori we would expect that the derivative expansion performs poorly if $\overline{g}\sim 1$ since the expansion parameter is $\frac{\alpha}{\overline{g}^2}$. And this is indeed what Figures \ref{fig:EFTVectorDeterminant} and \ref{fig:EFTVectorDeterminantPlot} shows. What's worse, the phase-transition is stronger for a smaller new-physics scale $\Lambda$ (bigger $c_6$ and $\lambda$)~\cite{Chala:2018ari,Camargo-Molina:2021zcp,Cai:2017tmh,deVries:2017ncy,Croon:2020cgk}; implying that $\overline{g}^2$ gets smaller for smaller $\Lambda$. This means that the derivative expansion gets successively worse for small $\Lambda$ values. Conversely, the derivative expansion gets better for larger $\Lambda$, that is, a larger $\overline{g}^2$. This can be seen in Figures \ref{fig:EFTVectorDeterminantLarge} and \ref{fig:EFTVectorDeterminantLargePlot}, for which $\overline{g}= 3$. The derivative expansion performs better still for $\overline{g}= 5$ as shown in Figures \ref{fig:EFTVectorDeterminantLarge2} and \ref{fig:EFTVectorDeterminantLargePlot2}.

All things considered, the derivative expansion is rather accurate for all $\overline{g}$ values used.

\begin{figure}[t!]
     \centering
     \captionsetup[subfigure]{oneside,margin={-0.9cm,0cm}}
     \begin{subfigure}[b]{0.48\textwidth}
         \centering
         \includegraphics[width=\textwidth]{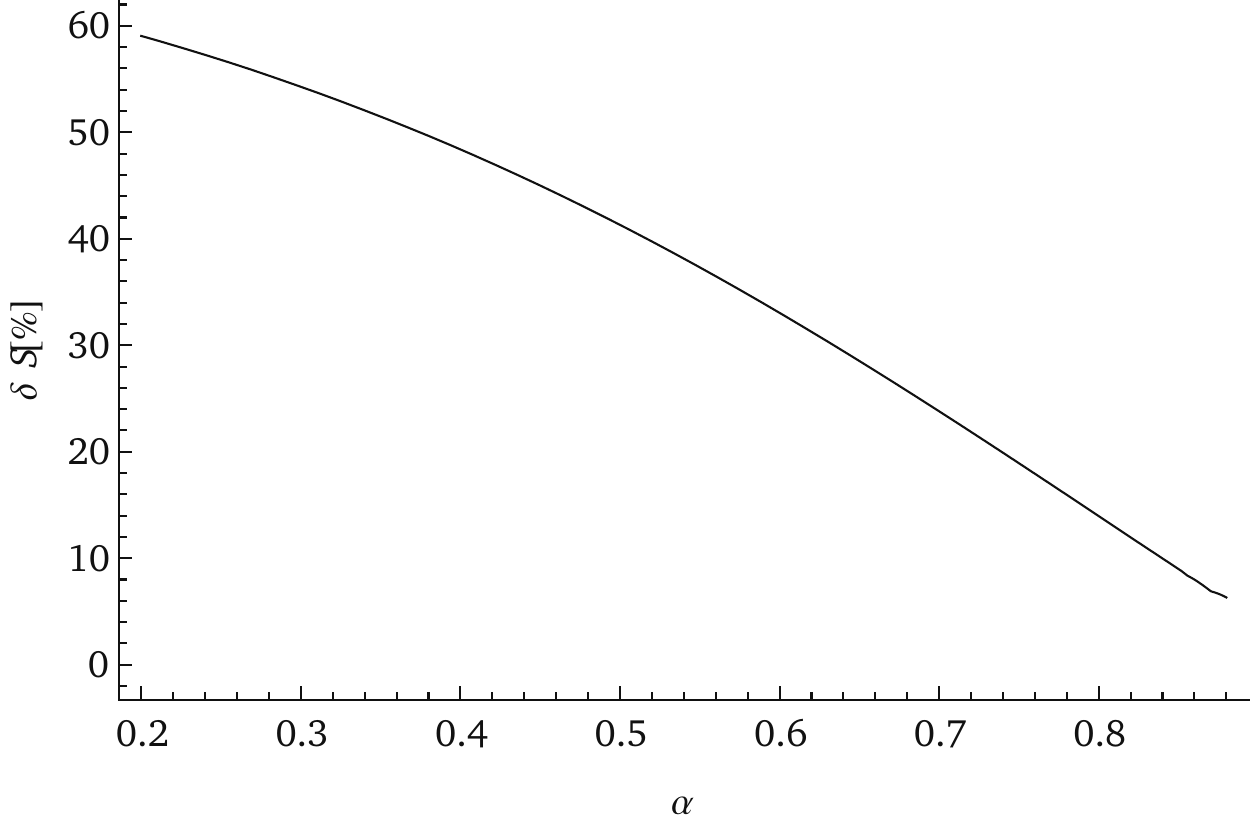}
         \caption{}
         \label{fig:EFT_Mixed}
     \end{subfigure}
     \hfill
      \captionsetup[subfigure]{oneside,margin={-0.1cm,0cm}}
     \begin{subfigure}[b]{0.48\textwidth}
         \centering
         \includegraphics[width=\textwidth]{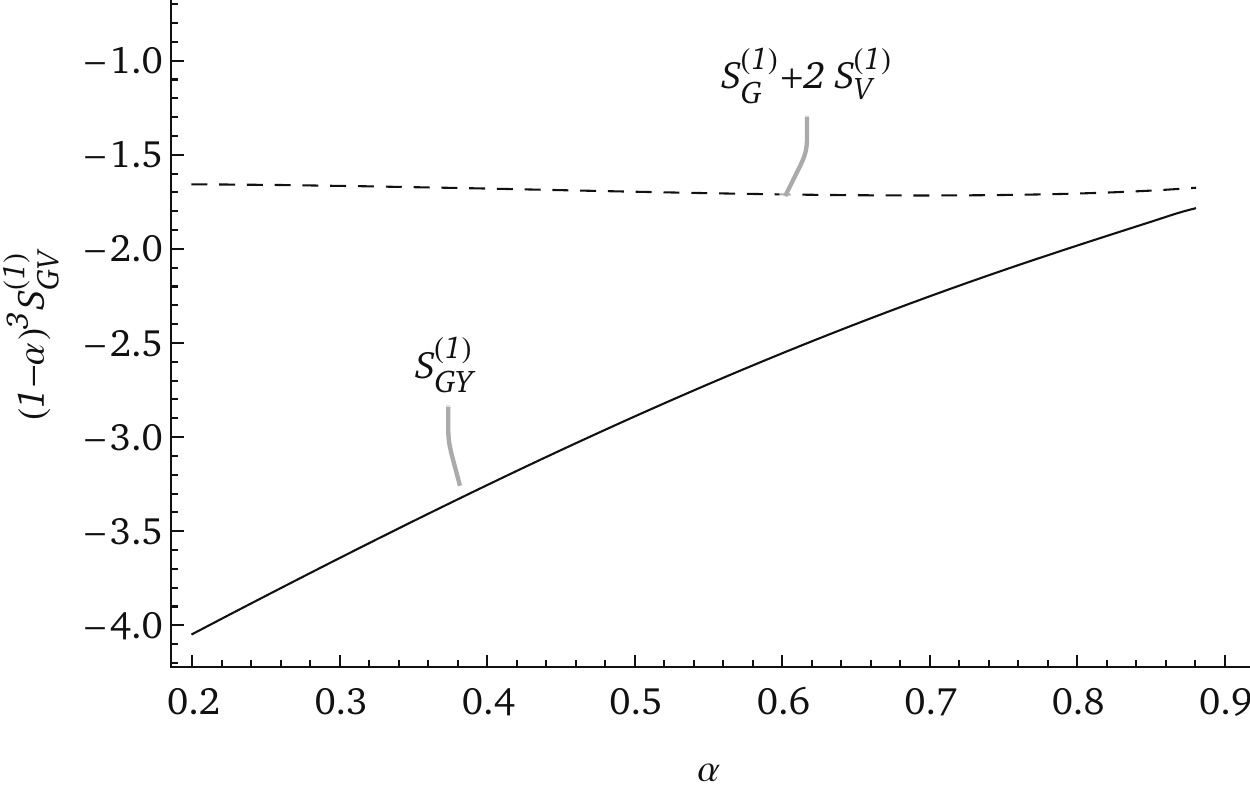}
         \caption{}
         \label{fig:EFT_MixedPlot}
     \end{subfigure}
\newline   
 \captionsetup[subfigure]{oneside,margin={-0.9cm,0cm}}
     \begin{subfigure}[b]{0.48\textwidth}
         \centering
         \includegraphics[width=\textwidth]{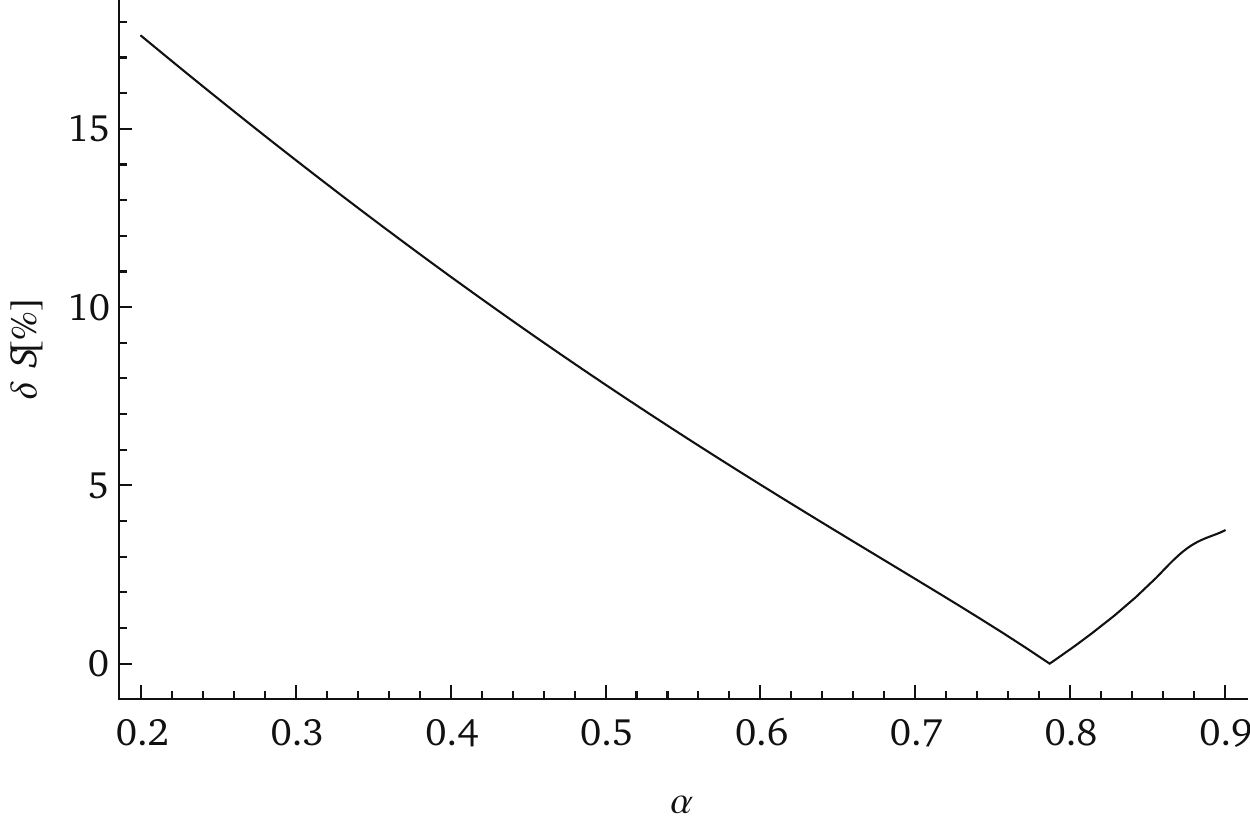}
         \caption{}
         \label{fig:EFT_MixedLarge}
     \end{subfigure}
     \hfill
      \captionsetup[subfigure]{oneside,margin={-0.1cm,0cm}}
     \begin{subfigure}[b]{0.48\textwidth}
         \centering
         \includegraphics[width=\textwidth]{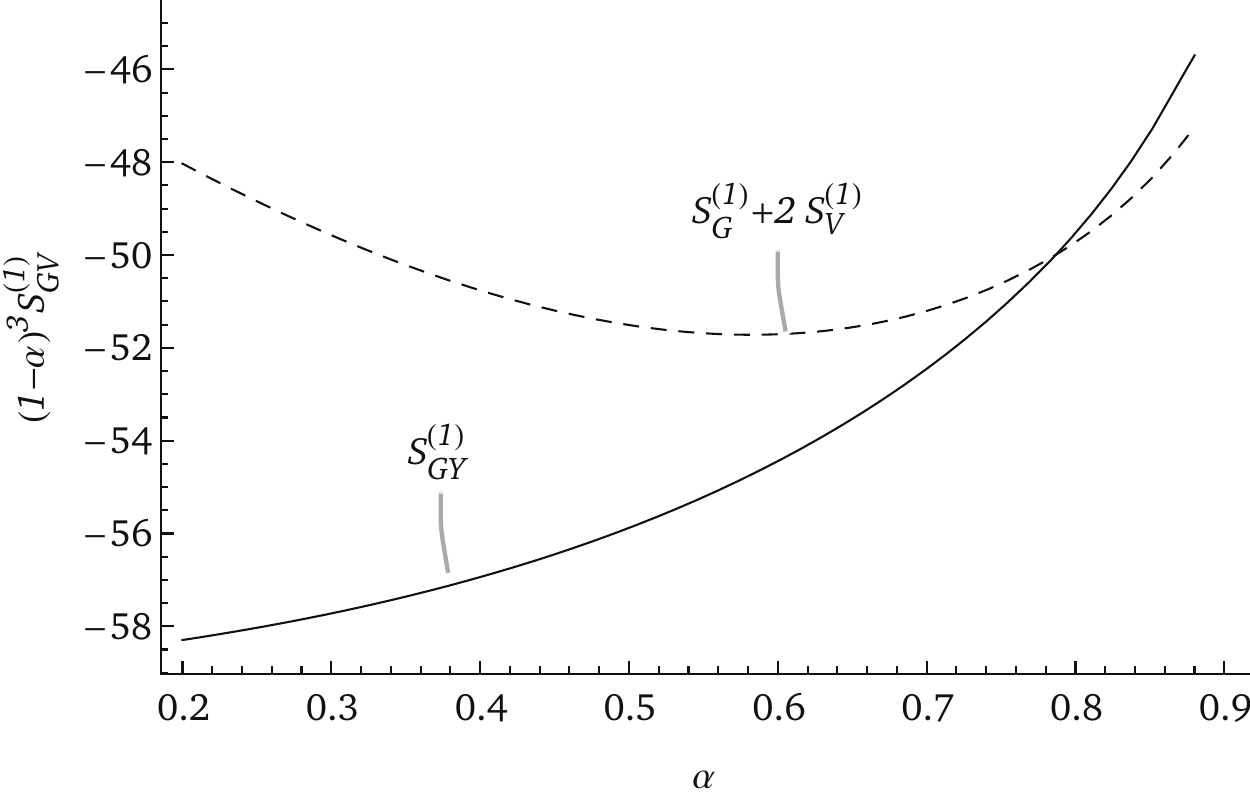}
         \caption{}
         \label{fig:EFT_MixedPlotLarge}
     \end{subfigure}  
     \newline   
 \captionsetup[subfigure]{oneside,margin={-0.9cm,0cm}}
     \begin{subfigure}[b]{0.48\textwidth}
         \centering
         \includegraphics[width=\textwidth]{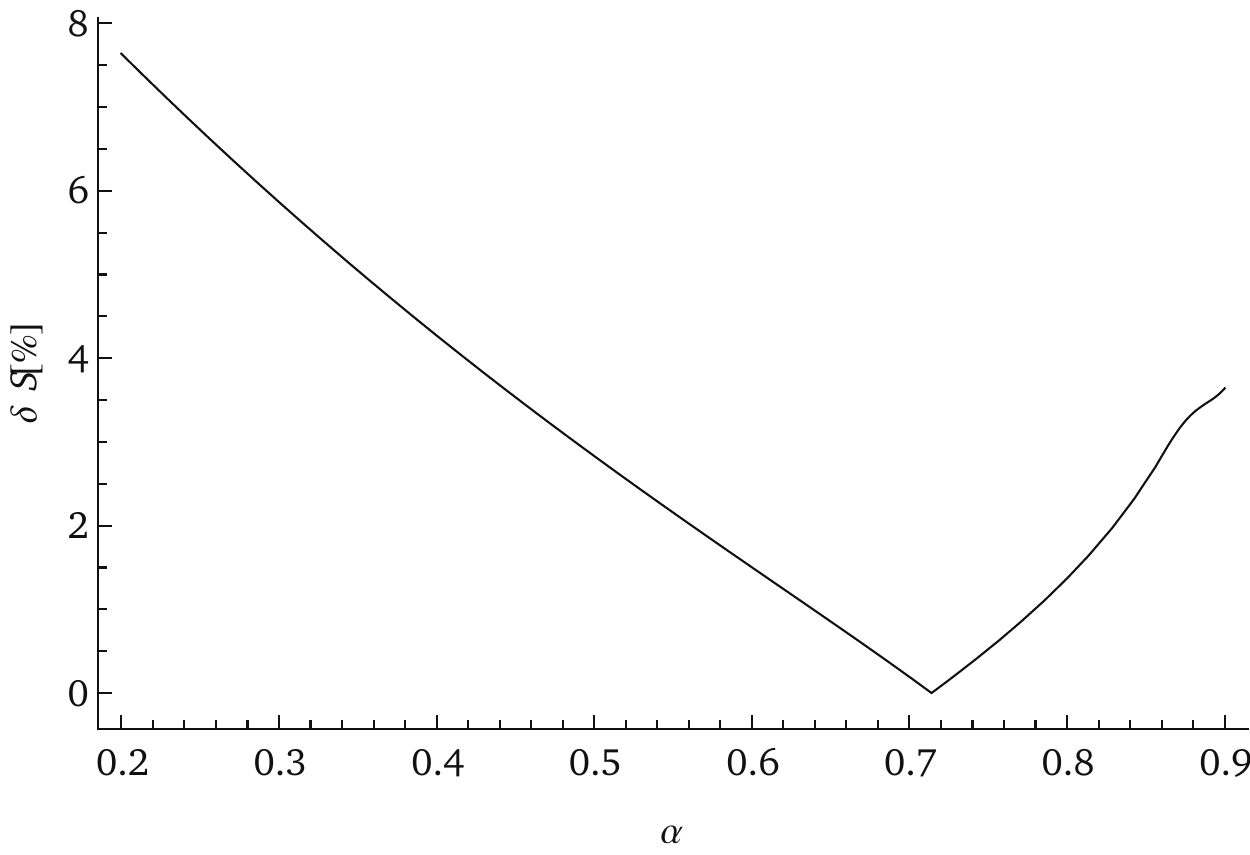}
         \caption{}
         \label{fig:EFT_MixedLarge2}
     \end{subfigure}
     \hfill
      \captionsetup[subfigure]{oneside,margin={-0.1cm,0cm}}
     \begin{subfigure}[b]{0.48\textwidth}
         \centering
         \includegraphics[width=\textwidth]{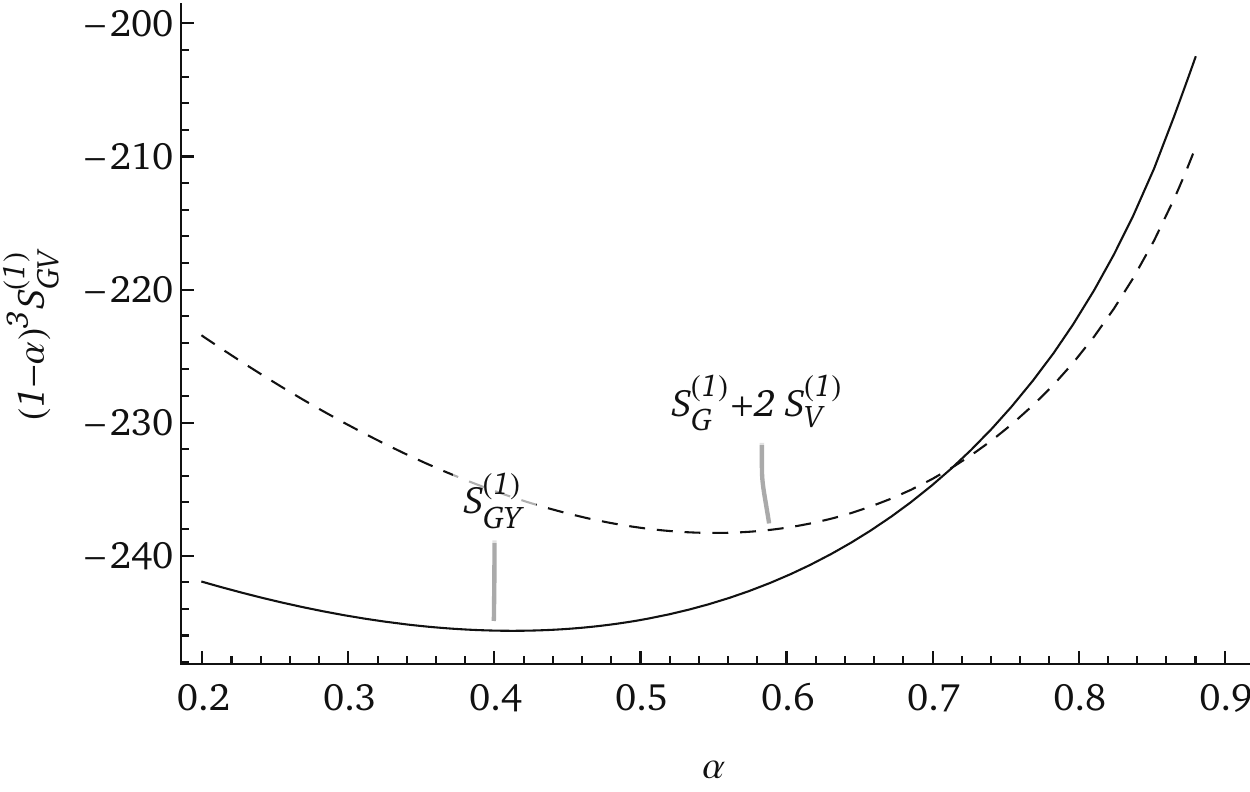}
         \caption{}
         \label{fig:EFT_MixedPlotLarge2}
     \end{subfigure}  
	\caption{\small Comparison of the full Goldstone-Vector determinant with the pure Goldstone and Vector contributions. The Gelfand-Yaglom method is used for all determinants. The Figures assumes that the vector-boson mass is $m_A= g \phi$. Figure (a) and (b) has $\overline{g}=\frac{g^2}{\lambda}=1$; Figure (a) shows  $\delta S \equiv \left|\frac{S^{(1)}_{GV}-S^{(1)}_G-2 S^{(1)}_V}{S^{(1)}_{GV}}\right|$, and Figure (b) compares $(1-\alpha)^3 S_{GV}^{(1)}(\alpha)$ with $(1-\alpha)^3\left[  S_{G}^{(1)}(\alpha)+2 S_{V}^{(1)}(\alpha) \right]$. Figures (c) and (d) are defined analogously but with $\overline{g}=3$. Likewise, Figures (e) and (f) use $\overline{g}=5$.}
	\label{fig:EFT_MixedDet}
\end{figure}

Now for the mixed determinant. Figure \ref{fig:EFT_MixedDet} shows the mixed determinant compared with treating the determinant as diagonal, that is, the comparison of $S^{(1)}_{GV}(\alpha)$ with $S^{(1)}_{G}(\alpha)+2S^{(1)}_{V}(\alpha)$. To be specific, with $S^{(1)}_{GV}(\alpha)$ I include the full contribution of Goldstones, all Vector polarizations, and ghosts. But I don't include more than one Goldstone-Vector contribution. That is, I don't sum over the representation. Also $S^{(1)}_V(\alpha)$ denotes the transverse-vector determinant.

A priori we would expect that $S^{(1)}_{GV}(\alpha)$ should approach $S^{(1)}_{G}(\alpha)+2S^{(1)}_{V}(\alpha)$ for large  $\overline{g}$. Reason being that the derivative expansion should be more-and-more apt the larger $\overline{g}$ gets. Specifically we expect corrections to $S^{(1)}_{GV}(\alpha)=S^{(1)}_{G}(\alpha)+2S^{(1)}_{V}(\alpha)$ to be of order $\overline{g}$, while $S^{(1)}_{GV}(\alpha)$ and $S^{(1)}_{V}(\alpha)$ are of order $\overline{g}^3$. And this is indeed what Figure \ref{fig:EFT_MixedDet} shows. In particular note that the error is significant for $\overline{g}=1$ as shown in Figure \ref{fig:EFT_Mixed}. While Figures \ref{fig:EFT_MixedLarge} and \ref{fig:EFT_MixedLarge2} show that the error rapidly decrease for larger $\overline{g}$.

By the by, both Figure \ref{fig:EFTVector} and \ref{fig:EFT_MixedDet} shows that the vector contribution grows as $\sim \overline{g}^3$, which is expected from the derivative expansion.

\subsection{Possible Issues with the SM-EFT Potential }
\label{sec:EFTvsLoopInduced}
The previous sections discussed models with a tree-level barrier and a loop-induced barrier, however, there is some overlap between these two cases.
And sometimes it's necessary to consider an amalgamation of the two. To see why, consider again the SM-EFT potential. For realistic parameters one finds $\beta=m \lambda^{-1}\sim 3-5$ while $\overline{g}\sim 1-5$. With $\beta$ and $\overline{g}$ being defined in Section \ref{sec:EFTPotential}. This means that vector-boson determinant can overpower the leading-order bounce determinant, and perturbativity is lost.

Let's see how. Define $c_6=c_6^{\star}+x_6$ so that $\lambda\sim  x_6$.{\interfootnotelinepenalty=10000 \footnote{Evaluating the (4-d) parameters at tree-level gives: $\lambda^{4d}=\frac{m_H^2}{2 v_H^2}-\frac{3}{8}c_6^{4d} v_h^2$~\cite{Croon:2020cgk,deVries:2017ncy,Cai:2017tmh,Chala:2018ari,Camargo-Molina:2021zcp}. Where $v_H\approx 246~\si{\giga\electronvolt}$. The dimensionally reduced parameters are to first-order (the minus-sign comes from my convention) $\lambda=-\lambda^{4d} T$ and $c_6= T^2 c_6^{4d}$. See \cite{Croon:2020cgk} for the details.}} From equation \ref{eq:EFTpotentialAction} we see that, close to the critical temperature, $\alpha =c_6 \beta^2 \sim 1$; other power-counting parameters scales as
\begin{align}
\beta \sim \frac{1}{\sqrt{c_6^\star}}\approx 4~\&~\lambda\sim m\sim x_6~\&~\overline{g}^2\sim\frac{1}{x_6}.
\end{align}

This is a problem. Not only is $\beta$ to first approximation a constant, but the vector-boson determinant surges for small $\lambda$.\footnote{Technically the 3-D $\lambda$ never vanishes even though the 4-D $\lambda$ does. However, for practical purposes this is irrelevant.} Rapidly. In fact, from the derivative expansion (this is also the case for Gelfand-Yaglom method) we expect the vector determinant to grow as $\log \det V\sim \overline{g}^3\sim \left(x_6\right)^{-3/2}$. Worse yet, as the temperature decrease and $\alpha$ gets smaller, so does $\beta$\te but not $\overline{g}$. This means that perturbation theory is doomed to break down for small $\lambda$, or equivalently, when the scale of new physics becomes large. This breakdown is hinted at in Figure 9 of~\cite{Croon:2020cgk}.

It should be stressed that the perturbative problems are not due to a breakdown of the derivative expansion. Quite the opposite. As shown in Figures \ref{fig:EFTVector} and \ref{fig:EFT_MixedDet}, the expansion works better for large $\overline{g}$ (small $\lambda$). Rather, the problem is that the assumed scaling $\lambda \sim g^2$ is incorrect. The same problems also appear for the effective potential with small quartics~\cite{Ekstedt:2020abj}. 

This hints at that it's possible to eliminate these issues by choosing a different power-counting. That is, to count $\lambda \sim g^3$. Because with such a power-counting the vector determinants should be expanded in powers of $\frac{\lambda}{g^2}$\te giving a loop-induced barrier.

So, does the loop-induced potential have the same perturbative problems for small $\lambda$? No. Consider the parameters for the loop-induced potential discussed in Section \ref{sec:LoopInduced}:
\begin{align}
&\alpha \sim m^2 \lambda \eta^{-2}\sim 1,
\\&\beta^2=\alpha^3\left(\frac{\eta^2}{\lambda^3}\right)\sim \frac{g^6}{\lambda^3},
\\&\overline{g}^2= \frac{m^2}{\eta^2}g^2=\alpha \frac{g^2}{\lambda}\sim \frac{g^2}{\lambda},
\end{align}
where I assumed $\eta \sim g^3$ as expected from the derivative expansion.

Now, the NLO contribution from vector bosons scales as $\overline{g}^4\beta^{-1}$. Comparing this with the leading-order-bounce action $\sim \beta$, gives a relative factor $\frac{\overline{g}^4}{\beta^2}\sim \frac{\lambda}{g^2}\sim g$. Actually, at $N$ loops, the vector-boson contribution scales as $\overline{g}^3\left(\frac{\overline{g}}{\beta}\right)^{N-1}\sim g^{-3}(g)^{N-1}$. So higher loops are suppressed by powers of $g$. 

Also, remarkable it's possible\te actually encouraged\te to include the two-loop effective potential as a sub-leading contribution. Only the pure-vector contribution mind you. Doing so would greatly reduce theoretical uncertainties for the nucleation rate.

Let's turn back to the SM-EFT case. As mentioned, perturbative problems pop-up for small $\lambda$s. The expansion can be improved by changing the power-counting to $\lambda \sim g^3$\te justifying a derivative expansion of the vector bosons. In short, the new leading-order potential is
\begin{align}
V(\phi)=\frac{1}{2}m^2\phi^2-\frac{1}{2}\eta \phi^3-\frac{\lambda}{4}\phi^4+\frac{c_6}{32}\phi^6,~ \eta \sim g^3.
\end{align}

The inclusion of the $\phi^3$ term at leading order cures the perturbative problems and gives a well-behaved expansion around the leading-order bounce action. Note that the inclusion of such a term can easily give an order one corrections to the gravitational wave spectrum. This is especially prevalent for small $\lambda$ values. An example of these considerations applied to the effective potential (physics at the critical temperature) can be found in~\cite{Camargo-Molina:2021zcp}.

All things considered, you are admonished to be wary when studying a model with a $\phi^6$ term. In practice, use the derivative expansion to gauge the convergence of perturbation theory. And if $\lambda$ is small, use the derivative expansion of the vector bosons (not scalars) to include a $\phi^3$ term at tree-level. The expansion is then well-behaved, and corrections to the leading-order-bounce action are equivalent to those discussed in Section \ref{sec:LoopInduced}.

\subsection{Two-Scalar Determinant}
\label{sec:MultiField}
Both the loop-induced and SM-EFT potentials involve just a single bounce. But the methods discussed so far works for more complicated models to be sure. To that end consider the potential~
\begin{align}
\label{eq:TwoScalarPotential}
V(\phi,\sigma)=-\frac{1}{2}m_\phi^2 \phi^2 -\frac{1}{2}m_\sigma^2 \sigma^2+\frac{1}{8}\lambda_\phi \phi^4+\frac{1}{8}\lambda_\sigma \sigma^4+\frac{1}{4}\lambda_{\phi\sigma} \phi^2 \sigma^2. 
\end{align}
This potential appears, mayhap with additional terms, in Two-Higgs-doublet models~\cite{Turok:1991uc,Land:1992sm}, singlet extensions~\cite{Huang:2018aja,Dev:2019njv,Niemi:2021qvp,Schicho:2021gca,Gould:2019qek,Alves:2018jsw,Barger:2007im}, and triplet extensions~\cite{Niemi:2018asa,Bell:2020hnr,Niemi:2020hto}. Note that in some of these models the scalar-determinant includes other fields than just $\phi$ and $\sigma$. To keep the discussion simple I will assume that the $\phi$ and $\sigma$ fields are the only fields that talk to each other; that is, I will not consider other Goldstone bosons or other scalars. This is a choice out of convenience, not necessity.

The above potential can be written in a dimensionless form through the redefinitions
\begin{align}
x\rightarrow m_\phi^{-1} x,~\phi \rightarrow \frac{m_\phi}{\sqrt{\lambda_\phi}}\phi,~\sigma \rightarrow \frac{m_\phi}{\sqrt{\lambda_\phi}}\sigma
\end{align}

This gives
\begin{align}
&S_3=\beta \int d^3x \left[\frac{1}{2}(\partial \phi)^2+\frac{1}{2}(\partial \sigma)^2-\frac{1}{2} \phi^2 -\frac{1}{2}\alpha_1 \sigma^2+\frac{1}{8}\phi^4+\frac{1}{8}\alpha_2 \sigma^4+\frac{1}{4}\alpha_3 \phi^2 \sigma^2\right]
\\& \beta=m_\phi \lambda_\phi^{-1},~\alpha_1=\frac{m^2_\sigma}{m^2_\phi},~\alpha_2=\frac{\lambda_\sigma}{\lambda_\phi},~\alpha_3=\frac{\lambda_{\phi\sigma}}{\lambda_\phi}.
\end{align}

This means that the determinants depend on three parameters: $\alpha_1$, $\alpha_2$, and $\alpha_3$. To make matters interesting I here focus on a two-step phase transition. An easy way to make this happen is to take all parameters as positive, choose $\alpha_3=\frac{2 \alpha_2}{\alpha_1}$ and $\alpha_1 \leq \alpha_2^2$. This choice of $\alpha_3$ ensures that one of the scalars has a mass equal to 1. In addition, the scalar potential is redefined so that the action evaluated on the false-vacuum solution is zero.

The scalar determinant can be found numerically by using the methods discussed in Sections \ref{section:MixedOperators} and \ref{subsec:ZeroMode}. Next, the leading term in the derivative expansion is
\begin{align}
S^{(1)}_H=-\frac{1}{12\pi}\sum_{i=1,2}\int d^3x \left[A_i^{3/2}(x)-A_{i,\text{FV}}^{3/2}\right],
\end{align}
where $A_1$ and $A_2$ are defined in Section \ref{section:MixedOperators}.

\begin{figure}[t!]
     \centering
     \captionsetup[subfigure]{oneside,margin={-0.9cm,0cm}}
     \begin{subfigure}[b]{0.48\textwidth}
         \centering
         \includegraphics[width=\textwidth]{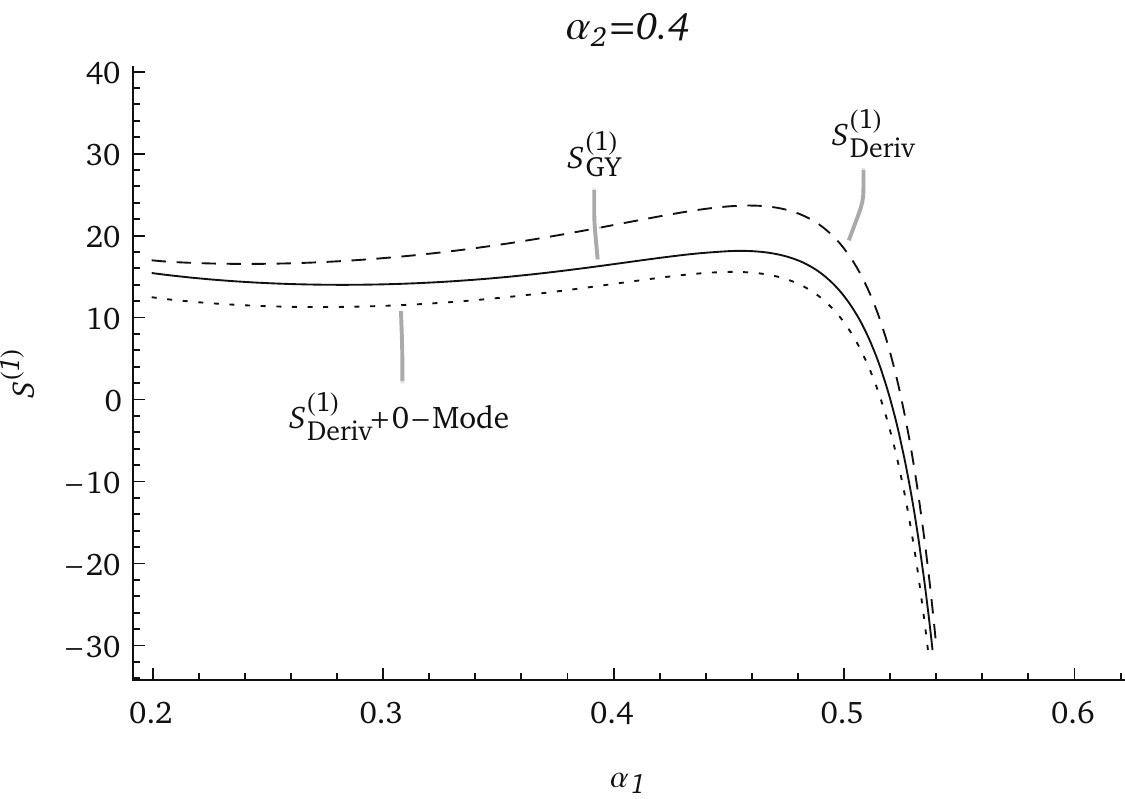}
         \caption{}
         \label{fig:Singlet1}
     \end{subfigure}
     \hfill
      \captionsetup[subfigure]{oneside,margin={-0.1cm,0cm}}
     \begin{subfigure}[b]{0.48\textwidth}
         \centering
         \includegraphics[width=\textwidth]{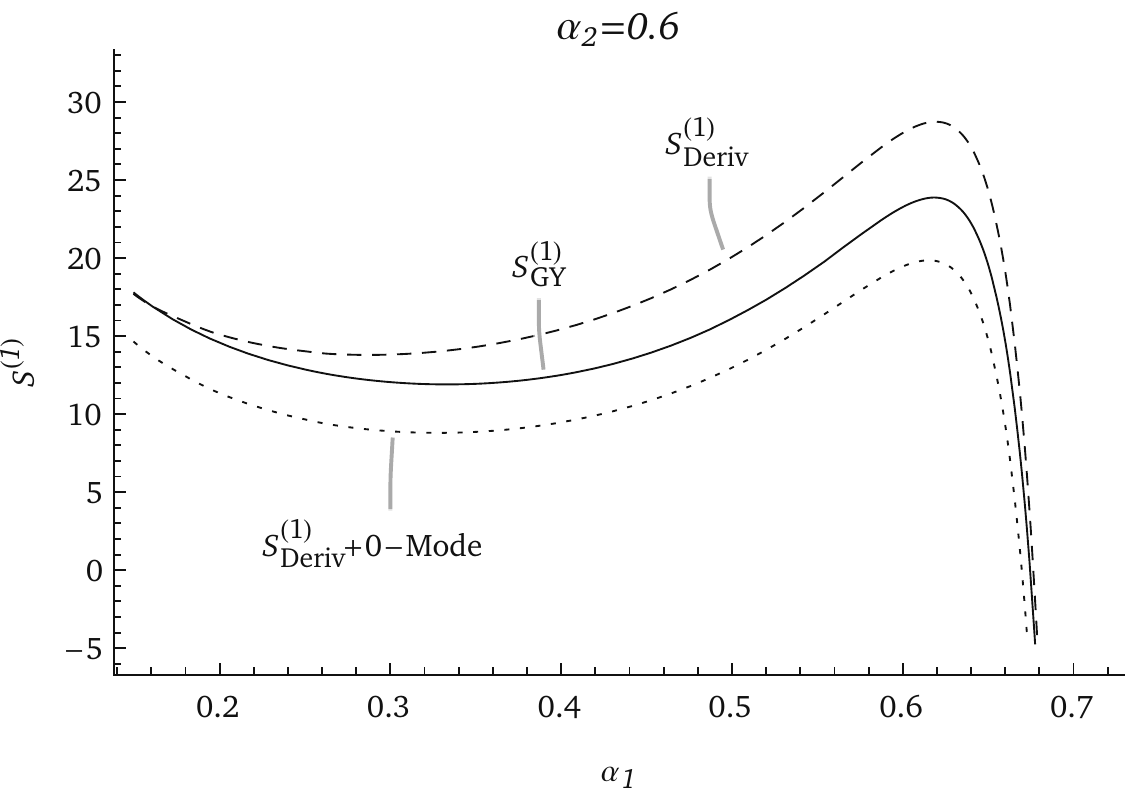}
         \caption{}
         \label{fig:Singlet2}
     \end{subfigure}
     \newline
        \captionsetup[subfigure]{oneside,margin={-0.9cm,0cm}}
    \begin{subfigure}[b]{0.48\textwidth}
         \centering
         \includegraphics[width=\textwidth]{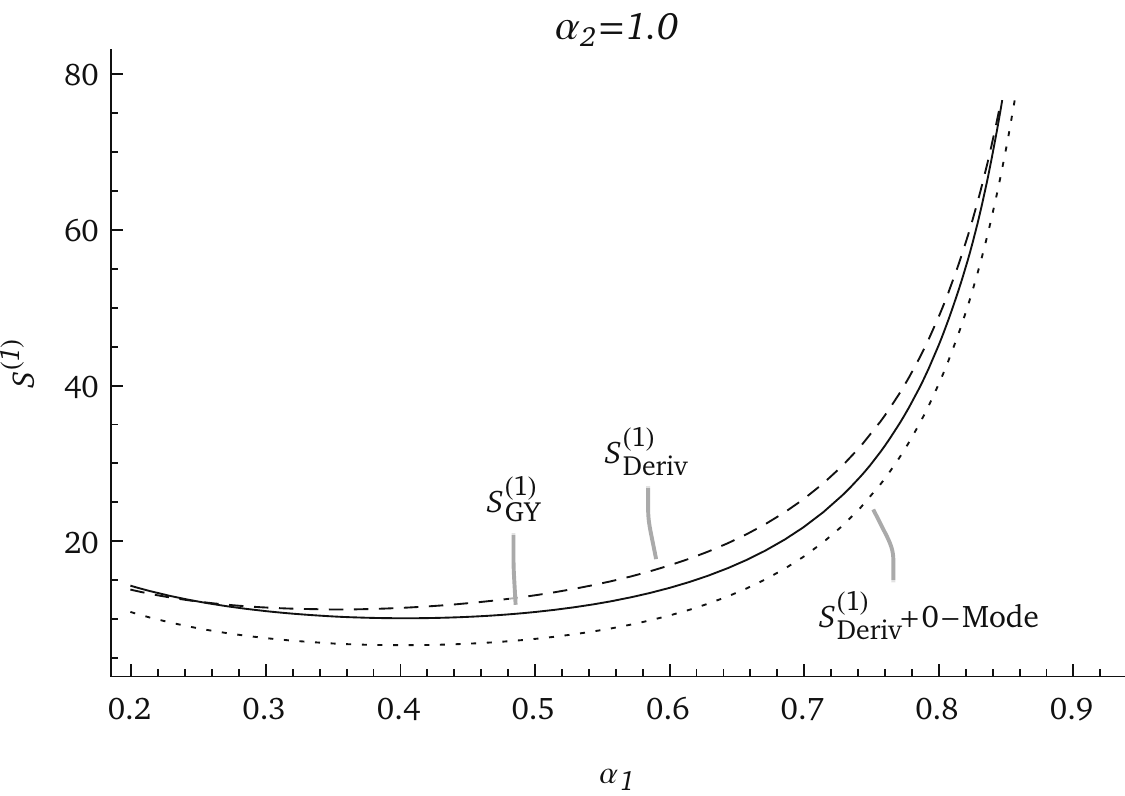}
         \caption{}
         \label{fig:Singlet3}
     \end{subfigure}
     \hfill
     \captionsetup[subfigure]{oneside,margin={-0.1cm,0cm}}
     \begin{subfigure}[b]{0.48\textwidth}
         \centering
         \includegraphics[width=\textwidth]{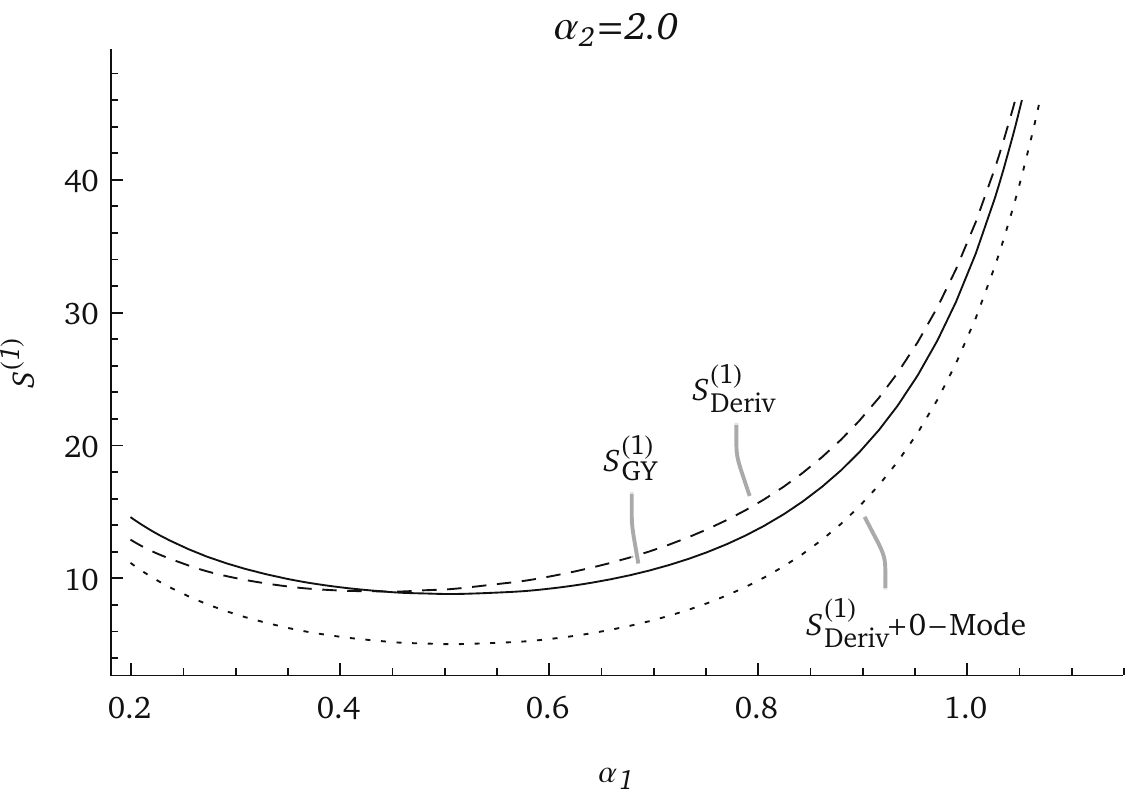}
         \caption{}
         \label{fig:Singlet4}
     \end{subfigure}
	\caption{\small Comparison of the derivative expansion with the Gelfand-Yaglom method for the scalar determinant. Figure (a) assumes $\alpha_2=0.4$; Figure (b) $\alpha_2=0.6$; Figure (c) $\alpha_2=1.0$; and Figure (d) $\alpha_2=2.0$. The $\alpha_1$ scales are different because the range is restricted to $\alpha_1^2 \leq \alpha_2$. The zero mode factor is $(-1)\frac{3}{2}\log \frac{\tilde{S}_3(\alpha)}{2\pi} $, where $S_3=\beta \tilde{S}_3(\alpha).$ 
}
	\label{fig:Singlet}
\end{figure}

 Below I focus on the first-order phase transition from $\phi=0$ to $\phi=\text{finite}$. Figure \ref{fig:Singlet} shows the scalar determinant for various value of $\alpha_2$. The range of $\alpha_1$ is different for each figure because $\alpha_1^2 \leq \alpha_2$. 
 
 From Figure \ref{fig:Singlet} we see that the error when including the zero-mode factor is comparable to not including it. And it does seem like the derivative expansion reproduces the general behaviour of the Gelfand-Yaglom method. Barring some quantitative differences. It should nonetheless be stressed that the uncertainty from using the derivative-expansion, with regards to gravitational waves, is quite small.

Finally, let's look at if perturbation theory is well-behaved. As before the vector-boson determinants are controlled by $\overline{g}^2=\frac{g^2}{\lambda_\phi}$. And the leading-order action scales as $\beta \sim \frac{m_\sigma}{\lambda_\sigma}$. So if $\lambda_\phi \sim g^2$, we require $\frac{m_\sigma}{\lambda_\sigma}\gg 1$ for a well-behaved expansion. Note that, just as for the SM-EFT case, the perturbative expansion is questionable when $\alpha_1 \rightarrow 0$.

Note however that the potential in equation \ref{eq:TwoScalarPotential} is minimal. And the potential is often augmented by additional terms that might modify these conclusions.

	\section{Discussion}\label{Sec:Discsussion}
In this paper I have calculated higher-order corrections to the nucleation rate for a variety of models. Doing so I have compared a numerical implementation of the Gelfand-Yaglom theorem with the derivative expansion. This comparison shows that the derivative expansion is decent for some models; for others not so much.

For example, it seems that the derivative expansion works quite well for the loop-induced potential discussed in Section~\ref{sec:LoopInduced}. The results also show that using the zero-mode factor in conjunction with the derivative expansion gives a worse fit than without. This is contrary to the tree-level barrier potential discussed in Section~\ref{sec:EFTPotential}; in this model including the zero-mode factor gave a better fit. While for the two-scalar-field model discussed in Section \ref{sec:MultiField}, adding the zero-mode factor doesn't change the results one way or another.

This suggests that there isn't any clear-cut way of using the derivative expansion in the presence of zero-modes. As such it is advisable to avoid using the derivative expansion save when it's strictly necessary, which is to say, only when it's forced by the chosen power-counting.

It should however be stressed that while using the derivative expansion introduces some uncertainties, these don't necessary affect observables, or gravitational wave predictions, too much. Seeing as the leading-order action should dominate the functional determinants. Nonetheless, for accurate predictions, such uncertainties are not desirable. So, I would venture that using the Gelfand-Yaglom method is just as easy as using the derivative expansion, and, with the added benefit of not having to worry about whether to include zero-mode factors or not. 

Not to say that the derivative expansion should be abandoned. In reality, as discussed in Section \ref{sec:DerivativeExpansion}, the derivative expansion can be essential. This is evidenced by the discussion in Section \ref{sec:EFTvsLoopInduced}. In this section I argued that the derivative expansion can ameliorate a badly converging perturbative expansion of the rate. Specifically in models with large vector-boson determinants. 

In short, when calculating the bubble-nucleation rate, one should be cautious of higher-order corrections; large corrections, for example estimated using the derivative expansion, can point to a flawed power-counting. In these cases the solution is to change the power-counting and apply perturbation theory accordingly.

In addition, the results show that leading-order calculations for bubble nucleation are insufficient. To actually assess the convergence, higher-order corrections are a must.

Furthermore, the three dimensional models studied in this paper maps to a wide variety of four dimensional models. As such, the parametrization of higher-order determinants in this paper can directly be applied to studying beyond the Standard Model physics. 

As for future prospects, it would be interesting to use the results of this paper to study the uncertainty of gravitational-wave calculations in detail. That is, to make in-depth studies of popular models; like Two-Higgs doublet models, singlet extensions, and real-triplet extensions. Such studies would undoubtedly further pin down uncertainties for the gravitational wave spectrum, and push the current understanding forward.
	\section*{Acknowledgments}
 I would like to thank Johan Löfgren for interesting discussions and for reading through the manuscript. 
This work has been supported by the Grant agency of the Czech Republic, project no. 20-17490S and from the Charles University Research Center UNCE/SCI/013.

\appendix
\section{Corrections to the Bounce}\label{app:BounceCorrection}

I here want to calculate corrections to the bounce using Green's functions.
The starting point is
\begin{align}
	&\frac{\delta S_\tNLO[\phi_\tLO]}{\delta \phi(x)}=\frac{
		1}{2}  G^H(x,x)V'''_\tLO\left[\phi_\tLO(x)\right],
	\\&(-\partial^2+V''_\tLO\left[\phi_\tLO(x)\right])G^H(x,y)=\delta^{(3)}(x-y).
\end{align}
This follows from 
\begin{align}
	S_\tNLO=\frac{1}{2}\tr\log\left[ -\nabla^2+V'''_\tLO\right].
\end{align}

So inverting equation \ref{eq:BounceCorrections} gives\footnote{Technically one of the Green's functions should have the boundary condition that $\partial_r G(x,x')=0$ when $|\vec{x}|=0$. As it happens this detail is irrelevant for the present discussion.}
\begin{align}
	\label{eq:BounceCorrection2}
	\phi_\tNLO(x)=-\frac{1}{2}\int d^3y  G^H(x,y)  G^H(y,y)V'''_\tLO\left[\phi_\tLO(y)\right].
\end{align}

Second, owing to spherical symmetry, the Higgs Green's function can be expanded in spherical harmonics. The radial Green's function then satisfies
\begin{align}
	\left(-\partial^2-2/r \partial +l(l+1)/r^2+V''_\tLO\left[\phi_\tLO(r)\right]\right)G^H_l(r,r')=-\delta(r-r')/r^2.
\end{align}
And the total Green's function is~\cite{1993FBS....14....1F}
\begin{align}
	G^H(x,x')=\frac{1}{4\pi}\sum_{l=0}^\infty (2l+1)C_l^{(1/2)}(\cos \alpha)G^H_l(r,r'),
\end{align} 
where $\cos \alpha=\frac{\vec{x}\cdot \vec{x'}}{|\vec{x}||\vec{x'}|}$, and $C_l^{(1/2)}(x)$ are Gegenbauer polynomials.

Now, take a look at the different pieces in equation \ref{eq:BounceCorrection2}: $G^H(y,y)$ involves the same space-time point and does not depend on any angles; $V'''_\tLO\left[\phi_\tLO(y)\right]$ only depends on $|\vec{y}|$; $G^H(x,y)$ only depends on the relative angle between $\vec{x}$ and $\vec{y}$. Thus $\phi_\tNLO(x)$ only depends on $|\vec{x}|$ as promised.

Everything has so far been exact, and in principal both $\phi_\tNLO(x)$ and $G^{H}(x,y)$ can be found numerically. Nevertheless, let's quit being circumspect. Assume that $S_\tNLO\left[\phi_\tLO\right]$ is known. Whether it be by numerical evaluation or some other approximation does not matter. The next correction is of order $\hbar^2$. Let's be pragmatic and replace  $S_\tNNLO\left[\phi_\tLO\right]$ with the leading-order term from the derivative expansion\te the effective potential. The derivative expansion does not converge, but we can still use the first term as a rough estimate. This has the added benefit of reducing the renormalization-scales dependence~\cite{Gould:2021oba,Endo:2015ixx}.

Second, use the derivative expansion to find $\phi_\tNLO(x)$. That is, treat $V''_\tLO\left[\phi_\tLO(x)\right]$ as a constant when finding the Green's function. This gives~\cite{Braaten:1995cm}
\begin{align}
	&G^H(x-y)=\frac{1}{4\pi|x-y|} \exp\left(-\kappa |x-y|  \right),
	\\& \kappa\equiv \sqrt{V''_\tLO\left[\phi_\tLO(x)\right]}.
\end{align}

We also have to ensure that $\phi_\tNLO(r)\rightarrow 0$ as $r\rightarrow \infty$. This condition is quite simple to enforce. It turns out that if we choose the counter-terms such that $\phi^{FV}_\tNLO=0$, then $\phi_\tNLO(r)$ automatically vanishes for large r.

Finally we need to evaluate
\begin{align}
	\int d^3y \frac{\delta S_\tNLO[\phi_\tLO]}{\delta \phi(y)}G^H(x-y).
\end{align}

To do this let's again use the Derivative expansion. To make the results lucid, consider the Higgs contribution. This gives\footnote{In the derivative expansion the leading-order term is $S_\tNLO^H[\phi_\tLO]=-\frac{1}{12\pi}\int d^3x \left(V''_\tLO\left[\phi_\tLO(x)\right]\right)^{3/2}$}
\begin{align}
	\frac{S_\tNLO^H[\phi_\tLO]}{\delta \phi(x)}=-\frac{1}{8\pi}\sqrt{V''_\tLO\left[\phi_\tLO(x)\right]}V'''_\tLO\left[\phi_\tLO(x)\right].
\end{align}

Putting everything together, and performing the angular integration, one finds
\begin{align}
	\label{eq:BounceCorrection3}
	\phi_\tNLO(x)&=\frac{V_\tLO'''(0)}{8\pi \sqrt{V_\tLO''(0)}}+\nonumber
	\\&\frac{1}{16\pi}\int d \rho \frac{\rho}{r}V_\tLO'''[\phi_\tLO(\rho)]\left(e^{-\sqrt{V''_\tLO[\phi_\tLO(\rho)]}|r-\rho|}-e^{-\sqrt{V_\tLO''[\phi_\tLO(\rho)]}|r+\rho|}\right),
	\\& r=|\vec{x}|.\nonumber
\end{align}
Given the tree-level bounce solution, the above integral is easy to evaluate numerically.
Note that $\phi_\tNLO(x)$ is in general complex valued. The same holds when evaluating the Higgs determinant with the derivative expansion. Hence, if comparing $\phi_\tNLO(x)$ with $\phi_\tLO(x)$ the real part should presumably be used.

In summary, it is in principle to get a handle on two-loop corrections to the nucleation rate by using equation \ref{eq:BounceActionHigherOrder} together with a derivative expansion for $S_\tNNLO[\phi]$ and $\phi_\tNLO(x)$. Though, keep in mind that this is just a rough estimate.

\section{Angular-Momentum Cutoff}\label{app:AngCutoff}

Let's look at this large $l$ behaviour. Following \cite{Langer:1937qr,Dunne:2005te,Dunne:2005rt}, the equation is first rewritten by defining $\psi^l=e^{-x/2} \Psi^l(x)$, where $r=e^x$. This gives
\begin{align*}
	\frac{d^2}{dx^2}\Psi^l(x)=Q(x)\Psi^l(x),~Q(x)=\left(\overline{l}^2+W(x) e^{2x}\right),~\overline{l}\equiv l+\frac{1}{2}.
\end{align*}

Using the WKB approximation~\cite{Kramers:1926njj,Wentzel:1926aor} and imposing the boundary conditions gives
\begin{align*}
	\log \frac{\Psi^l(\infty)}{\Psi^l_{FV}(\infty)}=\frac{1}{2\overline{l}}\int^{\infty}_0 dr r\left[W(r)-W(\infty)\right]+\mathcal{O}\left(\frac{1}{\overline{l}^3}\right).
\end{align*}

To find the full determinant we need to perform the sum
\begin{align*}
	\sum_{l=L}^\infty \frac{(2l+1)}{(l+\frac{1}{2})}.
\end{align*}
This sum is divergent. If we were using an angular cut-off to regularize the theory, this divergence would be cancelled by counter-terms from the leading-order Lagrangian. An angular cut-off would here restrict $l\in\left[0,L\right]$.
There are a number of other ways to regularize the sum. With zeta-function regularization the sum turns into~\cite{Dunne:2006ct}
\begin{align*}
	\left[\sum_{l=L}^\infty \frac{(2l+1)}{(l+\frac{1}{2})}\right]\rightarrow-2 L.
\end{align*}

Or equivalently
\begin{align}
	\sum_{l=0}^{\infty} (2l+1)\log\frac{\det \mathcal{M}^l}{\det \mathcal{M}^l_\text{FV}}=\sum_{l=0}^\infty(2l+1)\left[\log \frac{\Psi^l(\infty)}{\Psi^l_{FV}(\infty)}-\frac{1}{2(l+1/2)}\int^{\infty}_0 dr r\left[W(r)-W(\infty)\right]\right].
\end{align}

As a side-note, it's also possible to calculate everything in dimensional regularization. In that case the WKB approximation gives (up to $\mathcal{O}(\epsilon)$ terms)
\begin{align}
	\left[\sum_{l=L}^\infty \frac{(2l+1)}{(l+\frac{1}{2})}\right]\rightarrow \left[\sum_{l=L}^\infty \frac{2\Gamma(l-2\epsilon+1)}{\Gamma(l+1)\Gamma(2-2\epsilon)}\right]=-2L+\mathcal{O}(\epsilon).
\end{align}

\subsection{The Gelfand-Yaglom Theorem for Multiple Fields}
\label{section:MixedOperators}
Functional determinants for the Standard Model, barring Goldstone-Vector mixing, involve one-dimensional operators. For popular Standard-Model extensions, like adding a real/complex singlet~\cite{Huang:2018aja,Dev:2019njv,Niemi:2021qvp,Schicho:2021gca,Gould:2019qek} or real-triplet extensions~\cite{Niemi:2018asa,Bell:2020hnr,Niemi:2020hto}, there is usually a mixing between different fields. This necessitates calculating functional determinants of matrix operators. 

As an example, take a two-scalar theory with bounce equation
\begin{align}
	\label{eq:TwoScalarBounceEquation}
	& \partial^2 \phi(r)+\frac{2}{r}\partial \phi(r)=\partial_\phi V(\phi,\sigma),
	\\& \partial^2 \sigma(r)+\frac{2}{r}\partial \sigma(r)=\partial_\sigma V(\phi,\sigma),
	\\& \partial \phi\left.\right|_{r=0}=\partial \sigma\left.\right|_{r=0}=0,
	\\&\phi\left.\right|_{r=\infty}=\phi^{FV} \&~\sigma\left.\right|_{r=\infty}=\sigma^{FV}.
\end{align}

Where $\phi^{FV}$ and $\sigma^{FV}$ denote the false-vacuum solutions. The eigenvalue equation for the corresponding fluctuation operator is (for specific partial-wave $l$)
\begin{align}
	\label{eq:MixedHiggsOperator}
	& -\partial^2 \psi^l_1-2/r \partial \psi^l_1+\frac{l(l+1)}{r^2} \psi^l_1+V_{11} \psi^l_1+V_{12}\psi^l_2=\lambda  \psi^l_1,
	\\& -\partial^2 \psi_2^l-2/r \partial \psi^l_2+\frac{l(l+1)}{r^2} \psi^l_2+V_{22} \psi^l_2+V_{12}\psi^l_1=\lambda  \psi^l_2.
\end{align}

Or written more compactly (taking $i=1,2 \rightarrow \phi,\sigma$)
\begin{align}
	\label{eq:TwoScalarDeterminant}
	\mathcal{L}^l_{ij}\psi^l_j=\lambda \psi^l_i,~\mathcal{L}^l_{ij}=-\delta_{ij}\left(\partial^2+2/r\partial-l(l+1)/r^2\right)+V_{ij}.
\end{align}
Also, $V_{ij}=\partial_i \partial_j V,~i \in \left\lbrace\phi,\sigma\right\rbrace$, where $V$ is the leading-order potential.

Now for the functional determinant. Call the corresponding false-vacuum operator $\mathcal{L}^l_{FV}$. To use equation \ref{eq:GeneralizedGelfandYaglom}, we impose $\psi^l \sim r^l$ for both the bounce and false-vacuum determinant. Then use the false-vacuum solutions at $\lambda=0$ to define
\begin{align}
	T^l_\phi(r)=\frac{\psi^l_\phi(r)}{\psi^l_{\phi,FV}(r)}~\&~ T^l_\sigma(r)=\frac{\psi^l_\sigma(r)}{\psi^l_{\sigma,FV}(r)}.
\end{align}
Equation \ref{eq:MixedHiggsOperator} (for $\lambda=0$) can then be rewritten\footnote{The operator $\mathcal{M}^l_{ij}$ depends on the false-vacuum solutions. In general there are terms proportional to spherical Bessel functions.}
\begin{align}
	\mathcal{M}^l_{ij} T^l_{j}=0.
\end{align}
With boundary conditions $T^l_\phi(0)=T^l_\sigma(0)=1$ and $\partial T^l_\phi(0)=\partial T^l_\sigma(0)=0$. This corresponds to (see equation \ref{eq:BoundaryValue})

\begin{align*}
	M= \begin{pmatrix}
		0 &0 &0 &0\\
		0 & 0 &0 &0\\
		0 & 0 & 1 & 0\\
		0 & 0 & 0 &1
	\end{pmatrix},~
	N= \begin{pmatrix}
		1 &0 &0 &0\\
		0 &1 &0 &0\\
		0 & 0 & 0 & 0\\
		0 & 0 & 0 &0
	\end{pmatrix}.
	\\& 
\end{align*}

And the functional determinant is
\begin{align}
	\Det \mathcal{M}^l=\Det\left[M+N Y^l(\infty)\right] =y^l_{1;1}(\infty)y^l_{2;2}(\infty)-y^l_{1;2}(\infty)y^l_{2;1}(\infty),
\end{align}
with the fundamental solutions defined in equation \ref{eq:FundamentalSolutions}.

Mirroring the one-dimensional operator case, the ratio of determinants can be calculated numerically for any $l$ up to some large angular cut-off $L$. The idea is, as before, to solve the eigenvalue equation in powers of $L^{-1}$. Though, the WKB approach is tricky for the mixed-operator case. 

To save ink I'll use a trick. Consider how renormalization comes in with an angular cut-off. Terms that diverge for large $l$ must be the same whether we use the effective potential (derivative expansion) or the Gelfand-Yaglom method. But for the effective potential we treat all $V_{ij}$ elements as constants. Therefore, the contribution to the effective action is for large $l$(the $-2$ comes from the definition of the effective action)
\begin{align*}
	(-2)\delta S_\text{eff,deriv}^l=\frac{1}{2\overline{l}}\int dr r A_1(r)+\frac{1}{2\overline{l}}\int dr r A_2(r),
\end{align*}
where $A_1$ and $A_2$ are the eigenvalues of $V_{ij}$, and as before $\overline{l}=l+1/2$. The sum 
$$\sum_{l=0}^L (2l+1)\delta S_\text{eff,deriv}^l$$ behaves as $\sim L$, and needs to be made finite by counter-terms. Yet these counter-terms must cancel the same divergences regardless if we use the effective potential or the Gelfand-Yaglom method. Thus for large $l$ it must be that 
\begin{align}
	\log \frac{\Det \mathcal{L}^l}{\Det \mathcal{L}^l_{FV}}=\frac{1}{2\overline{l}}\int dr r A_1(r)+\frac{1}{2\overline{l}}\int dr r A_2(r)+\mathcal{O}\left(\frac{1}{l^3}\right).
\end{align}
This result is generic. I have also confirmed this large $l$ behaviour numerically for a variety of models.

In the present case
\begin{align*}
	A_{1,2}=\frac{1}{2}\left(V_{\phi \phi}+V_{\sigma \sigma}\pm\sqrt{(V_{\phi \phi}-V_{\sigma \sigma})^2+4 V_{\phi \sigma}^2}\right).
\end{align*}

All in all, the numerical method using the Gelfand-Yaglom method is fast, accurate, and doesn't suffer from the same pathologies as the derivative expansion.

\subsection{Vector-Goldstone Mixing}
\label{subsec:VectorGoldstoneMixing}
Fluctuations are much simpler around a constant background Higgs field rather than around a spatially varying one. For one, vector fields generally mix with Goldstone fields. The mixing involves terms of the form
\begin{align}
	\Phi_G(x)\phi(x)\partial_\mu A_\mu(x),~\Phi_G(x) \partial_\mu \phi(x) A_\mu(x),~\partial_\mu  \Phi_G(x) A^\mu(x)\phi(x).
\end{align}
Given a gauge-choice, the equations of motion can be separated in partial waves using the ansatz~\cite{Andreassen:2017rzq,Endo:2017tsz,Isidori:2001bm,Baacke:1999sc}

\begin{align}
	&A_\mu(x)=\sum_{l=0}^\infty\left[a_S(r) n_\mu +a_L(r) \frac{r}{\sqrt{l(l+1)}}\partial_\mu +a_T(r)\epsilon_{\mu\nu\alpha}x_\nu \partial_\alpha\right]Y_{lm}(\phi,\theta),
	\\& \Phi_G(x)=a_G(r) Y_{lm}(\phi,\theta),~
	n_\mu=\frac{x_\mu}{r}.
\end{align}

The determinants then decouple for each partial wave $l$. See \cite{Isidori:2001bm,Baacke:1999sc} for the details, the relevant formulas are also summarized in appendix \ref{app:VectorGoldstoneMixing}. The transverse fluctuation, $a_T$, is independent from the others. Hence the remaining determinant is written in the  $(a_S,~a_L,~a_G)$ basis. In Fermi gauges, assuming a gauge-fixing term $\frac{1}{2\xi}\left(\partial_\mu A_\mu\right)^2$ and a vector-boson mass $m_A=e\phi$, the fluctuation matrix is
\begin{align}
	&\begin{pmatrix}
		-\nabla^2_l+\frac{2}{r^2}+e^2 \phi^2& -2\frac{L}{r^2}& e \phi'-e \phi \partial \\
		-2\frac{L}{r^2}& -\nabla^2_l+e^2\phi^2&-e \phi \frac{L}{r}\\
		e\phi \frac{2}{r}+2 e \phi'+e \phi \partial & -e\phi\frac{L}{r}&-\nabla^2_l+G
	\end{pmatrix}
	\\& +\left(1-\frac{1}{\xi}\right)\begin{pmatrix}
		\partial^2+\frac{2}{r}\partial-\frac{2}{r^2}&-\frac{L}{r}\left(\partial-\frac{1}{r}\right)&0 \\
		\frac{L}{r}\left(\frac{2}{r}+\partial\right)& -\frac{L}{r^2}&0\\
		0 & 0 &0
	\end{pmatrix}
\end{align}
I here use the shorthand notation $G(x)=\frac{1}{\phi(x)}V'\left[\phi(x)\right]$.
While for $R_\xi$ gauges the fluctuation matrix is of the form
\begin{align}
	&\begin{pmatrix}
		-\nabla^2_l+\frac{2}{r^2}+e^2 \phi^2& -2\frac{L}{r^2}& e \phi'-e \phi \partial \\
		-2\frac{L}{r^2}& -\nabla^2_l+e^2\phi^2&-e \phi \frac{L}{r}\\
		e\phi \frac{2}{r}+2 e \phi'+e \phi \partial & -e\phi\frac{L}{r^2}&-\nabla^2_l+G
	\end{pmatrix}
	\\&+\left(1-\frac{1}{\xi}\right)\begin{pmatrix}
		\partial^2+\frac{2}{r}\partial-\frac{2}{r^2}&-\frac{L}{r}\left(\partial-\frac{1}{r}\right)&0 \\
		\frac{L}{r}\left(\frac{2}{r}+\partial\right)& -\frac{L}{r^2}&0\\
		0 & 0 &0
	\end{pmatrix}.
	\\& +\frac{1}{\xi} \begin{pmatrix}
		0&0&e\phi'+e\phi\partial \\
		0&0&e\phi\frac{L}{r}\\
		-e\phi \partial-e\phi\frac{2}{r} & e \phi \frac{L}{r} &e^2 \phi^2.
	\end{pmatrix}
\end{align}
Where $\nabla^2_l=\partial^2+\frac{2}{r}\partial-\frac{L^2}{r^2}$ and $L^2=l(l+1)$. For $R_\xi$ gauges there is also a Ghost determinant associated to the operator
\begin{align}
	\mathcal{L}_\text{gh}=\left(-\nabla^2+e^2 \phi^2 \frac{1}{\xi}\right).
\end{align}
For both of these gauge choices the transverse determinant decouples. Also, the determinant is gauge invariant so-long as $\phi$ is the leading-order bounce~\cite{Baacke:1999sc,Endo:2017gal}.

To evaluate the mixed determinant numerically it is felicitous to choose $R_\xi$ gauges with $\xi=1$, because the fluctuation matrix is then rather simple~\cite{Isidori:2001bm,Andreassen:2017rzq}
\begin{align}
	&\begin{pmatrix}
		-\nabla^2_l+\frac{2}{r^2}+e^2 \phi^2& -2\frac{L}{r^2}& 2e \phi' \\
		-2\frac{L}{r^2}& -\nabla^2_l+e^2\phi^2&0\\
		2 e \phi' & 0&-\nabla^2_l+G+e^2\phi^2
	\end{pmatrix}
\end{align}

Further, it is useful to make a change basis with the matrix
\begin{align}
	\mathcal{O}=\frac{1}{\sqrt{2l+1}}\begin{pmatrix}
		\sqrt{l}&-\sqrt{l+1}&0\\
		\sqrt{l+1}&\sqrt{l}&0\\
		0&0&1
	\end{pmatrix}
\end{align}

The fluctuation determinant in the new basis is 
\begin{align}
	\mathcal{M}^l=&\begin{pmatrix}
		-\nabla^{-}+e^2\phi^2&0& 2 e\sqrt{\frac{l}{2l+1}}\phi'\\
		0& -\nabla^{+}+e^2 \phi^2 &-2 e\sqrt{\frac{l+1}{2l+1}}\phi' \\
		2 e\sqrt{\frac{l}{2l+1}}\phi'&-2 e \sqrt{\frac{l+1}{2l+1}}\phi'&-\nabla^2_l+G+e^2\phi^2
	\end{pmatrix},
\end{align}
where $\nabla^{\pm}=\partial^2+2/r\partial-\frac{(l\pm1)(l+1\pm1)}{r^2}$.

The Gelfand-Yaglom theorem described in Section \ref{sec:NumericalDeterminants} can then be used to find the functional determinant for each partial wave $l$. The case $l=0$ requires special attention and is discussed in Section \ref{subsec:VectorGoldstoneMixingZeroMode}.

A couple of observations is in order. First, in the chosen gauge the ghosts contribution is minus two times the transverse contribution. Second, the arguments from Section \ref{section:MixedOperators} implies that for large $l$
\begin{align}
	&\log \frac{\Det \mathcal{M}^l}{\Det \mathcal{M}^l_{FV}}=\frac{1}{2(l+\frac{1}{2}+1)}\int dr r e^2 \phi(r)^2+\frac{1}{2(l+\frac{1}{2}-1)}\int dr r e^2 \phi(r)^2
	\\&+\frac{1}{2(l+\frac{1}{2})}\int dr r \left[e^2 \phi(r)^2+G(r)\right]+\mathcal{O}\left(\frac{1}{l^3}\right)
	\\&=\frac{3}{2(l+\frac{1}{2})}\int dr r e^2 \phi(r)^2+\frac{1}{2(l+\frac{1}{2})}\int dr r G(r)+\mathcal{O}\left(\frac{1}{l^3}\right).
\end{align}
That is, for large $l$ the functional determinants separate into three times the pure transverse result plus the pure Goldstone result. For some models this holds exactly for all $l$ (besides $l=0$)~\cite{Andreassen:2017rzq}. So if numerical resources are limited, one can approximate the determinant of $\mathcal{M}^l$ as $\left[\Det(-\nabla^2_l+e^2\phi^2)\right]^3\times\Det(-\nabla^2_l+G)$. The error of this approximation is on the percent level if $l 	\gtrsim 50$ and $e\sim 1$.

\section{Vector-Goldstone Mixing}\label{app:VectorGoldstoneMixing}

To decouple the Vector-Goldstone determinants we need various combinations of derivatives acting on

\begin{align}
&A_\mu(x)=\sum_{l=0}^\infty\left[a_S(r) n_\mu +a_L(r) \frac{r}{\sqrt{l(l+1)}}\partial_\mu +a_T(r)\epsilon_{\mu\nu\alpha}x_\nu \partial_\alpha\right]Y_{lm}(\phi,\theta),
\\& G(x)=a_G(r) Y_{lm}(\phi,\theta),
\\& n_\mu=\frac{x_\mu}{r}.
\end{align}

The relevant identities are
\begin{align}
&\nabla^2_l\equiv \partial^2+\frac{2}{r}\partial-\frac{L^2}{r^2},~L^2=l(l+1),
\\& \partial_\mu \partial^\mu \left(a_S n_i Y_{lm}\right)=n_i Y_{lm} \left(\nabla^2_l-\frac{2}{r^2}\right) a_S+a_S\frac{2}{r}\partial_i Y_{lm},
\\&\partial_\mu \partial^\mu\left( a_L r\partial_i Y_{lm}\right)=r \partial_i Y_{lm}  \nabla^2_l a_L+2\frac{ L^2}{r}n_i Y_{lm},
\\& \partial_\mu \partial^\mu\left(a_T \epsilon_{ijk} x_j \partial_k Y_{lm}\right)=(\nabla^2_l a_T) \epsilon_{ijk} x_k \partial_k Y_l,
\\& \partial_i\left(a_S n_i Y_{lm}\right)=a'_S Y_{lm}+a_S Y_{lm}\frac{2}{r},
\\&\partial_i\left(a_L r \partial_i Y_{lm}\right)=-\frac{L^2}{r} a_L Y_{lm},
\\& \epsilon_{ijk}\partial_i\left( a_T x_j \partial_k Y_{lm}\right)=0,
\\& \partial_a \partial_i \left(a_S n_i Y_{lm}\right)=n_a Y_{lm}\left(a_S''+\frac{2}{r}a_S'-\frac{2}{r^2}a_S\right)+\partial_a Y_{lm}\left( a_S'+\frac{2}{r}a_S\right),
\\& \partial_a \partial_i\left(a_L r\partial_i Y_{lm}\right)=\frac{L^2}{r^2}n_a a_L Y_{lm}-\frac{L^2}{r}a_L \partial_a Y_{lm}-\frac{L^2}{r}a' n_a Y,
\\& n_i  a_L r \partial_i Y_{lm}=0,
\\& \partial_\mu \partial^\mu \left(a_G Y_{lm}\right)=Y_{lm}\nabla^2_l a_G,
\\& \partial_i \left(a_G Y_{lm}\right)=n_i a_G' Y_{lm}+a_G \partial_i Y_{lm}.
\end{align}

\section{General Zero-mode Contribution}\label{app:ZeroMode}
Consider an operator of the form

\begin{align}
&\mathcal{L}^l_{ij}\equiv-(\partial^2+2/r \partial-l(l+1)/r^2)\delta_{ij}+V_{ij},
\\& V_{ij}=\left.\frac{\partial^2}{\partial \phi_i \partial \phi_j}V(\vec{\phi})\right\vert_{\vec{\phi}=\vec{\phi}^{FV}}.
\end{align}
Where I assume $n$ tunneling scalars: $i=1,\ldots,n$.

For $l=1$ there's a normalizable zero mode of the form $\vec{\psi}(x)=\partial \vec{\phi}(x).$ This means that $\Det ~\mathcal{L}^1=0$. In terms of fundamental solutions this is the statement (assuming the boundary conditions $\vec{\psi}(0)=\vec{0}$)

\begin{align}
\Det \left( y_{n+1}~\ldots~y_{2n}\right)=0,
\end{align}
where $y_{n+1}$ is a short-hand notation for $y_{n+1}=(y_{1;n+1},~\ldots,y_{n;n+1})^t$.
In addition, equation \ref{eq:FundamentalSolutionMatrix} implies that

\begin{align}
\label{eq:ZeroModeRelation}
\partial \psi_i(x)=\sum_{a=1}^n \partial^2\psi_a(0)y_{i;n+a}(x).
\end{align}

To remove the zero-eigenvalue I follow the procedure set out in Section \ref{sec:ZeroModes}. First modify the eigenvalue equation for the fundamental solutions to
\begin{align}
\mathcal{L}^\epsilon_{ij}y_{j;a}^\epsilon \equiv \left[\mathcal{L}_{ij}+\epsilon \delta_{ij}\right] y^\epsilon_{j;a}=0.
\end{align}
Second, define the inner product $ \left\langle f g\right\rangle\equiv \sum_i \int d^3x  f_i(x) g_i^{\star}(x)$. Then we can use the relations $0=\left\langle \partial\phi \mathcal{L}^\epsilon y^\epsilon_{n+1}\right\rangle=\ldots=\left\langle \partial\phi \mathcal{L}^\epsilon y^\epsilon_{2n}\right\rangle$ to deduce

\begin{align}
\label{eq:ZeroModeRelationRegularized}
& 8\pi R^2 \sum_{i=1}^n \partial^2 \phi_i(R) y_{i;n+1}(R)=-\epsilon  \left\langle \partial\phi y^\epsilon_{n+1}\right\rangle
\\& \hspace*{3.5cm}\vdots
\\&  8\pi R^2 \sum_{i=1}^n \partial^2 \phi_i(R) y_{i;2n}(R)=-\epsilon  \left\langle \partial\phi y^\epsilon_{2n}\right\rangle.
\end{align}

Then using $\Det' \mathcal{L}^1=\lim_{\epsilon\rightarrow 0}\frac{\Det \left( y^\epsilon_{n+1}~\ldots~y^\epsilon_{2n}\right)}{\epsilon}$, equations \ref{eq:ZeroModeRelation} and \ref{eq:ZeroModeRelationRegularized}, we find\footnote{In deriving this formula I threw away terms that vanish when normalizing with the false-vacuum determinant. I also assume that all solutions are either identically zero, or behave as $\partial\phi_i(R)\sim e^{-a_i R}/R$ for large $R$. In particular, the formulas need to be modified for scaleless potentials, for example $V(\phi)\sim -\lambda \phi^4$. }

\begin{align}
\label{eq:GeneralZeroModeFormula}
\Det \mathcal{L}'=\frac{\left\langle \partial\phi \partial \phi\right\rangle~M_{11}}{-8\pi R^2 \partial^2\phi_1(R) \partial^2 \phi_1(0)},
\end{align}
where $M_{11}$ is the $(1,1)$ minor of the matrix $\left( y_{n+1}~\ldots~y_{2n}\right)$, to wit 

\begin{align}
M_{11}=\Det 
 \begin{pmatrix}
  y_{2,n+2} & y_{2,n+3} & \cdots & y_{2,2n} \\
  y_{3,n+2} & y_{3,n+3} & \cdots & y_{3,n+2} \\
  \vdots  & \vdots  & \ddots & \vdots  \\
  y_{n,n+2} & y_{n,n+2} & \cdots & y_{n,2n}
 \end{pmatrix}.
\end{align}
Also note that $\left\langle \partial\phi \partial \phi\right\rangle=3 S_3$ where $S_3$ is the leading-order action evaluated on the bounce. The determinant in equation \ref{eq:GeneralZeroModeFormula} should, of-course, be normalized by the false-vacuum determinant $\Det \mathcal{L}^1_{FV}=\prod_{i=1}^{n} y^{FV}_{i;n+i}(R)$.

The two-field case given in equation \ref{eq:ZeroModeFormulaTwoFields} is a special case of equation \ref{eq:GeneralZeroModeFormula}. Similar formulas for Goldstone-boson zero-modes can be derived for any n. However, this has to be done on a case-by-case basis since the zero-eigenfunctions in general doesn't vanish exponentially as $R\rightarrow \infty$. The above procedure still works, but the formulas are more complicated.

Equation \ref{eq:GeneralZeroModeFormula} also holds when some of the components vanish identically; for example a Two-Higgs doublet model where the zero-mode is of the form $\partial\phi(x)_i=\left(\partial v_1(x)~\partial v_2(x)~\vec{O}\right)^t$. The only requirement is that the $\partial^2 \phi_1(x)$ field doesn't identically vanish.

 \bibliographystyle{utphys}

{\linespread{0.6}\selectfont\bibliography{Bibliography}}
\end{document}